\documentclass[twocolumn,amsmath,amssymb,longbibliography,noeprint,nofootinbib]{revtex4-1}

\usepackage{bbm}
\usepackage[caption=false]{subfig}
\usepackage{overpic}
\usepackage{amsmath}
\usepackage{epsfig,psfrag}
\usepackage{pstricks}
\usepackage{dcolumn}
\usepackage{bm}
\usepackage{graphicx}
\usepackage{color}
\usepackage{version}
\usepackage{hyperref}
\usepackage{blindtext}
\definecolor{darkblue}{RGB}{46,48,146}
\hypersetup{
   colorlinks=true,
   citecolor={darkblue},
   linkcolor={darkblue},
   urlcolor={darkblue},
   pdfhighlight=/I
}

\usepackage{times}
\usepackage{xcolor}
\usepackage{multirow}

\newcommand{\ket}[1]{\left|#1\right>}

\begin{document}

\title{Quantum Computing with Majorana Fermion Codes}
\author{Daniel Litinski and Felix von Oppen}
\affiliation{Dahlem Center for Complex Quantum Systems and Fachbereich Physik, Freie Universit\"at Berlin, Arnimallee 14, 14195 Berlin, Germany}

\begin{abstract}


We establish a unified framework for Majorana-based fault-tolerant quantum computation with Majorana surface codes and Majorana color codes. All logical Clifford gates are implemented with zero time overhead. This is done by introducing a protocol for Pauli product measurements with tetrons and hexons which only requires local 4-Majorana parity measurements. An analogous protocol is used in the fault-tolerant setting, where tetrons and hexons are replaced by Majorana surface code patches, and parity measurements are replaced by lattice surgery, still only requiring local few-Majorana parity measurements. To this end, we discuss twist defects in Majorana fermion surface codes and adapt the technique of twist-based lattice surgery to fermionic codes. Moreover, we propose a new family of codes that we refer to as Majorana color codes, which are obtained by concatenating Majorana surface codes with small Majorana fermion codes. Majorana surface and color codes can be used to decrease the space overhead and stabilizer weight compared to their bosonic counterparts.


\end{abstract}

\maketitle

\section{Introduction}\label{sec:intro}

In an effort to construct robust qubits for quantum computation, Majorana zero modes (Majoranas) in topological superconductors~\cite{Kitaev2001,Kitaev2003,Alicea2012,Beenakker2013,Lutchyn2017,Aguado2017} are currently being explored as potential building blocks of topological qubits. Even though no such topological qubit has been built to date, a range of precursor experiments exists~\cite{Mourik2012,Albrecht2016a,Deng2016,Zhang2017}, providing various signatures of Majorana zero modes. While it remains uncertain whether Majorana-based qubits will offer higher coherence times compared to other solid-state qubits, Majorana-based qubits have one distinct feature that sets them apart from non-topological qubits. Due to the nonabelian statistics of Majoranas, these qubits can be measured in all three Pauli bases $X$, $Y$, and $Z$, and not just in a single computational basis as is the case with conventional qubits. As a consequence, Majorana-based qubits implement robust single-qubit Clifford gates with zero time overhead. These gates are products of Hadamard gates $H$ and phase gates $S$, and are equivalent to the operations implemented by braiding Majoranas.

The coherence times of Majorana-based qubits are predicted to be long, but still limited~\cite{Aasen2016,Knapp2018}. Thus, a quantum error-correcting code is necessary for large-scale quantum computation with arbitrarily long qubit survival times. Quantum error correction combines many physical qubits into more error-resilient logical qubits~\cite{TerhalRMP}. Errors are detected and corrected by measuring certain stabilizer operators. Two-dimensional (2D) topological codes are of particular interest, since their local stabilizers are compatible with the constraints of solid-state architectures. 

Topological codes come as bosonic or fermionic codes. Bosonic codes are defined on a 2D lattice where vertices correspond to qubits and faces define stabilizers, products of Pauli operators with support on all qubits of the face. 
Fermionic codes are defined on a lattice where vertices correspond to Majoranas and stabilizers are products of all Majorana operators of a face. A bosonic code maps onto a fermionic code by replacing each qubit with four Majoranas in a fixed parity sector~\cite{Bravyi2010}, but not all fermionic codes can be straightforwardly mapped back onto a bosonic code. This implies that Majorana-based qubits admit a wider range of topological codes than conventional qubits. 

One practical problem is that fault-tolerant quantum computing requires logical operations on encoded qubits, which may be entirely different from operations on physical qubits. Indeed, various recent proposals for Majorana-based implementations of codes exist, based on, e.g., bosonic surface codes~\cite{Vijay2016,Landau2016,Plugge2016,Li2016}, bosonic color codes~\cite{Litinski2017,Litinski2017a}, fermionic surface codes~\cite{Bravyi2010,Vijay2015,Li2017}, and small Majorana fermion codes~\cite{Vijay2017,Hastings2017}, each featuring individual protocols for a universal set of logical gates. Moreover, except for twist-based encodings in surface codes~\cite{Bombin2010,Hastings2015,Litinski2017b} and transversal gates of bosonic color codes~\cite{Bombin2006,Litinski2017,Litinski2017a}, logical single-qubit Clifford gates require a series of code operations in all aforementioned proposals, and the Majoranas' advantage of zero-overhead single-qubit Clifford gates is lost.

According to the Gottesman-Knill theorem~\cite{Gottesman1999}, not only single-qubit Clifford gates, but all Clifford gates including the two-qubit controlled-NOT (CNOT) gate can be tracked using a classical computer. Therefore, all Clifford gates can, in principle, be implemented with zero time overhead. Operations on logical qubits need to be appropriately designed in order to take full advantage of the Gottesman-Knill theorem.


\renewcommand{\arraystretch}{1.2}

\begin{table*}[t]

\begin{tabular}{c|c|c|c|c|c|c|}
\cline{1-7}
\multicolumn{1}{|c|}{\raisebox{-.1em}{\rotatebox[origin=c]{90}{Order}}} & \begin{tabular}{c} Code family \\ (with minimal-overhead Clifford gates) \end{tabular} & \begin{tabular}{c} Majorana overhead \\ for code distance $d$ \end{tabular} & Building blocks & \begin{tabular}{c} Maximum \\ stabilizer weight \end{tabular} & \begin{tabular}{c} Bulk \\ weights \end{tabular} & \begin{tabular}{c} Max. stabilizer wt. \\ for lattice surgery \end{tabular} \\ \cline{1-7}
\multicolumn{1}{|c|}{} & \multicolumn{6}{c|}{\begin{tabular}{c} surface codes \\ (non-overlapping 2D-local topological codes) \end{tabular}}  \\ \cline{1-7} 
\multicolumn{1}{|c|}{\multirow{5}{*}{1}} & 4.6.12 Majorana surface code & $12d^2 + \mathcal{O}(d)$ & dodecons & 6 & 4, 6 & 6 \\
\multicolumn{1}{|c|}{} & bosonic subsystem surface code\hyperlink{ref1}{$^{1}$} & $12d^2 + \mathcal{O}(d)$ & qubits* & 6 & 6 & 10 \\
\multicolumn{1}{|c|}{} & 6.6.6 Majorana surface code\hyperlink{ref2}{$^{2}$} & $6d^2 + \mathcal{O}(d)$ & hexons / octons& 6 & 6 & 6 \\
\multicolumn{1}{|c|}{} & 4.8.8 Majorana surface code\hyperlink{ref3}{$^{3}$} & $4d^2 + \mathcal{O}(d)$ & tetrons / hexons & 8 & 8 & 8 \\
\multicolumn{1}{|c|}{} & bosonic surface code\hyperlink{ref4}{$^{4}$} & $4d^2 + \mathcal{O}(d)$ & qubits* & 8 & 8 & 10 \\ \cline{1-7}
\multicolumn{1}{|c|}{} & \multicolumn{6}{c|}{\begin{tabular}{c} color codes \\ (multiple layers of surface codes obtained by concatenation) \end{tabular}}  \\ \cline{1-7}
\multicolumn{1}{|c|}{\multirow{4}{*}{2}}& 4.8.8 $\left([[6,2,2]]_m\right)$ Majorana color code& $3d^2 + \mathcal{O}(d)$ & hexons & 8 & 8 & 8 \\
\multicolumn{1}{|c|}{} & bosonic 6.6.6 color code\hyperlink{ref5}{$^{5}$} & $3d^2 + \mathcal{O}(d)$ & qubits* & 12 & 12 & 12 \\
\multicolumn{1}{|c|}{} & 6.6.6 $([[20,4,4]]^{\text{\hyperlink{ref6}{$^{6}$}}}_m)$ Majorana color code& $2.5d^2 + \mathcal{O}(d)$ & decons & 12 & 10, 12 & 12 \\
\multicolumn{1}{|c|}{} & bosonic 4.8.8 color code\hyperlink{ref5}{$^{5}$} & $2d^2 + \mathcal{O}(d)$ & qubits* & 16 & 8, 16 & 12 \\ \cline{1-1}
\multicolumn{1}{|c|}{\multirow{2}{*}{3}} & 4.8.8  $([[8,3,2]]_m~)$ Majorana color code & $\frac{8}{3} d^2 + \mathcal{O}(d)$ & octons & 10 & 8, 10 & 10\\  
\multicolumn{1}{|c|}{}  & 4.8.8  $([[16,3,4]]_m^{\text{\hyperlink{ref6}{$^{6}$}}})$ Majorana color code & $\frac{4}{3}d^2 + \mathcal{O}(d)$ & octons & 16 & 8, 16 & 12  \\ \cline{1-1}
\multicolumn{1}{|c|}{4} & 4.8.8 $([[20,4,4]]^{\text{\hyperlink{ref6}{$^{6}$}}}_m)$ Majorana color code & $1.25d^2 + \mathcal{O}(d)$ & decons & 16 & 10, 16 & 12 \\ \cline{1-7}
\multicolumn{1}{|c|}{\multirow{3}{*}{$k$}} & \begin{tabular}{c}4.8.8 Majorana surface code \\ concatenated with $[[n_m,k,d_m]]_m$ \end{tabular}& $\frac{\displaystyle 4 n_m}{\displaystyle kd_m^2} d^2 + \mathcal{O}(d)$ &  & at least $4d_m$ & &  \\
\multicolumn{1}{|c|}{} & \begin{tabular}{c} bosonic surface code \\ concatenated with $[[n,k,d_b]]$ \end{tabular}& $\frac{\displaystyle 4n}{\displaystyle kd_b^2} d^2 + \mathcal{O}(d)$ & qubits*  & at least $8d_b$ & &  \\ \cline{1-7} 
\multicolumn{7}{r}{first mentioned in: \hypertarget{ref1}{1)} \cite{Bravyi2013}, \hypertarget{ref2}{2)} \cite{Vijay2015}, \hypertarget{ref3}{3)} \cite{Landau2016}, \hypertarget{ref4}{4)} \cite{Bravyi1998},  \hypertarget{ref5}{5)}~\cite{Bombin2006},  \hypertarget{ref6}{6)}~\cite{Hastings2017} (albeit without minimal-overhead Clifford gates)} \\ 
\multicolumn{7}{r}{*for Majorana-based hardware: tetrons} 

\end{tabular}

\caption{Comparison of different two-dimensional topological codes with respect to the following characteristics: number of Majoranas necessary to encode a logical qubit with code distance $d$, proposed number of Majoranas in a fixed parity sector used as a building block (tetrons/hexons/octons/decons/dodecons refer to 4/6/8/10/12 Majoranas in a fixed parity sector), maximum Majorana weight of the stabilizers that need to be measured, stabilizer weights in the bulk of the code (excluding the boundaries), and the maximum stabilizer weight for lattice surgery operations. For Majorana color codes, the $[[n_m,k,d_m]]$ code in parentheses is the code that is concatenated with the corresponding surface code in order to obtain the color code.  The order of each code indicates how many layers of Majorana surface codes it corresponds to. All codes listed in the table implement logical Clifford gates with zero time overhead.}
\label{tab:results}
\end{table*}

\textbf{Overview and main results.} In this work, we establish a unified framework for Majorana-based quantum computation with bosonic and fermionic surface and color codes, in all of which logical Clifford gates, including CNOT gates, are implemented with zero time overhead. To this end, we first discuss the construction of Majorana surface code patches in Sec.~\ref{sec:surfacecodes}. These surface code patches are essentially fault-tolerant versions of tetrons and hexons which can correct a certain number of errors. While tetrons and hexons were introduced in Refs.~\cite{Plugge2016a,Karzig2016} as quantum-wire-based~\cite{Lutchyn2010,Oreg2010} constructions, we use these terms for any physical system with four or six Majoranas in a fixed parity sector.  Next, in Sec.~\ref{sec:majoranagates}, we describe a protocol to implement minimal-overhead Clifford gates with \textit{physical} tetrons and hexons by measuring arbitrarily nonlocal Pauli product operators using only local operations. This is not fault-tolerant in the sense that the protocol assumes perfect measurements and error-free qubits. However, it is entirely analogous to the fault-tolerant protocol discussed in Sec.~\ref{sec:twists}, where we extend the protocol for minimal-overhead Clifford gates to \linebreak \textit{logical} tetrons and hexons, i.e., surface code patches. This is done by describing twist defects in Majorana surface codes and adapting twist-based lattice surgery~\cite{Litinski2017b} to fermionic codes. Finally, in Sec.~\ref{sec:colorcodes}, we propose a construction procedure for Majorana color codes, i.e., the fermionic equivalent of bosonic color codes. Since bosonic color codes can be obtained by concatenating bosonic surface codes with small non-topological codes~\cite{Criger2016}, we describe code concatenation for fermionic codes and construct Majorana color codes by concatenating Majorana surface codes with small Majorana fermion codes.

With fermionic and bosonic versions of surface and color codes, a plethora of topological codes is available, all of which can implement logical Clifford gates with zero time overhead. To decide which code to use for Majorana-based quantum computation, we show a comparison of 2D topological codes in Tab.~\ref{tab:results}. The main differences between these codes lie in their Majorana overhead, i.e., the number of Majoranas required to encode a qubit with code distance $d$, and in their stabilizer weight, i.e., the number of Majorana operators contained in the stabilizers that need to be measured for error correction. It is desirable to keep both of these figures low, as a lower Majorana overhead implies a higher encoding rate and more efficient space usage, and a lower stabilizer weight implies easier implementation due to lower hardware requirements. However, the general trend seen in bosonic surface and color codes indicates that codes featuring lower-weight stabilizers tend to have a higher space overhead (or Majorana overhead in the case of Majorana-based qubits).

\begin{figure*}
\centering
\def\svgwidth{0.97\linewidth}
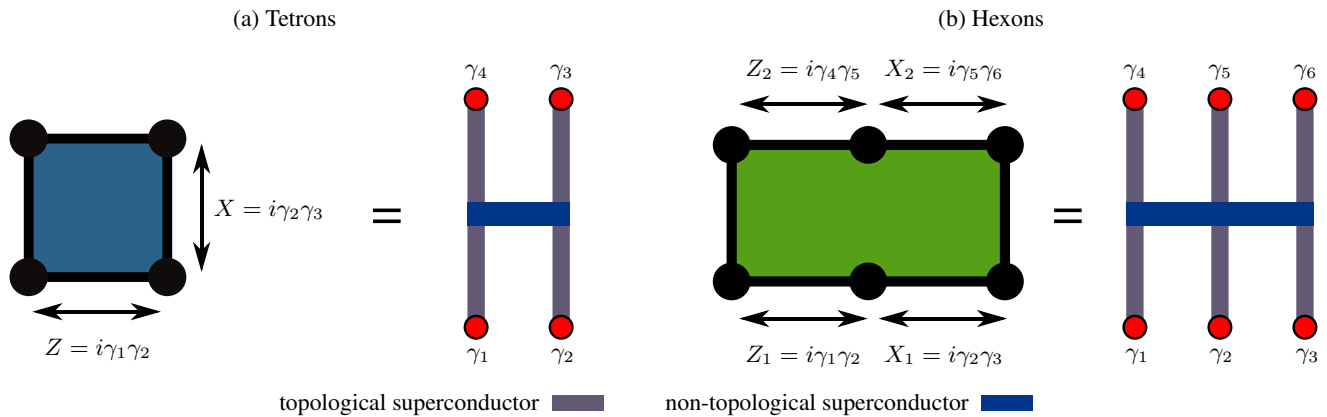
\caption{Tetrons and hexons as the two smallest Majorana fermion codes, and their correspondence to four or six Majoranas in a fixed parity sector, e.g., via the quantum-wire construction in Ref.~\cite{Karzig2016}.}
\label{fig:tetronshexons}
\end{figure*}

Out of all known bosonic 2D topological codes, surface codes have the lowest stabilizer weight of four qubits. Topological subsystem codes~\cite{Bombin2010a} can further reduce the weight of the operators that need to be measured, with a subsystem variant of the surface code~\cite{Bravyi2013} reducing the weight to three qubits. With Majorana-based hardware, this corresponds to 8-Majorana and 6-Majorana stabilizers. In order to implement minimal-overhead Clifford gates, twist-based lattice surgery requires the measurement of higher-weight 5-qubit operators corresponding to twist defects, which corresponds to 10-Majorana operators.
Majorana surface codes decrease the stabilizer weight below the weight of bosonic surface codes, such that only 4- and 6-Majorana operators need to be measured. This might be useful if higher-weight stabilizers turn out to be difficult to measure in a given implementation. Majorana color codes, on the other hand, can encode quantum information with a lower space overhead compared to bosonic color codes with the same stabilizer weight.

In general, bosonic surface codes can be concatenated with any $[[n,k,d]]$ code, where $n$ is the number of physical qubits, $k$ is the number of encoded logical qubits, and $d$ is the code distance, i.e., the minimum qubit weight of all logical operators. Similarly, Majorana surface codes can be concatenated with any $[[n_m,k,d_m]]_m$ Majorana fermion code, where $n_m$ is the number of Majoranas, and $d_m$ is the Majorana distance, i.e., the minimum Majorana weight of all logical operators. Concatenating surface codes with high-distance codes can be used to obtain topological codes with arbitrarily low space overhead. As one uses higher-distance codes for concatenation, one increasingly sacrifices stabilizer weight and locality in favor of lower Majorana overhead. From a pragmatic point of view, if a given Majorana-based quantum computer has a maximum stabilizer weight that it can measure, a comparison in the spirit of Tab.~\ref{tab:results} can be used to determine a suitable encoding, even though we note that our collection of codes is not exhaustive. In particular, it does not include Majorana surface codes based on non-uniform tilings.


\section{Majorana Surface Codes}

\label{sec:surfacecodes}

\begin{figure*}
\centering
\def\svgwidth{\linewidth}
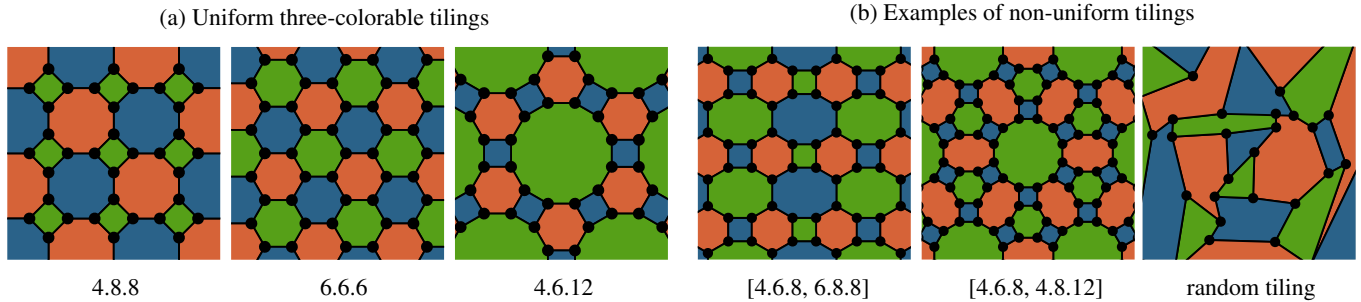
\caption{The three uniform three-colorable tilings of the 2D plane (a), and three examples of non-uniform tilings (b).}
\label{fig:tilings}
\end{figure*}

A Majorana fermion code~\cite{Bravyi2010} encodes logical information in a set of Majorana zero modes placed on the vertices of 2D lattices, which are described by self-adjoint operators $\gamma_i = \gamma_i^\dagger$, and fulfill the fermionic anticommutation relations $\{\gamma_i,\gamma_j\} = 2\delta_{i,j}$. The code is defined by its stabilizers~--~more precisely, by its mutually commuting stabilizer generators~--~which are products of all Majorana operators 
\begin{equation}
	\mathcal{O}_{\rm face} = \prod\limits_{j \in \mathrm{face}} i^{1/2}~\gamma_j
\end{equation}
associated with a face and have eigenvalues $\pm 1$. A code with $n$ Majoranas and $m$ stabilizers encodes $n/2-m$ logical qubits in the degenerate ground-state manifold of the Hamiltonian
\begin{equation}
	H_{\rm code} = - \sum\limits_{i=1}^m \mathcal{O}_i \, .
\end{equation}
Due to fermion parity superselection, the product of all $n$ Majoranas is always a stabilizer, such that $n$ Majoranas can at most encode $n/2-1$ qubits.

\subsection{Smallest Majorana fermion codes}

The smallest fermion code is the four-Majorana tetron code shown in Fig.~\ref{fig:tetronshexons}a, which involves a single stabilizer $\mathcal{O}_{\rm tetron}~=~- \gamma_1\gamma_2\gamma_3\gamma_4$. The qubit is encoded in the doubly degenerate ground-state space of the Hamiltonian
\begin{equation}
	H_{\rm tetron} = \gamma_1\gamma_2\gamma_3\gamma_4 \, .
\end{equation}
While it is possible to define a qubit in the Schr\"odinger picture by defining two computational states $\ket{0}$ and $\ket{1}$, it is more convenient to use the Heisenberg picture. Since any single-qubit unitary operator can be written using the $X$ and $Z$ Pauli operators via the Euler decomposition, a qubit can be defined through these two operators. The $X$ and $Z$ operators need to square to the identity, $X^2=Z^2=\mathbbm{1}$, anticommute, $XZ = -ZX$, and commute with all stabilizers of the code. For a tetron, one choice is $Z=i\gamma_1\gamma_2=i\gamma_3\gamma_4$ and \linebreak $X=i\gamma_2\gamma_3=i\gamma_1\gamma_4$. The remaining Pauli operator follows as the product $Y=iXZ$.

There are two ways to implement stabilizer terms such as $\mathcal{O}_{\rm tetron}$. One way is to interpret stabilizers as physical parity-fixing constraints realized in the laboratory, e.g., via the charging energy of two topological superconducting nanowires. Alternatively, they can be interpreted as measurement prescriptions. Measuring all stabilizers of the code projects the system into the ground-state space as long as the measurement outcome is $+1$ for all stabilizers. For tetrons, it is easy to see that both approaches are susceptible to errors. In particular, an error process described by the operator $i\gamma_i\gamma_j$ involving two Majorana fermion operators will lead to errors that neither violate the parity-fixing constraint nor change the measurement outcome of the parity measurement. 
The fact that stabilizers can be interpreted either as physical constraints or measurement prescriptions will become important when we discuss larger codes with more than one stabilizer.

The second smallest Majorana fermion code is the six-Majorana hexon code shown in Fig.~\ref{fig:tetronshexons}b. It encodes two logical qubits in the ground-state space of the Hamiltonian
\begin{equation}
	H_{\rm hexon} = -i\gamma_1\gamma_2\gamma_3\gamma_4\gamma_5\gamma_6 \, .
\end{equation}
The logical Pauli operators of the two qubits can be chosen as $Z_1 = i\gamma_1\gamma_2$, $X_1=i\gamma_2\gamma_3$, $Z_2=i\gamma_4\gamma_5$, $X_2=i\gamma_5\gamma_6$. 

\subsection{Logical tetrons and hexons}

Tetrons and hexons encode qubits, but cannot correct any errors. We now describe a procedure to obtain \textit{logical} tetrons and hexons based on 2D Majorana surface codes. 

We define Majorana surface codes as fermionic topological codes with non-overlapping stabilizers. Thus, they can be defined by a tiling of the 2D plane with vertices corresponding to Majoranas and faces (or tiles) corresponding to stabilizers. Since stabilizers need to commute, neighboring faces should share two vertices (or, in fact, any even number). This implies that any valid tiling needs to be three-colorable~\cite{Bravyi2010}, in the sense that faces can be colored in three colors with neighboring faces having different colors. Moreover, each stabilizer needs to contain an even number of Majoranas. Remarkably, these requirements are identical to the restrictions of bosonic color codes, such that all valid color code tilings with vertices as qubits are also valid Majorana surface code tilings with vertices as Majoranas. 

There exist three uniform three-colorable tilings that can be used to define a Majorana surface code: 4.8.8~\cite{Vijay2015,Landau2016,Plugge2016,Li2017}, 6.6.6~\cite{Vijay2016}, and 4.6.12, as shown in Fig.~\ref{fig:tilings}a. The numbers refer to the three polygons that meet at each vertex, e.g., for the 4.6.12 tiling, a square, a hexagon, and a dodecagon. Uniform means that the same types of polygons meet at each vertex. There exist also non-uniform three-colorable tilings, which have at least two different vertex types. Examples are shown in Fig.~\ref{fig:tilings}b, such as the [4.6.8,~6.8.8] tiling~\cite{Bravyi2010} which has two types of vertices, namely 4.6.8 and 6.8.8. In this work, we restrict ourselves to the three uniform tilings shown in Fig.~\ref{fig:tilings}a, although our constructions can be straightforwardly extended to non-uniform tilings. We stress that the colors bear no physical meaning, but are merely a useful bookkeeping tool.

\begin{figure*}[t]
\centering
\def\svgwidth{\linewidth}
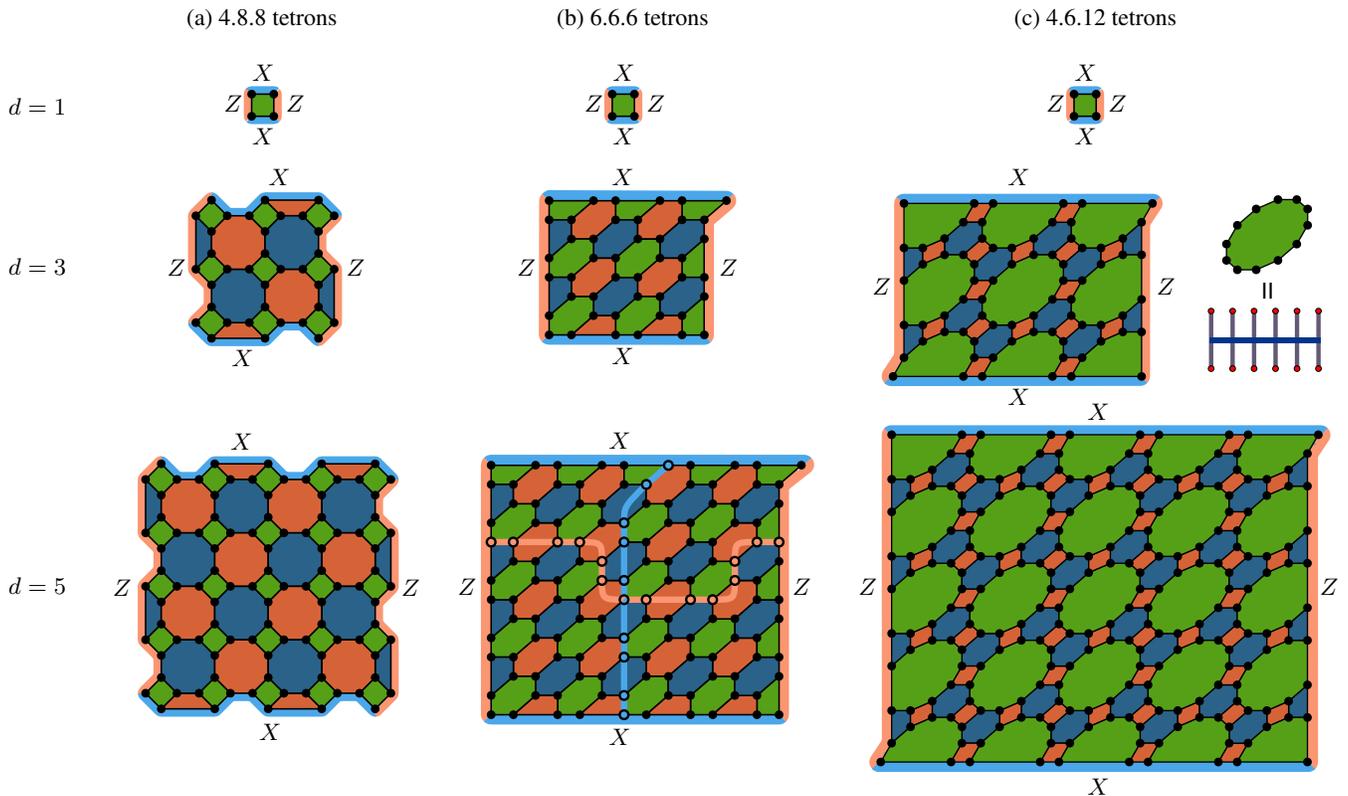
\caption{Logical tetrons of the 4.8.8 (a), 6.6.6 (b), and 4.6.12 (c) Majorana surface code with code distances $d=1$, 3, and 5. The 6.6.6 Majorana surface code with $d=5$ shows an example of strings of red and blue edges through the bulk of the code. Products of Majoranas along the red and blue strings are equivalent to logical $X$ and $Z$ operators, respectively.}
\label{fig:logicaltetrons}
\end{figure*}

\begin{figure*}[t]
\centering
\def\svgwidth{\linewidth}
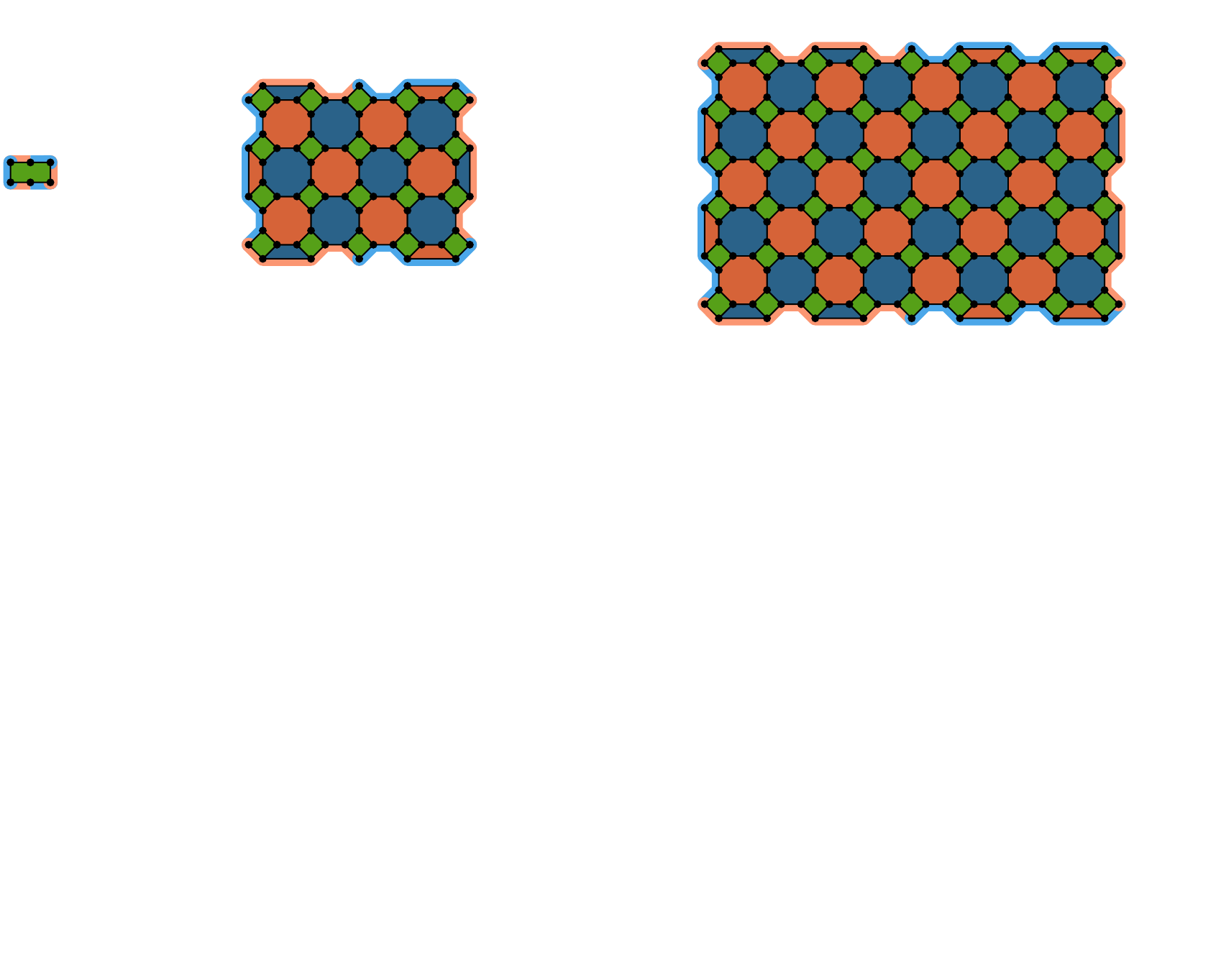
\caption{Logical hexons of the 4.8.8 (a), 6.6.6 (b), and 4.6.12 (c) Majorana surface code with code distances $d=1$, 3, and 5.}
\label{fig:logicalhexons}
\end{figure*}

A logical tetron encodes one logical qubit in $n$ Majoranas with $n/2-1$ stabilizers. It is constructed by appropriately introducing boundaries to a three-colorable tiling, as shown in Fig.~\ref{fig:logicaltetrons}. At a boundary, Majoranas are no longer part of three differently-colored stabilizers, but only two. We then assign the remaining color to the boundary, e.g., on a red boundary, Majoranas are part of blue and green stabilizers. Two differently-colored boundaries meet at a corner, where a Majorana is part of only one stabilizer. A logical tetron is obtained by terminating the tiling by four boundaries, a pair of opposing red and blue boundaries each (or any other pair of colors). This procedure always yields a code with $n$ Majoranas and $n/2-1$ stabilizers. 

One can also assign colors to the edges of faces, i.e., to the two-Majorana operators that lie between two faces. The color of an edge is then the third remaining color. For instance, an edge that lies between a red and a blue face is referred to as a green edge. The logical $X$ ($Z$) operators of a logical tetron are strings of red (blue) edges that connect the red (blue) boundaries, as shown in Fig.~\ref{fig:logicaltetrons}b. In particular, the product of all Majoranas along a blue boundary is a string of red edges, i.e., an $X$ operator. Similarly, the product of all Majoranas along a red boundary is a $Z$ operator. Logical $Z$ and $X$ operators anticommute, since they always share an odd number of Majoranas. The red and blue boundary operators, specifically, share one Majorana in the corner where the boundaries meet.

The four-corner surface codes depicted in Fig.~\ref{fig:logicaltetrons} are indeed the higher-distance equivalents of four-Majorana tetrons. In Ref.~\cite{Brown2017}, it was pointed out that the corners of surface code qubits correspond to twist defects, which feature the same nonabelian statistics as Majoranas~\cite{Bombin2010}. This can be understood from the observation that since one type of surface code boundary can absorb $e$ anyons, and the other type of boundary can absorb $m$ anyons, corners can absorb $\epsilon$ anyons. Therefore, according to Ref.~\cite{Bombin2010}, corners are twist defects. Alternatively, in Ref.~\cite{Wen2003}, edges of surface codes are identified as flat bands of uncoupled Majoranas. The boundary stabilizers gap out the edge Majoranas, leaving unpaired Majoranas in the corners. Thus, the identification of four-boundary Majorana surface codes with tetrons is indeed justified, as their corners can be interpreted as \textit{logical} Majoranas, i.e., twist defects. 

Similarly, six-boundary Majorana surface codes can be identified as logical hexons, as shown in Fig.~\ref{fig:logicalhexons}. Logical hexons encode two logical qubits on $n$ Majoranas with $n/2-2$ stabilizers. This can be achieved by introducing three red and blue boundaries each in an alternating fashion. Logical hexons require approximately twice as many Majoranas as logical tetrons for a given code distance, but also encode twice as many qubits~--~one qubit at the two bottom and one qubit at the two top boundaries.

The error processes considered in this work correspond to quasiparticle poisoning and are described by the action of single Majorana operators $\gamma_i$. We refer to these errors as Majorana errors. A Majorana error flips the measurement outcome of the stabilizers that the affected Majorana is part of, which allows for detection of errors and their subsequent correction. Corrections do not need to be applied explicitly, but merely tracked by a classical computer.
We define the Majorana code distance $d_m$ as the minimum number of Majorana errors necessary to introduce a logical error that will go undetected, i.e., $d_m$ is the number of Majorana operators contained in the shortest logical operator. 
Codes with Majorana distance $d_m$ tolerate $d_m/2-1$ Majorana errors, as the syndrome of error strings that cover more than half the Majorana distance will be misinterpreted and lead to a correction that introduces a logical Pauli error.
Since only even-number products of Majoranas are physically measurable, $d_m$ is always even. For this reason, $d_m$ is not the same quantity as the code distance $d$ of bosonic codes. For a fair comparison between fermionic and bosonic codes, we label fermionic codes by their bosonic code distance $d = d_m/2$. This is justified, because Pauli operators of physical Majorana-based qubits, such as tetrons or hexons, correspond to products of two Majorana operators, and therefore single-qubit Pauli errors always correspond to \textit{two} Majorana errors. Thus, the Majorana surface code tetrons shown in Fig.~\ref{fig:logicaltetrons} have Majorana distances $d_m=2$, 6, and 10, but code distances $d=1$, 3, and 5.

To minimize the number of operators that need to be measured, we implement stabilizers of one color as parity-fixing constraints, while the other two colors are interpreted as measurement prescriptions. The parity-fixed color should be the one color that is not a boundary. For instance, the green four-Majorana stabilizers of the 4.8.8 Majorana surface codes in Fig.~\ref{fig:logicaltetrons}a could be interpreted as physical tetrons, whereas the red and blue stabilizers are operators that need to be measured in order to actively detect and correct errors. This can be done with any three-colorable tiling, such that, say, red and blue stabilizers need to be measured, while green stabilizers are interpreted as Majoranas in a fixed parity sector. We then refer to the green stabilizers as the building blocks of the code. Note that Majorana errors that respect the parity-fixing constraints set by the green stabilizers always come in pairs.

If green stabilizers are not measured, but implemented as parity-fixing constraints, then single-Majorana errors violating these constraints are leakage errors, i.e., errors that take the qubit out of the computational subspace and may go undetected by the measurement of red and blue stabilizers. For quantum-wire-based architectures, the time scales of dephasing and depolarizing errors that are detectable by red and blue stabilizer measurements were calculated to range from hundreds of nanoseconds to several minutes for achievable experimental parameters~\cite{Knapp2018}. On the other hand, leakage errors caused by single-Majorana errors are suppressed exponentially with $\exp(-E_C/T)$, where $E_C$ is the charging energy fixing the parity, and $T$ is the temperature. Already for $E_C \sim 100~\mathrm{\mu eV}$ and $T \sim 10~\mathrm{mK}$, $E_C/T \approx 100$, which suggests that leakage may be negligible~\cite{Knapp2018}. In implementations where leakage errors cannot be neglected, they can be treated by also measuring the green stabilizers.

4.8.8 Majorana surface codes can be defined on a square lattice of tetrons in the spirit of Ref.~\cite{Landau2016}, such that the \linebreak 8-Majorana stabilizers need to be measured. One qubit with code distance $d$ requires $4d^2$ Majoranas. Similarly, 6.6.6 Majorana surface codes can be defined on a square lattice of hexons. While they require more Majoranas per qubit ($6d^2-4d+2$), the weight of the operators that need to be measured reduces to 6 Majoranas. We note that even though the codes in Fig.~\ref{fig:logicaltetrons}b display green tetrons at the top and right boundaries, these codes can indeed be defined on a square lattice of hexons, as we show in Appendix~\ref{app:squarelattice}. Finally, 4.6.12 Majorana surface codes can be defined on a square lattice of dodecons, i.e., 12 Majoranas in a fixed parity sector. In the context of color codes, 4.6.12 codes are often ignored, since they feature both a high space overhead and high-weight stabilizers. However, by interpreting the 12-Majorana operators as building blocks that satisfy parity-fixing constraints, only 4- and 6-Majorana operators need to be measured. Thus, the stabilizer weight has decreased to an average of 5 Majoranas, albeit at the price of an increased space overhead of $12d^2-12d+4$ Majoranas per qubit.

\begin{figure*}
\centering
\def\svgwidth{\linewidth}
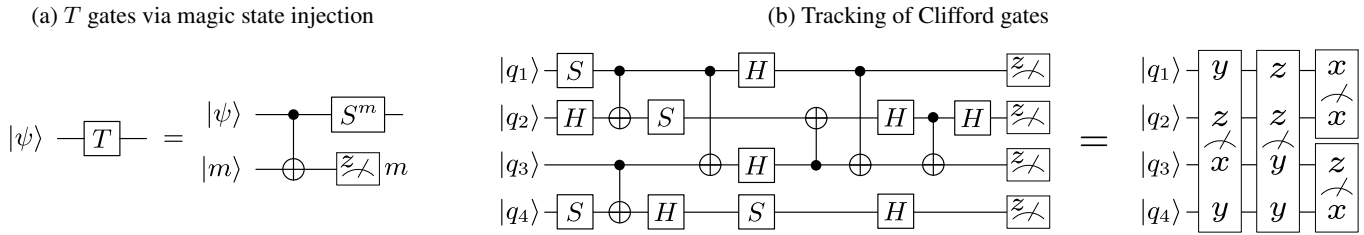
\caption{(a) Applying a $T$ gate to a qubit $\ket{\psi}$ is equivalent to a CNOT between the qubit and a magic state $\ket{m} = \frac{1}{\sqrt 2}(\ket{0} + e^{i\pi/4} \ket{1})$, a measurement of the magic state with outcome $m \in \{0,1\}$, and a corrective operation $S^m$ on the qubit. (b) Example of a circuit of Clifford gates and four $Z$ measurements that is compressed into four Pauli product measurements.}
\label{fig:circuitswide}
\end{figure*}

An important part of error correction is decoding. While the error syndrome is known from the stabilizer measurements, the most likely error configuration that causes this syndrome needs to be determined by a classical program called a decoder. Since Majorana surface codes are indeed surface codes, they can be decoded by any bosonic surface code decoder,\footnote{We note that this is only true, if building blocks are interpreted as parity-fixed, such that, say, green stabilizers cannot be violated. If this is not the case, then the decoder needs to match all three colors, similar to the decoding of bosonic color codes.} either by interpreting blue and red stabilizers as two types of anyons that need to be matched (analogous to $e$ and $m$ anyons in bosonic surface codes), or by first mapping the Majorana code onto a bosonic code and then decoding the bosonic surface code. As we show in Appendix~\ref{app:bosonicmap}, Majorana surface codes can be mapped onto bosonic codes, but the mapping is not unique. In the case of 4.8.8 codes, replacing each tetron with a qubit yields exactly the bosonic surface code on a square lattice. Replacing each hexon of the 6.6.6 code with two qubits also yields a bosonic code on a square lattice, but the lattice is rotated. With a 4.6.12 code, replacing each of the four-Majorana operators with a qubit yields a bosonic surface code on a tri-hexagonal lattice.


\section{Minimal-overhead Clifford gates \\ with tetrons and hexons}
\label{sec:majoranagates}

In this section, we show how to perform universal quantum computation with \textit{physical} tetrons and hexons. This is not fault-tolerant, since the protocols assume perfect measurements and error-free qubits. Closely analogous fault-tolerant protocols based on logical tetrons and hexons will be discussed in Sec.~\ref{sec:twists}.

We start with general considerations concerning the classical tracking of Clifford gates.
These gates map Pauli operators onto other Pauli operators and are products of Hadamard gates $H$, phase gates $S$, and controlled-NOT gates CNOT. Any quantum circuit can be expressed in terms of Clifford gates, $T$ gates (where $T=\sqrt{S}$) and single-qubit measurements in the $Z$ basis~\cite{Boykin2000}. If magic states $\ket{m} = (\ket{0} + e^{i\pi/4} \ket{1})/\sqrt{2}$ are available as a resource, $T$ gates can be rewritten as a CNOT gate and a measurement of the magic-state qubit, see Fig.~\ref{fig:circuitswide}a. In realistic architectures, only faulty magic states are available. Using magic state distillation~\cite{Bravyi2005}, multiple faulty magic states can be converted into fewer low-error magic states. This is yet another circuit of Clifford gates and measurements. Thus, any quantum computation can be expressed as a quantum circuit consisting only of Clifford gates and measurements.

According to the Gottesman-Knill theorem, gate operations for Clifford gates need not be performed, as they can be tracked classically. Mapping Pauli operators onto other Pauli operators, they merely change the basis of the $Z$ measurements: $H$ gates map $Z \rightarrow X$ and $X \rightarrow Z$, $S$ gates map $Z \rightarrow Z$ and $X \rightarrow Y$, and CNOTs map $\mathbbm{1} \otimes Z \rightarrow Z \otimes Z$, $Z \otimes 1 \rightarrow Z \otimes \mathbbm{1}$, $X\otimes \mathbbm{1} \rightarrow X \otimes X$, and $\mathbbm{1} \otimes X \rightarrow \mathbbm{1} \otimes X$. The action of Clifford gates on all other Pauli operators can be inferred from these rules. Consequently, any circuit of Clifford gates and $n$ measurements can be contracted to $n$ measurements of Pauli product operators.  
Thus, apart from the generation of (faulty) magic states, a quantum computer merely needs to be able to fault-tolerantly measure nonlocal products of Pauli operators. An example of such a contraction is shown in Fig.~\ref{fig:circuitswide}b. In this sense, a quantum computer that can measure arbitrary Pauli products implements Clifford gates with zero time overhead, as these gates can be tracked classically and do not require any quantum operations.

The state-injection circuit in Fig.~\ref{fig:circuitswide}a then corresponds to a $Z \otimes Z$ measurement between the qubit $\ket{\psi}$ and the magic state, and a tracked $S$ gate correction. In order to discard the magic state qubit, it needs to be disentangled via an $X$ measurement and a subsequent $Z$ correction on the qubit $\ket{\psi}$, as we discuss in Appendix~\ref{app:clifftracking}. We note that if one prefers to explicitly perform CNOT operations, this can also be done using Pauli product measurements via the circuit identity shown in Appendix~\ref{app:clifftracking}.

\begin{figure}[b]
\centering
\def\svgwidth{\linewidth}
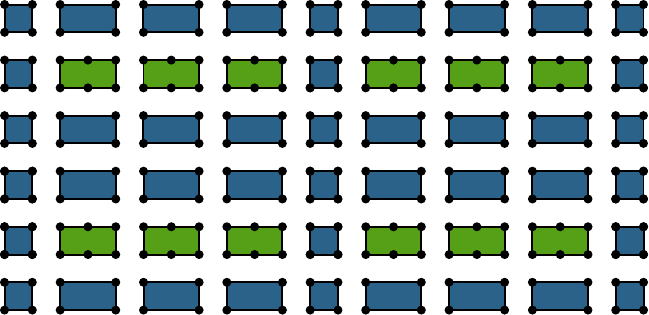
\caption{Array of tetrons and hexons, where hexons are used to encode data qubits, and tetrons are used as ancillas for Pauli product measurements.}
\label{fig:qubitarr}
\end{figure}

\subsection{Pauli product measurements with tetrons and hexons}

\begin{figure}[t]
\centering
\def\svgwidth{0.95\linewidth}
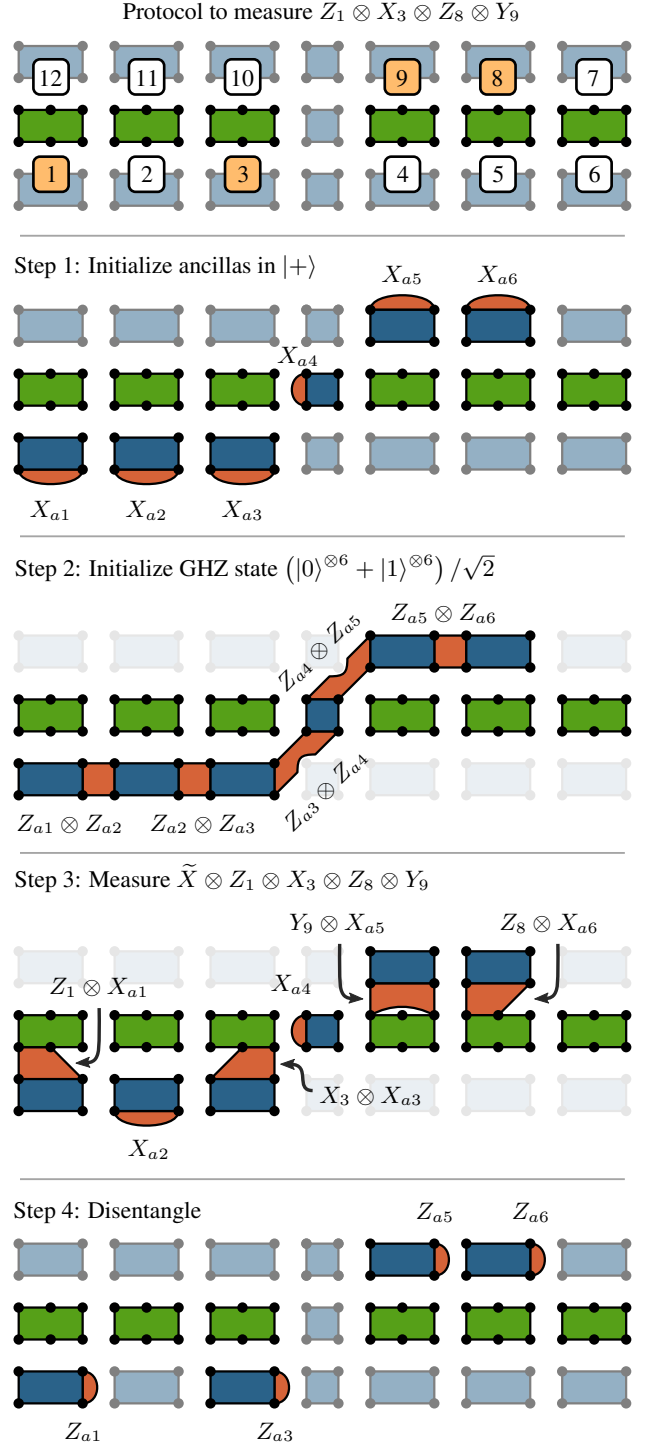
\caption{Protocol for the measurement of the Pauli product operator $Z_1 \otimes X_3 \otimes Z_8 \otimes Y_9$.}
\label{fig:pauliprodprotocol}
\end{figure}

We now describe a protocol for Pauli product measurements with tetrons and hexons using only local two-qubit operations. Our scheme uses hexons to encode qubits used for quantum computation (referred to as data qubits), and tetrons to encode ancilla qubits that enable long-range communication. The hexons form blocks of data qubits in a 2D array, with tetrons as ancillas between blocks. A possible arrangement of 24 data qubits is shown in Fig.~\ref{fig:qubitarr}. 
We describe the protocol using the example in Fig.~\ref{fig:pauliprodprotocol}.

The goal is to measure the nonlocal $n$-qubit Pauli product $Z_1 \otimes X_3 \otimes Z_8 \otimes Y_9$ without measuring any of the $n$ individual Pauli operators $Z_1$, $X_3$, $Z_8$ or $Y_9$.
The first step is to initialize an $n+m$-ancilla GHZ state $\ket{0}^{\otimes n+m} + \ket{1}^{\otimes n+m}$, where $m$ ancillas are used to bridge long distances. In our example, we use $n+m=4+2$ ancilla qubits. Each ancilla is measured in the $X$ basis by measuring the corresponding two-Majorana operator (labelled $X_{a1} \cdots X_{a6}$) and thereby prepared in the $X$ eigenstate $\ket{+} = (\ket{0} + \ket{1})/\sqrt{2}$. For measurement outcomes $-1$, a corrective $Z$ operation is necessary, which can also be tracked. Measuring $Z \otimes Z$ of neighboring pairs of $\ket{+}$ ancillas (corresponding to the red four-Majorana operators shown in step 2 of Fig.~\ref{fig:pauliprodprotocol}) prepares the ancillas in the desired GHZ state, assuming that all $Z \otimes Z$ measurements yield $+1$. Again, measurement outcomes $-1$ require corrective Pauli $X$ operations.

The GHZ state is an entangled state with known global properties, but unknown local properties. It is a $+1$ eigenstate of $\widetilde{X} = X^{\otimes n+m}$, but the measurement outcome of each individual $X$ operator is entirely random. Thus, we can use the GHZ state to measure the desired Pauli product $\widetilde{X} \otimes Z_1 \otimes X_3 \otimes Z_8 \otimes Y_9$ without measuring any of the individual operators. By measuring the six operators shown in step 3 of Fig.~\ref{fig:pauliprodprotocol}, the desired operator is obtained as the product of the six measurements.

In the general case of the measurement of a Pauli product $P_1 \otimes \cdots \otimes P_n$, the $n$ ancilla qubits adjacent to the corresponding data qubits are used to measure $P_n \otimes X_{a, n}$, while the remaining $m$ ancilla qubits are read out in the $X$ basis. The product of all $n+m$ measurements is $\widetilde{X} \otimes P_1 \otimes \cdots \otimes P_n$. Because $\widetilde{X} = 1$, but each individual $X_{a,n}$ is random, this measures the Pauli product without measuring any of the individual Pauli operators.

The $n$ ancilla qubits adjacent to data qubits still need to be disentangled from the data qubits in step 4 of the protocol, before they can be discarded. This is done by measuring the ancilla qubits in the $Z$ basis with outcome $m \in \{0, 1\}$, leading to a $P^m$ Pauli correction on the adjacent data qubit, where $P$ is the Pauli operator that was part of the two-qubit measurement in step 3, e.g., $Z_1$ for qubit 1 in Fig.~\ref{fig:pauliprodprotocol}.

When combined with the preparation of (faulty) magic states, this is sufficient to implement universal quantum computation. Specifically, this protocol can be straightforwardly used with the quantum-wire constructions of Ref.~\cite{Karzig2016}. In Appendix~\ref{app:wireimplementation}, we explicitly show the tunnel coupling configurations to implement Fig.~\ref{fig:pauliprodprotocol} in a network of topological superconducting nanowires.

To summarize, the following operations are necessary to implement a Majorana-based universal quantum computer with minimal-overhead Clifford gates: \pagebreak
\begin{enumerate}
	\item[Op1] Measurements of tetrons in the bases $X$ and $Z$,
	\item[Op2] Measurements of $Z \otimes Z$ between adjacent tetrons,
	\item[Op3] Measurements of $X \otimes X$, $Y \otimes X$, and $Z \otimes X$ between adjacent hexons and tetrons,
	\item[Op4] Application of (potentially faulty) $T$ gates on tetron qubits.
\end{enumerate}
In the following section, we show how to implement these four operations fault-tolerantly with \textit{logical} tetrons and hexons.


\section{Twists in Majorana Surface Codes}
\label{sec:twists}

Remarkably, a completely analogous scheme can be implemented with logical tetrons and hexons, allowing for fault-tolerant quantum computing with tracked Clifford gates. Specifically, we implement the four operations Op1-Op4 with the logical tetrons and hexons from Figs.~\ref{fig:logicaltetrons} and \ref{fig:logicalhexons}. The first operation is the measurement of logical tetrons in the $X$ and $Z$ basis. With bosonic surface codes, logical qubits are read out in the $X$ or $Z$ basis by measuring all physical qubits in the $X$ or $Z$ basis, and performing classical error correction. Similarly, Majorana surface code tetrons are read out in the $X$ or $Z$ basis by measuring all two-Majorana operators corresponding to red or blue edges, and performing classical error correction. The measurement outcome corresponds to the product of all Majoranas along the corresponding boundary of the tetron. In analogy to bosonic surface codes, Majorana surface code tetrons can be initialized in the $\ket{+}$ state by measuring all red edges, and then measuring the stabilizers and correcting the errors.

The second and third operations require measurements of two-qubit Pauli products. For logical tetrons with code distance $d$, this corresponds to products of $4d$ Majoranas, which are highly nonlocal operators. In order to measure these operators using only measurements of low-weight local operators, we adapt the technique of lattice surgery~\cite{Horsman2012}, and in particular twist-based lattice surgery~\cite{Litinski2017b}, to fermionic codes. While we only explicitly show lattice surgery protocols for the three uniform Majorana surface codes, similar constructions can also be used for non-uniform tilings.

\subsection{4.8.8 Majorana surface codes}

\begin{figure}[t]
\centering
\def\svgwidth{\linewidth}
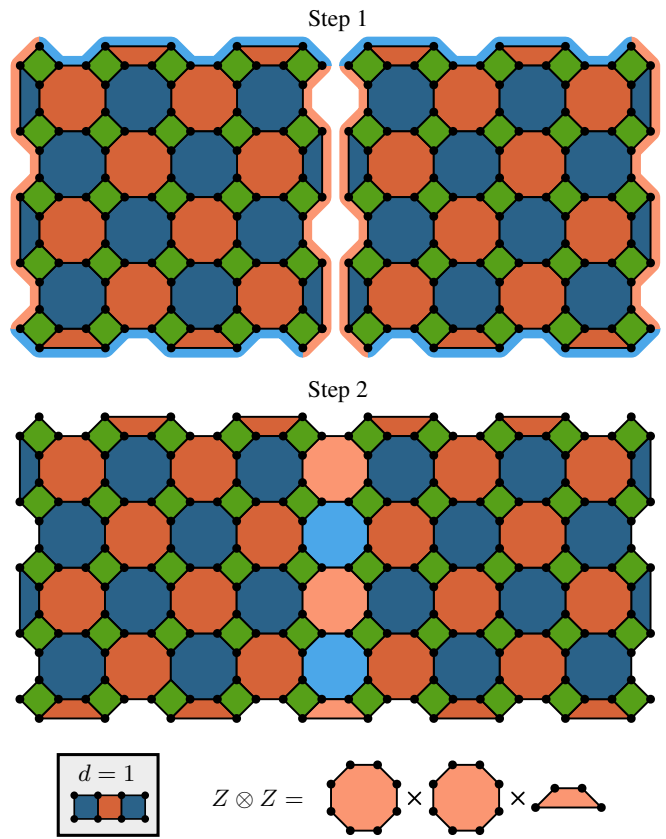
\caption{Lattice surgery between two 4.8.8 Majorana surface code tetrons with code distance $d=5$ measuring $Z \otimes Z$. The box in the bottom left corner shows the corresponding measurement with $d=1$, where $Z \otimes Z$ is the red 4-Majorana operator.}
\label{fig:488surgery1}
\end{figure}

We start with 4.8.8 Majorana surface codes. The second operation Op2 requires the measurement of $Z \otimes Z$ between two tetrons, such as the two $d=5$ tetrons in Fig.~\ref{fig:488surgery1}. In this example, the operator $Z \otimes Z$ is the product of the 20 Majorana fermions along the two neighboring red boundaries. Lattice surgery is a protocol that temporarily changes the stabilizer configuration in order to measure this operator fault-tolerantly using only local measurements. Pairs of blue 4-Majorana stabilizers are merged to form 8-Majorana stabilizers (light blue in step 2). The measurement outcome of these 8-Majorana stabilizers is trivial, as it is simply the product of previously known stabilizers. In addition, new stabilizers are introduced along the boundary (light red). In this new stabilizer configuration in step 2, all stabilizers commute, but the number of stabilizers is increased by one compared to step 1. Thus, the number of degrees of freedom is reduced by one, and one bit of information is measured. Since the light blue stabilizers are trivial, this bit of information is given by the product of the light red stabilizers. Because the stabilizers involve each red-boundary Majorana exactly once, their product is exactly $Z \otimes Z$. This is the fault-tolerant generalization of the 4-Majorana parity measurement between two physical tetrons (with $d=1$) that is shown in the inset in Fig.~\ref{fig:488surgery1}. Not only is this protocol fault-tolerant in the sense that it can correct qubit errors such as quasiparticle poisoning, but by repeating rounds of syndrome measurement, erroneous stabilizer measurements can also be corrected.

\begin{figure}
\centering
\def\svgwidth{0.93\linewidth}
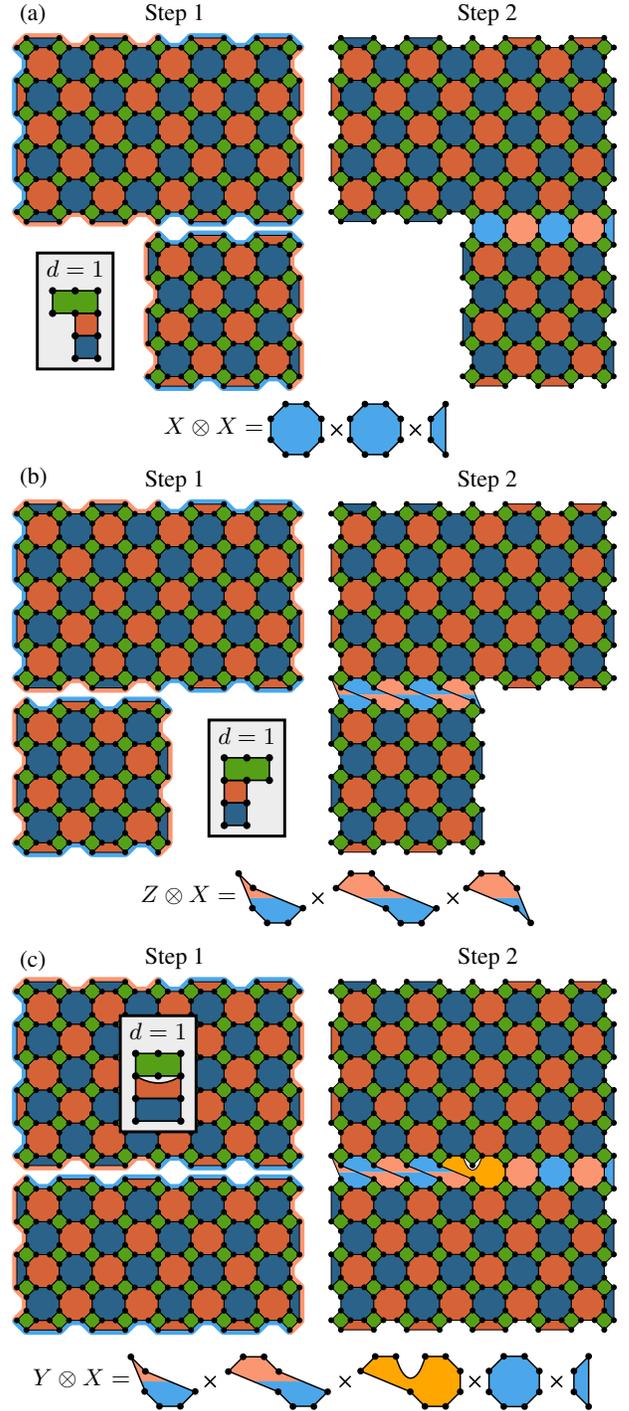
\caption{Lattice surgery protocols for the measurements of the operators $X \otimes X$ (a), $Z \otimes X$ (b), and $Y \otimes X$ (c) between a 4.8.8 Majorana surface code hexon and tetron. The insets show the equivalent operation with $d=1$.}
\label{fig:488surgery2}
\end{figure}

A very similar protocol can be used for the third operation to measure $X \otimes X$, $Y \otimes X$, and $Z \otimes X$ between a tetron and a hexon. In Fig.~\ref{fig:488surgery2}a, an $X \otimes X$ measurement is shown for $d=5$. Here, $X \otimes X$ is the product of the 20 Majoranas along the two adjacent blue boundaries. Again, the 4-Majorana operators are merged to 8-Majorana operators (light red), and the product of the new light blue operators yields precisely $X \otimes X$. The protocol for the $Z \otimes X$ measurement in Fig.~\ref{fig:488surgery2}b measures the product of 20 Majoranas that are located on a red boundary on the hexon, but on a blue boundary on the tetron. While again stabilizers are merged and new stabilizers are introduced to yield $Z \otimes X$, the stabilizer configuration is different from the situation in Fig.~\ref{fig:488surgery2}a.
Because the two boundaries have different colors, the stabilizers in step 2 form what is called a \textit{dislocation line}. If flipped red and blue stabilizers are interpreted as anyons in analogy to $e$ and $m$ anyons of bosonic surface codes, then anyons passing this dislocation line will change color.

The measurement of $Y \otimes X$ is shown in Fig.~\ref{fig:488surgery2}c. Because $Y=iXZ$, the $Y$ operator of the (bottom) hexon qubit is the product of 18 Majoranas on both the red and blue bottom boundary, excluding the center Majorana where the two boundaries meet (since it is part of both boundaries and $\gamma^2=\mathbbm{1}$). Thus, $Y \otimes X$ is the product of 36 Majoranas on the long blue boundary of the tetron, and the adjacent red and blue boundaries of the hexon. The product is again measured by merging stabilizers and introducing new ones whose product is precisely the product of the 36 Majoranas. The stabilizer configuration in step 2 corresponds to a dislocation line that is terminated by a 10-Majorana stabilizer (orange). This 10-Majorana stabilizer is what is called a twist defect~\cite{Bombin2010}. Twists are found at the ends of dislocation lines. Incidentally, corners of surface code qubits~--~i.e., the meeting points of two different boundaries~--~can also be interpreted as ends of dislocation lines, and therefore as twists~\cite{Brown2017}. The protocols in Figs.~\ref{fig:488surgery1} and \ref{fig:488surgery2} further illustrate the equivalence between twists and Majoranas: What for $d=1$ is a measurement of a product of four Majoranas, for higher code distances becomes a measurement of the four twists in the corners that are part of the lattice surgery protocols. Twist defects become visible when the twist is not located in a corner during lattice surgery. This is what happens in Fig.~\ref{fig:488surgery2}c for the $Y \otimes Z$ measurement, where the 10-Majorana twist defect is revealed as it is in the center of the qubit corresponding to the one Majorana that is not part of the parity measurement in the $d=1$ case.

\begin{figure*}[t]
\centering
\def\svgwidth{\linewidth}
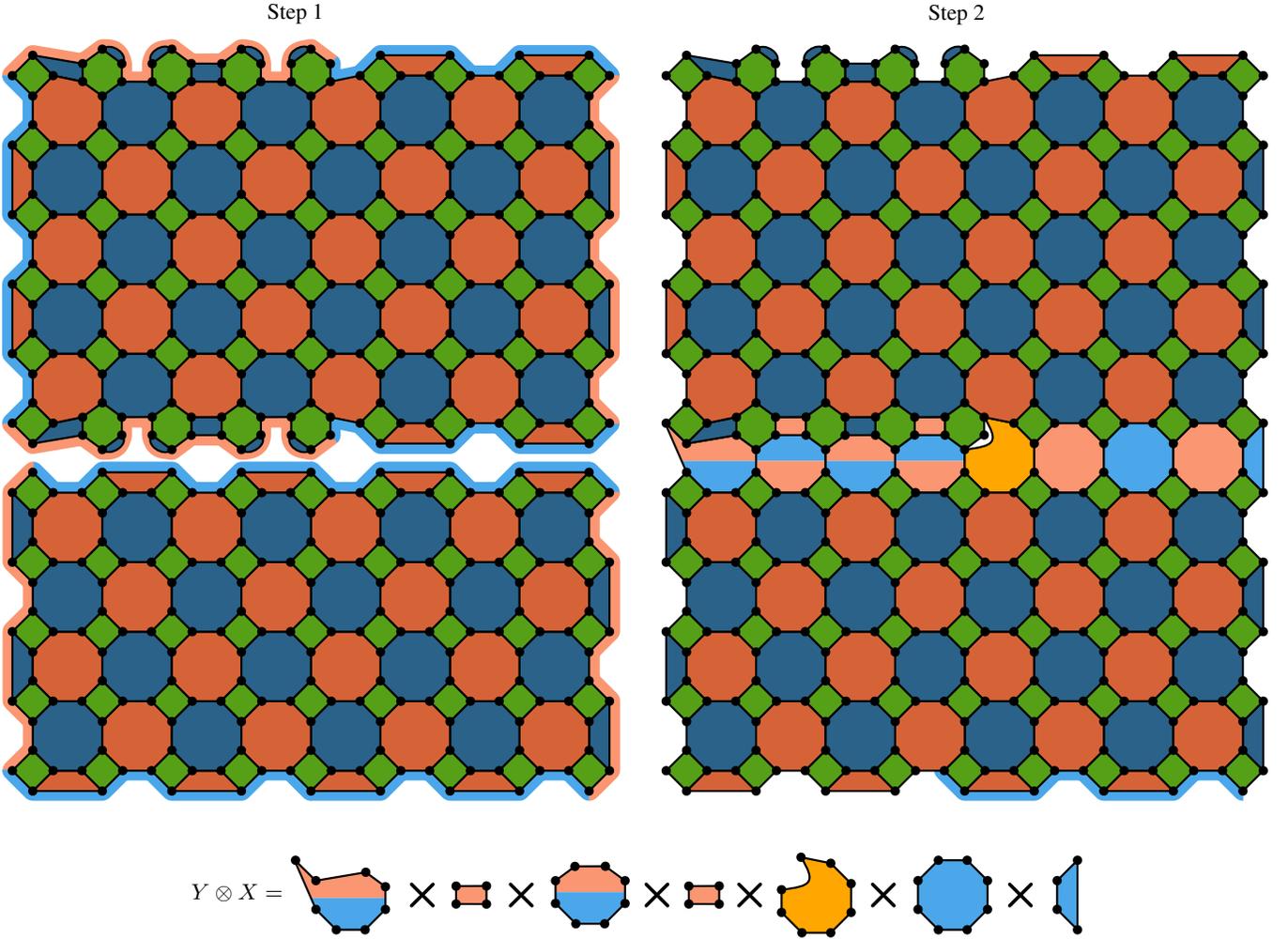
\caption{Lattice surgery protocol for the measurement of $Y \otimes X$ between a 4.8.8 Majorana surface code hexon and tetron. The hexon is modified to feature hexons instead of tetrons at the bottom and top red boundaries. As a consequence, the weight of the twist defect (orange) reduces to 8 Majoranas.}
\label{fig:488surgeryhexons}
\end{figure*}

The 10-Majorana twist defect of the 4.8.8 Majorana surface code corresponds to the 5-qubit  twist defect of bosonic surface codes. What is more, all of the aforementioned lattice surgery protocols with 4.8.8 Majorana surface codes are exactly equivalent to the protocols with bosonic surface codes~\cite{Litinski2017b}. In fact, no Majorana fermion code that is purely based on tetron building blocks (such as 4.8.8 codes) can display any features that are different from bosonic codes, as any bosonic code can be mapped onto a fermionic code by replacing each qubit with a tetron~\cite{Bravyi2010}. However, by replacing only a few of the tetrons with hexons, the 4.8.8 Majorana surface code can display some Majorana-specific characteristics. In particular, the maximum stabilizer weight necessary for lattice surgery can be reduced from 10 Majoranas (corresponding to 5 qubits in the bosonic case) to 8 Majoranas. 

Consider the modified $d=5$ hexons shown in Fig.~\ref{fig:488surgeryhexons}. The tetrons along the bottom and top red boundaries have been replaced by hexons. Since two of the Majoranas of the hexons are in a fixed parity sector due to the two-Majorana plaquettes, they are equivalent to tetrons. In order to measure $Y \otimes X$ between the hexon and a tetron via lattice surgery, stabilizers are merged and new stabilizers are introduced to yield the desired operator. The stabilizer configuration again features a dislocation line that is terminated by a twist defect, but the weight of the twist defect has decreased to just 8 Majoranas. Due to the three-colorability of Majorana surface code lattices, their twist defects closely resemble the so-called color twists of bosonic color codes~\cite{Kesselring2017}.

Thus, 4.8.8 Majorana surface codes feature the same Majorana overhead as bosonic surface codes of $\sim 4d^2$ Majoranas per qubit, but a lower stabilizer weight for twist-based lattice surgery. As we show in the remainder of the section, 6.6.6 and 4.6.12 Majorana surface codes can be used to reduce the stabilizer weight even further, albeit at the cost of a higher Majorana overhead.

Note that, in contrast to the Pauli product measurement protocol for physical tetrons, the fault-tolerant protocol features tetrons of different lengths. Specifically, the tetrons used for $Y \otimes X$ measurements are almost twice as long as tetrons used for $X \otimes X$ or $Z \otimes X$ measurements. The protocol of Fig.~\ref{fig:pauliprodprotocol} can still be used in the fault-tolerant setting, but additional tetrons may be required to bridge distances between ancillas. For concreteness, we show an implementation of the protocol of Fig.~\ref{fig:pauliprodprotocol} using 4.8.8 Majorana surface codes with $d=3$ in Appendix~\ref{app:logicalpauliprod}.

\begin{figure*}[t]
\centering
\def\svgwidth{0.93\linewidth}
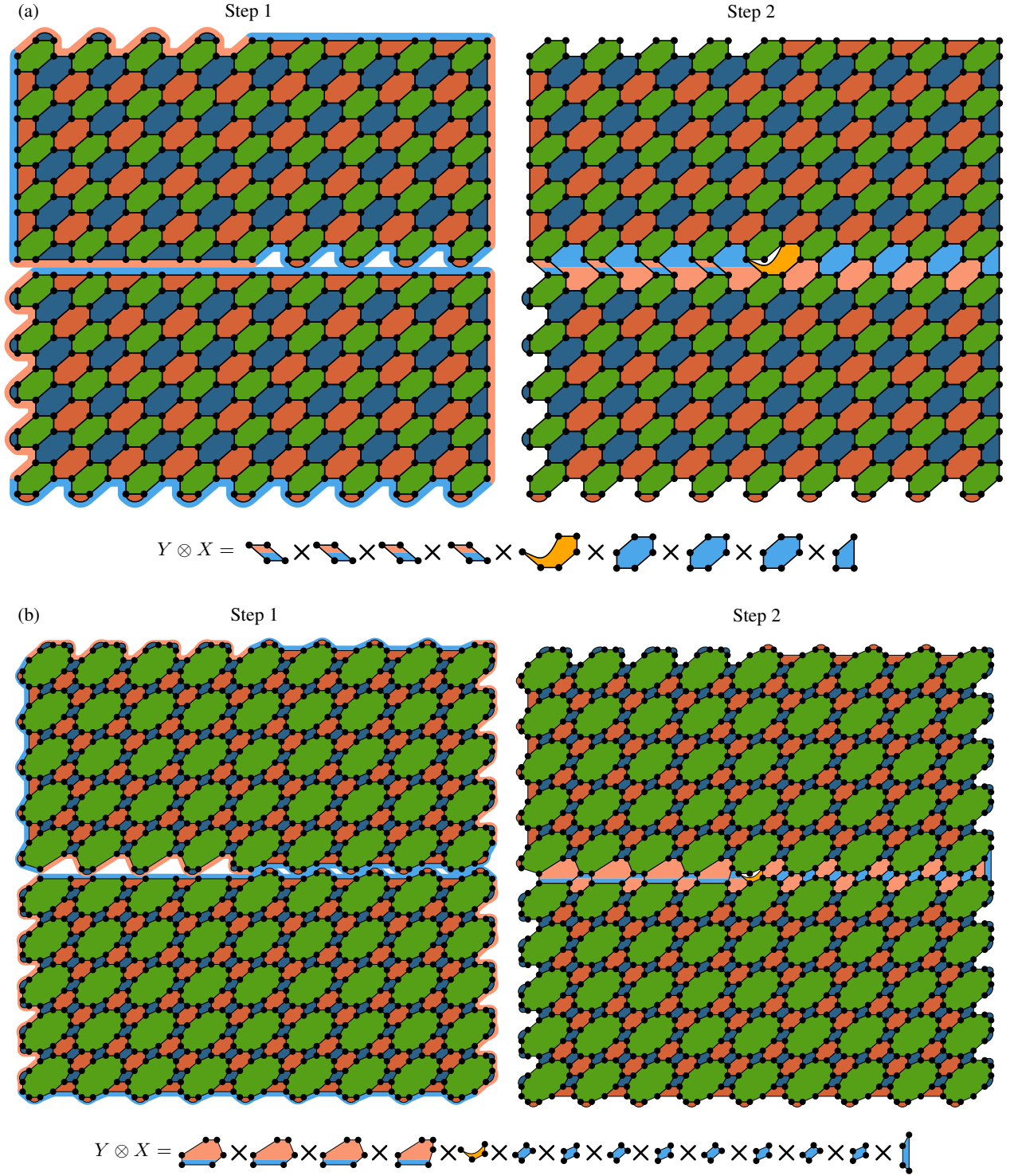
\caption{Lattice surgery protocol for the measurement of $Y \otimes X$ between 6.6.6 (a) and 4.6.12 (b) Majorana surface code hexons and tetrons. The maximum weight of the operators during lattice surgery is 8 Majoranas for 6.6.6 codes, and 6 Majorana for 4.6.12 codes.}
\label{fig:surgerycombined}
\end{figure*}

\begin{figure*}[t]
\centering
\def\svgwidth{\linewidth}
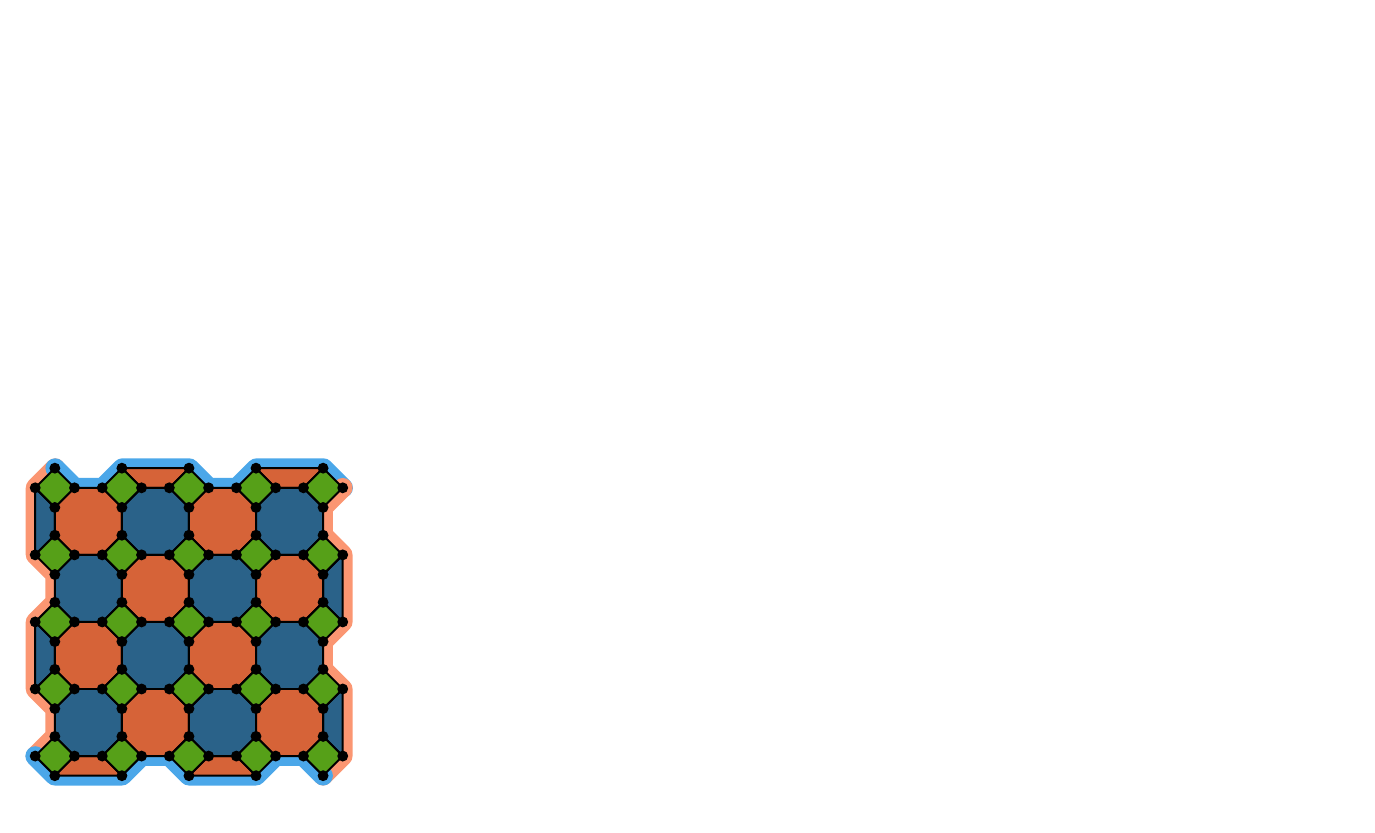
\caption{State injection protocols to convert a state encoded in a physical tetron into a logical 4.8.8 (a), 6.6.6 (b), or 4.6.12 (c) tetron for the example of $d=5$.}
\label{fig:injection}
\end{figure*}

\subsection{6.6.6 and 4.6.12 Majorana surface codes}

6.6.6 codes only require the measurement of 6-Majorana operators in the bulk of the code, as opposed to 8-Majorana operators for 4.8.8 codes.  In Fig.~\ref{fig:surgerycombined}a, we show the protocol to measure $Y \otimes X$ between a 6.6.6 hexon and tetron. This case is the most instructive, as it features standard lattice surgery, a dislocation line, and a twist defect. The measurement protocols for the other operators are shown in Appendix~\ref{app:surgery}. Even though 6.6.6 codes reduce the stabilizer weight to 6 Majoranas for most stabilizers, twist-based lattice surgery requires the measurement of some 8-Majorana operators that are part of dislocation lines. However, we can use the same trick as for 4.8.8 Majorana surface codes to reduce the weight of the lattice surgery operators. As we show in Appendix~\ref{app:666octons}, the weight of the dislocation line operators can be reduced to 6 Majoranas by replacing some of the boundary hexons with octons (8 Majoranas in a fixed parity sector), such that at most 6-Majorana parity measurements are required for quantum computing with 6.6.6 Majorana surface codes.

Figure \ref{fig:surgerycombined}b shows the same scenario for 4.6.12 codes. Here, both the dislocation line and the twist defect consist of operators that have a weight of six Majoranas. While the green building blocks along the dislocation line are drawn as decons (10-Majorana building blocks) in Fig.~\ref{fig:surgerycombined}b, as opposed to dodecons, this protocol can in fact be realized in a square lattice of dodecons. This is shown in Appendix~\ref{app:surgery}, where we also present the remaining lattice surgery protocols with 4.6.12 codes.

Thus, 4.6.12 Majorana surface codes require at most 6-Majorana parity measurements for quantum computation. Compared to bosonic surface codes, this comes at the price of a threefold increased space overhead of $\sim 12d^2$ Majoranas per qubit. This is reminiscent of bosonic subsystem surface codes~\cite{Bravyi2013}, which only require 3-qubit operator measurements (corresponding to 6 Majoranas), and also feature an increased space overhead of $\sim 4d^2$ physical qubits (corresponding to $\sim 12d^2$ Majoranas) per logical qubit. However, since these are bosonic codes, twist-based lattice surgery features a 5-qubit twist defect, corresponding to 10-Majorana operators. Since there exists no work on lattice surgery with these codes, we show the twist-based lattice surgery protocol with subsystem surface codes in Appendix~\ref{app:subsystem}, even though it is exactly the same as with standard bosonic surface codes.

While all of our Majorana surface code constructions can be applied to any Majorana platform, they can in particular be implemented using the quantum-wire constructions of Ref.~\cite{Karzig2016}. We show nanowire implementation of the three uniform Majorana surface codes in Appendix~\ref{app:wirecodes}.

\subsection{State injection}
So far, we have shown how to implement the operations Op1, Op2, and Op3 with 4.8.8, 6.6.6, and 4.6.12 Majorana surface codes, which allow tracking of Clifford gates, thus implementing them with zero time overhead. The remaining operation required for universality is the generation of (faulty) magic states that are encoded in tetrons. Unfortunately, even under the assumption of error-free $d=1$ tetrons, there is no robust way to initialize a magic state. One possibility to prepare a faulty magic state is to initialize a $\ket{+} = \frac{1}{\sqrt{2}}(\ket{0} + \ket{1})$ state by measuring $i\gamma_2\gamma_3$, and to split the degeneracy between the two states $\ket{0}$ and $\ket{1}$ by $\Delta E$ for a time $\tau = \pi/4 \cdot \hbar/\Delta E$. The dynamical phase difference between the states $\ket{0}$ and $\ket{1}$ will yield a magic state $\ket{m} = \frac{1}{\sqrt{2}}(\ket{0} + e^{i\pi/4}\ket{1})$. The degeneracy splitting can be implemented by coupling the two Majoranas $\gamma_1$ and $\gamma_2$, as is done for the measurement of $Z=i\gamma_1\gamma_2$. This protocol is obviously not robust against perturbations, as it requires precise control of the coupling parameters to generate an exact phase difference of $\pi/4$.
While there exist more sophisticated protocols for $\pi/8$ rotations in Majorana-based architectures~\cite{Bonderson2010,Karzig2015,Clarke2016}, magic state distillation is still required to obtain a low-error magic state.

Majorana magic gates or the coupling of two Majoranas can be used to obtain a noisy magic state encoded in a ($d=1$) tetron. Fault-tolerant quantum computing requires us to convert this magic state into a magic state encoded in a \textit{logical} tetron with a higher code distance. This is done via a protocol called state injection. There exist several protocols to inject an arbitrary state into a surface code~\cite{Horsman2012,Landahl2014,Li2015}. In particular, Ref.~\cite{Landahl2014} describes a two-step protocol that minimizes the time that the magic state spends with code distance $d=1$. Here, we adapt this protocol to fermionic codes.

Consider the stabilizer configuration shown in the top panel of Fig.~\ref{fig:injection}a. The number of stabilizers is the same as in a 4.8.8 Majorana surface code with $d=5$, i.e., exactly one qubit is encoded. Even though the four segments in the corners look like small surface codes, they do not encode any qubits, since they only have two different boundaries each and therefore only two corners. The encoded qubit is in fact the physical tetron in the center. Notice that the product of the Majoranas highlighted in red would correspond to a logical $X$ operator in an actual $d=5$ code, since it is a string connecting the two red boundaries that commutes with all stabilizers. In the stabilizer configuration in the top panel, this string also commutes with all stabilizers. Since the product of the Majoranas that are not located at the center tetron is fixed to be $+1$ by the stabilizers, the product of red Majoranas is equal to the $X$ information encoded in the center tetron. Similarly, the product of blue Majoranas is equivalent to the $Z$ information encoded in the center tetron, and would correspond to a blue-to-blue string of Majoranas in a $d=5$ code. By switching the stabilizer configuration from the top configuration to the bottom configuration, the state encoded in the physical tetron is converted into a logical tetron. As the highlighted logical operators commute with the stabilizers in both cases, they remain unchanged in the process. However, the measurement outcomes of the new stabilizers straddling the corner segments are random. The errors need to be corrected in such a way that the correction operation commutes with the highlighted logical operators. Note that since we start out with a $d=1$ qubit, state injection is not fault-tolerant, which further emphasizes the need for magic state distillation.

The injection procedures for 6.6.6 and 4.6.12 Majorana surface codes are very similar. In the 6.6.6 case in Fig.~\ref{fig:injection}b, the logical information is initially encoded in four of the six Majoranas of the center hexon. Again, the highlighted Majorana strings commute with both stabilizer configurations and correspond to the logical operators. For the 4.6.12 code in Fig.~\ref{fig:injection}c, the logical information is initially encoded in four of the 12 Majoranas of the center dodecon.

Thus, we have implemented all four operations necessary for universal quantum computation with tetrons and hexons with minimal-overhead Clifford gates. The Pauli product measurement protocol of Fig.~\ref{fig:pauliprodprotocol} can be implemented with \textit{logical} tetrons and hexons exactly the same way as was described in Sec.~\ref{sec:majoranagates} by replacing the measurements of four Majoranas by (twist-based) lattice surgery.

\section{Majorana color codes}
\label{sec:colorcodes}

In the previous section, we have shown that Majorana surface codes can be used to decrease the weight of the operators that need to be measured for fault-tolerant quantum computing. However, they come with a higher space overhead compared to bosonic surface codes. What about the opposite direction? How does one design topological codes with higher stabilizer weights, but, in turn, lower space overhead?

The answer lies in color codes. Bosonic color codes~\cite{Bombin2006,Landahl2011} are known to feature higher stabilizer weights compared to bosonic surface codes, but they can encode qubits more compactly, using fewer physical qubits per logical qubit of a given code distance. Color codes are closely related to surface codes, as they can be obtained by concatenating surface codes with small non-topological codes~\cite{Criger2016}. In an alternative construction, color codes are obtained by folding surface codes in half~\cite{Pastawski2015}. Here, we adapt the concatenation scheme to Majorana surface codes, in order to obtain Majorana color codes.\footnote{Note that in Ref.~\cite{Bravyi2010}, the term \textit{Majorana color code} was used to refer to Majorana surface codes.}

\begin{figure}[t]
\centering
\def\svgwidth{\linewidth}
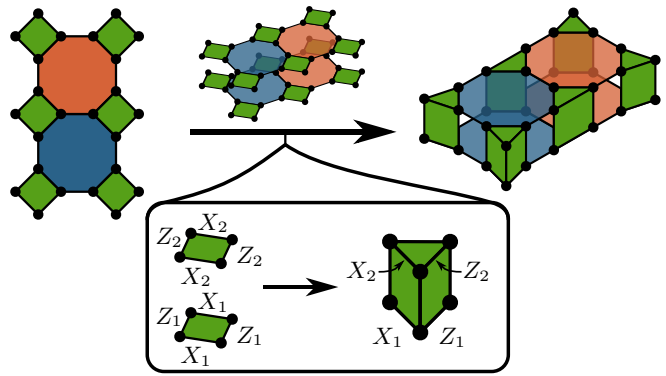
\caption{Scheme to obtain a 4.8.8 ($[[6,2,2]]_m$) Majorana color code by concatenating the 4.8.8 Majorana surface code with a $[[6,2,2]]_m$ code, i.e., by stacking two surface codes and replacing each pair of tetrons with a hexon.}
\label{fig:smallcolorcode}
\end{figure}

\begin{figure*}[t]
\centering
\def\svgwidth{\linewidth}
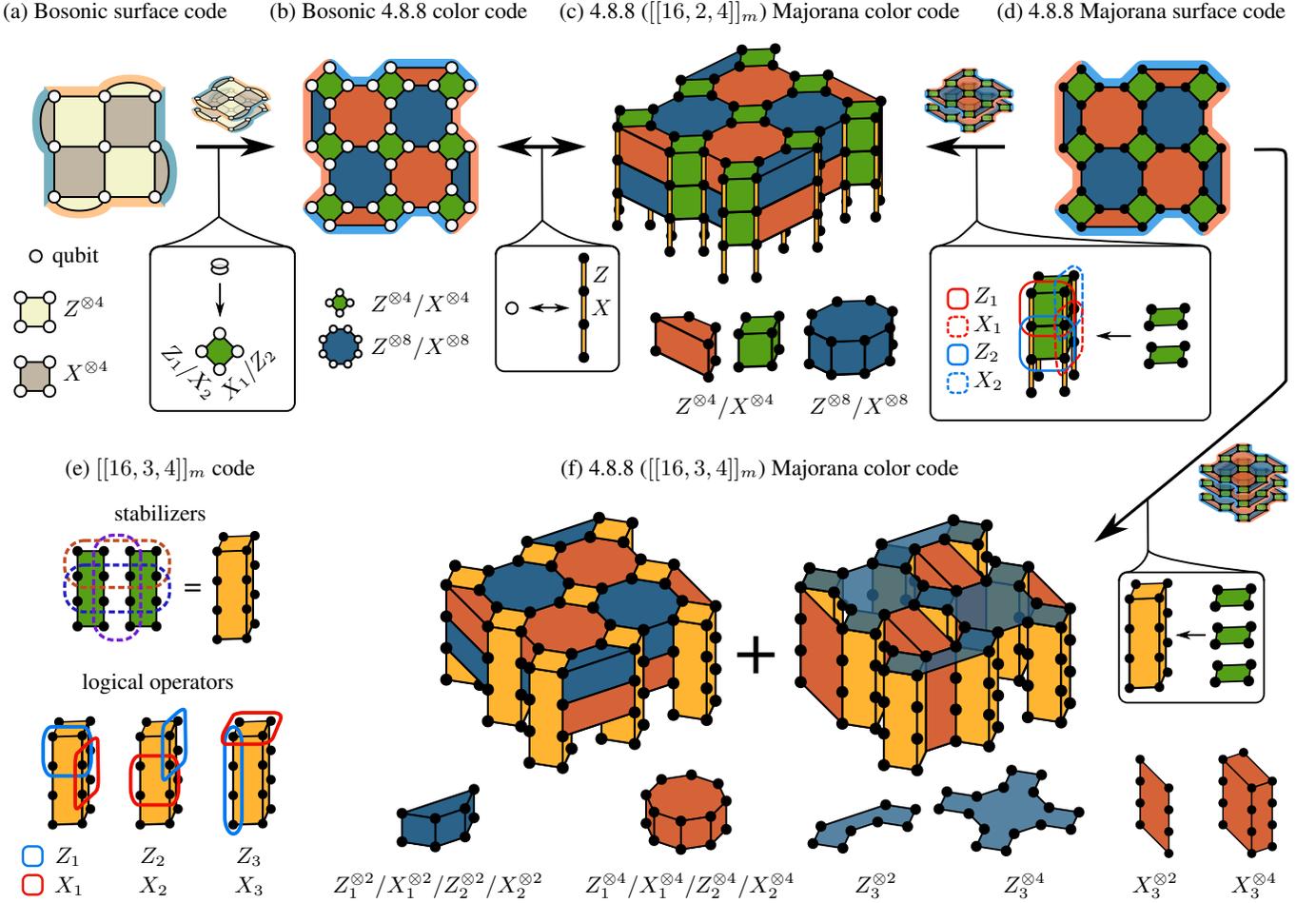
\caption{A bosonic surface code (a) concatenated with a $[[4,2,2]]$ code yields a bosonic 4.8.8 code (b). This can also be drawn as a Majorana fermion code (c) by replacing each qubit with a tetron (yellow lines). This Majorana color code can also be obtained by replacing each tetron of the 4.8.8 Majorana surface code with a $[[16,2,4]]_m$ code. (e) Stabilizers and logical operators of the $[[16,3,4]]_m$ code presented in Ref.~\cite{Hastings2017}. (f) Order-3 Majorana color code obtained by concatenating a 4.8.8 Majorana surface code with the $[[16,3,4]]_m$ code. Each yellow 16-Majorana box corresponds to a $[[16,3,4]]_m$ code. Since the stabilizers overlap, they are shown in two separate figures.}
\label{fig:colorcode}
\end{figure*}

For bosonic codes, concatenation means that each physical qubit of a code is replaced by another logical qubit. For instance, a 4.8.8 bosonic color code can be obtained by concatenating a bosonic surface code with a $[[4,2,2]]$ code~\cite{Criger2016}. This means that two bosonic surface codes are stacked on top of each other, and pairs of stacked qubits are replaced by $[[4,2,2]]$ codes, as shown in Fig.~\ref{fig:colorcode}a/b. Similarly, we propose that Majorana color codes can be obtained by concatenating Majorana surface codes with $[[n_m,k,d_m]]_m$ codes, where $[[n_m,k,d_m]]_m$ labels Majorana fermion codes that use $n$ Majoranas to encode $k$ qubits with Majorana distance $d_m$. 

We suggest to concatenate fermionic codes by stacking Majorana surface codes on top of each other and replacing stacked \textit{building blocks} of the code by $[[n_m,k,d_m]]_m$ codes. The simplest example of a Majorana color code is a 4.8.8 Majorana surface code concatenated with a $[[6,2,2]]_m$ code, which is shown in Fig.~\ref{fig:smallcolorcode}. The $[[6,2,2]]_m$ code uses six Majoranas to encode two qubits with Majorana distance $d_m=2$, which means that it is a hexon. In the concatenation procedure shown in Fig.~\ref{fig:smallcolorcode} for a small segment of a 4.8.8 surface code, two Majorana surface codes are placed on top of each other. Next, each stack of two tetrons is replaced by a single hexon. Hexons encode twice as many qubits as tetrons, but use only 1.5 times as many Majoranas to do so. Thus, a logical tetron built from the Majorana color codes shown in Fig.~\ref{fig:smallcolorcode} encodes two qubits as two layers of surface codes on top of each other, 
but uses only $3d^2$ Majoranas per qubit with distance $d$, while the stabilizer weight remains the same. Note that even though we draw Majorana color codes as three-dimensional stacks, they can also be implemented in 2D architectures. In particular, the 4.8.8 ($[[6,2,2]]_m$) Majorana color code can be implemented in a square lattice of hexons, but, in contrast to Majorana surface codes, the stabilizers now overlap, as we show in Appendix~\ref{app:2dcolorcode}.


\begin{figure*}[t]
\centering
\def\svgwidth{\linewidth}
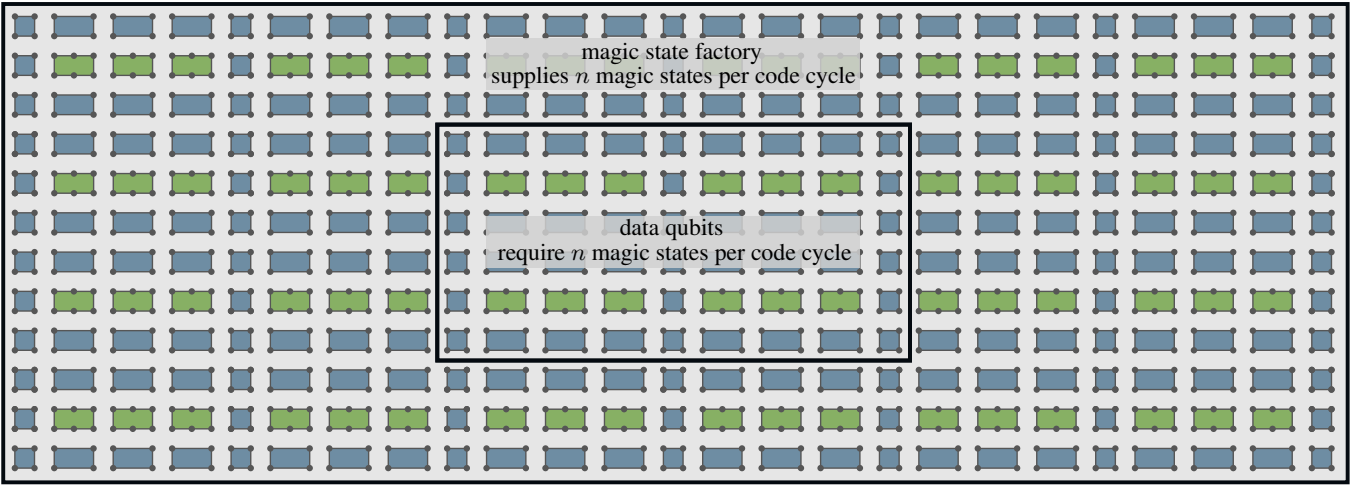
\caption{Scheme for quantum computation in an array of tetrons and hexons, where the hexons are used to encode qubits, and tetrons are used as ancillas for Pauli product measurements. One part of the array is used to encode the data qubits for quantum computation, whereas the rest of the quantum computer is used to distill magic states.}
\label{fig:quantumcomp}
\end{figure*}


More generally, we define Majorana color codes as any number of Majorana surface code layers obtained by concatenating Majorana surface codes with any other code. We refer to a code corresponding to $n$ layers of surface codes as an order-$n$ code. Stacking $k$ 4.8.8 Majorana surface codes and replacing each stack of $k$ tetrons with an $[[n_m,k,d_m]]_m$ code yields an order-$k$ Majorana color code. For instance, the previously discussed 4.8.8 ($[[6,2,2]]_m$) Majorana color code is an order-2 Majorana color code. Majorana surface codes are order-1 codes. One way to obtain an order-3 code is to place three 4.8.8 Majorana surface codes on top of each other, and replace each stack of three tetrons with an octon (an $[[8,3,2]]_m$ code). Since one of the logical operators of the octon is a 4-Majorana operator, the maximum stabilizer weight increases to 10 Majoranas, as we discuss in Appendix~\ref{app:octoncode}.

Bosonic color codes in their usual definition~\cite{Bombin2006} are \linebreak order-2 codes, as they correspond to two surface code layers. One such code is the previously mentioned bosonic 4.8.8 color code, which is obtained by replacing each qubit of a bosonic surface code (Fig.~\ref{fig:colorcode}a) with a $[[4,2,2]]$ code. This is a code that uses four qubits and has two stabilizers $Z \otimes Z \otimes Z \otimes Z$ and $X \otimes X \otimes X \otimes X$. It encodes two logical qubits with logical operators $Z_1 = Z \otimes Z \otimes \mathbbm{1} \otimes \mathbbm{1}$, $X_1 = \mathbbm{1} \otimes X \otimes X \otimes \mathbbm{1}$, and $Z_2 = \mathbbm{1} \otimes Z \otimes Z \otimes \mathbbm{1}$, $X_2 = X \otimes X \otimes \mathbbm{1} \otimes \mathbbm{1}$. Thus, a bosonic 4.8.8 color code tetron (see Fig.~\ref{fig:colorcode}b) is equivalent to two surface codes on top of each other.  It can be converted to a Majorana fermion code by replacing each qubit with a tetron. In Fig.~\ref{fig:colorcode}c, these tetron building blocks are the yellow strings of four Majoranas. Another way of obtaining this Majorana fermion code is by stacking two  4.8.8 Majorana surface codes and replacing pairs of tetrons with $[[16,2,4]]_m$ codes, which is the Majorana representation of the bosonic $[[4,2,2]]$ code. The bosonic 4.8.8 color code uses four times as many physical qubits (or Majoranas) to encode twice as many logical qubits with twice the code distance. Thus, the Majorana overhead of bosonic 4.8.8 color codes is $2d^2$ per qubit, which is half the overhead of bosonic surface codes. However, the stabilizer weight doubles from 8 Majoranas to 16 Majoranas.

The $[[16,2,4]]_m$ code used to obtain the bosonic 4.8.8 color code by concatenation is a bosonic code in the sense that its building blocks are tetrons. With Majoranas, there is also the possibility to use Majorana fermion codes for concatenation. These codes can encode information more compactly. In Ref.~\cite{Hastings2017}, a collection of small Majorana fermion codes is presented. The smallest $d_m=4$ Majorana fermion code is the $[[16,3,4]]_m$ code shown in Fig.~\ref{fig:colorcode}e, which uses 16 Majoranas to encode three instead of just two logical qubits with a Majorana distance of $d_m=4$. It is based on two octons with three additional 8-Majorana operators as stabilizers. All of its logical operators are weight-4 Majorana operators. The \linebreak order-3 Majorana color code obtained by concatenating the 4.8.8 surface code with the $[[16,3,4]]_m$ code is shown in Fig.~\ref{fig:colorcode}f. It essentially corresponds to three layers of Majorana surface codes and has a reduced Majorana overhead of $\frac{4}{3}d^2$ with octons as building blocks, while still featuring a maximum stabilizer weight of 16 Majoranas.

In general, concatenating a bosonic surface code with an $[[n,k,d_b]]$ bosonic code increases the number of physical qubits by a factor of $n$, the number of logical qubits by a factor of $k$, and the code distance by a factor of $d_b$. Thus, the number of physical qubits per logical qubit goes from $d^2$ to $\frac{n}{kd_b^2}d^2$. The maximum stabilizer weight of the code may increase from 4 to $4w_{\rm max}$, where $w_{\rm max}$ is the maximum weight of the logical operators of the $[[n,k,d_b]]$ code. Similarly, a 4.8.8 Majorana surface code concatenated with an $[[n_m,k,d_m]]_m$ code decreases the Majorana overhead from $4d^2$ to $\frac{4 n_m}{k d_m^2} d^2$ Majoranas for a logical qubit with code distance $d$.

The second smallest $d_m=4$ Majorana fermion code is the $[[20,4,4]]_m$ code. Its stabilizers and logical operators are shown in Appendix~\ref{app:2044code}. The code is based on two decons (10 Majoranas in a fixed parity sector) and four 10-Majorana stabilizers, and encodes four qubits whose logical operators all have a weight of 4 Majoranas. Thus, the order-4 Majorana color code obtained by concatenating a 4.8.8 surface code with a $[[20,4,4]]_m$ has a Majorana overhead of $1.25d^2$ while still only featuring a maximum stabilizer weight of 16 Majoranas. Since some of the logical operators of the third-smallest $d_m = 4$ code, the $[[24,6,4]]_m$ code, have a weight of 6 Majoranas, the stabilizer weight of a Majorana code obtained by concatenating this code with a 4.8.8 Majorana surface code would exceed 16 Majoranas. In general, Majorana surface codes can be concatenated with arbitrary Majorana fermion codes to obtain codes with a smaller Majorana overhead, while increasingly sacrificing stabilizer weight and locality of the code.

Majorana color codes can be used for quantum computation the same way as Majorana surface codes, i.e., by encoding tetrons and hexons, and performing lattice surgery. In fact, the stabilizer weight and space overhead of the lattice surgery protocol can be reduced by using surface-to-color code lattice surgery~\cite{Litinski2017a}, such that the ancilla tetron is not a color code, but a surface code. For the previously discussed codes with 16-Majorana stabilizers, surface-to-color code lattice surgery has a maximum stabilizer weight of 12 Majoranas. We show an example of this in Appendix~\ref{app:colortosurface}.

In summary, we have described Majorana color codes, which are multiple layers of Majorana surface codes obtained by concatenation. The advantage of Majorana color codes compared to bosonic color codes is that they can encode logical qubits more compactly with the same stabilizer weight. In particular, the Majorana color code obtained by concatenating the 4.8.8 Majorana surface code with the $[[20,4,4]]_m$ code has a Majorana overhead of $1.25d^2$, compared to the Majorana overhead of $2d^2$ of the bosonic 4.8.8 color code.

\section{Conclusion}
\label{sec:conclusion}

We have described a unified framework for fault-tolerant quantum computation with Majorana-based qubits. A full quantum computer could look like the array of tetrons and hexons shown in Fig.~\ref{fig:quantumcomp}. Hexons are used to encode qubits, while tetrons are ancillas that are used during the Pauli product measurement protocol. One part of the quantum computer consists of the data qubits that are used for quantum computation, which requires a certain number of magic states $n$ per code cycle as a resource for $T$ gates, where $n$ depends on the given quantum computation. In order for magic state distillation not to be a bottleneck of the quantum computer, the resources devoted to magic state distillation need to be large enough to supply $n$ magic states per code cycle. The main operation in addition to the preparation of faulty magic states is the measurement of Pauli product operators according to the protocol in Fig.~\ref{fig:pauliprodprotocol}. Taking full advantage of the Gottesman-Knill theorem, this enables the classical tracking of all Clifford gates, which means that these gates require zero time overhead.

In principle, the scheme with $d=1$ tetrons and hexons shown in Fig.~\ref{fig:quantumcomp} can be implemented in a nanowire array in the spirit of Ref.~\cite{Karzig2016}, as shown in Appendix~\ref{app:wireimplementation}. However, this only allows for quantum computation on time scales of the order of the coherence times of the physical tetrons and hexons. For fault-tolerant quantum computation, each tetron and hexon needs to be replaced by a \textit{logical} tetron and hexon with an appropriate code distance, and 4-Majorana parity measurements need to be replaced by lattice surgery.

There is a wide variety of codes that one can choose for this purpose, a collection of which is shown in Tab.~\ref{tab:results}. Surface codes only require the measurement of low-weight stabilizers, but feature a high space overhead. Majorana surface codes feature lower-weight stabilizers compared to bosonic surface codes. In particular, 4.8.8 Majorana surface codes have the same bulk stabilizer weights as bosonic surface codes, but a lower stabilizer weight for lattice surgery of 8 Majoranas compared to 10 Majoranas. 6.6.6 and 4.6.12 Majorana surface codes reduce the bulk stabilizer weights to an average of 6 and 5 Majoranas, respectively, but at the same time increase the Majorana overhead by a factor of 1.5 and 3, respectively.

Color codes can reduce the space overhead at the price of higher-weight stabilizers. Majorana color codes encode qubits more compactly compared to bosonic color codes. The 4.8.8 ($[[6,2,2]]_m$) Majorana color code has the same stabilizer weights as a bosonic surface code, but features a lower Majorana overhead. Similarly, the 4.8.8 ($[[20,4,4]]_m$) Majorana color code has the same stabilizer weights as the bosonic 4.8.8 color code, but a lower Majorana overhead by a factor of 5/8.

Our entire scheme can also be realized with non-topological (e.g., superconducting) qubits, but is then limited to the use of bosonic codes. From the perspective of Majorana fermion codes, non-topological qubits can only implement logical tetrons and hexons with physical tetrons as building blocks, whereas Majorana-based qubits can also implement codes with other building blocks such as physical hexons, octons, decons, or dodecons.

There are many questions that remain unanswered. Even though our protocols allow the classical tracking of CNOT gates, it remains uncertain whether tracking is always advantageous compared to actually executing theses gates. After all, if a layer of Clifford gates is followed by a layer of $T$ gates, all of these $T$ gates can be executed simultaneously. However, if CNOTs are tracked, these $T$ gates require spatially overlapping Pauli product measurements. While our measurement protocol shows how arbitrary Pauli products can be measured, it is unclear how multiple (commuting) Pauli products can be read out simultaneously. If it turns out that it is more advantageous to execute CNOT gates explicitly, our Pauli product measurement protocol can be straightforwardly used for multi-target CNOT gates using the protocol in Appendix~\ref{app:clifftracking}.

Moreover, while bosonic surface-code decoders can be used to decode Majorana surface codes, one could also devise decoding schemes that are tailored towards Majorana fermion codes. In particular, in a setting where parity-fixing constraints can be violated and stabilizers of all colors are measured, a decoder needs to be able to match three types of anyons, as opposed to just two. Furthermore, one could adapt the scheme of Majorana-based fermionic quantum computation~\cite{Akhmerov2017} to the fault-tolerant setting by replacing arrays of Majoranas with arrays of twist defects in Majorana surface codes. Based on twists in Majorana surface codes, it would also be interesting to see whether it is possible to come up with a Majorana version of the twist-based triangle code presented in Ref.~\cite{Yoder2017}. Also, we did not investigate the performance of Majorana fermion codes, i.e., the logical error rate under the assumption of a realistic error model. While it is known that the bit-flip error thresholds of bosonic surface and color codes without measurement errors are the same~\cite{Katzgraber2009,Andrist2016}, the error thresholds of codes corresponding to an arbitrary number of surface-code layers are unknown. We hope that the framework presented in this work will prove useful for Majorana-based quantum computation.

\section*{Acknowledgments}
The authors would like to thank Markus Kesselring, Jens Eisert, Theodore Yoder, Michael Beverland, Benjamin Brown, and Aleksander Kubica for insightful discussions. This work has been supported by the Deutsche Forschungsgemeinschaft (Bonn) within the network CRC TR 183.

\begin{figure}[t]
\centering
\def\svgwidth{0.85\linewidth}
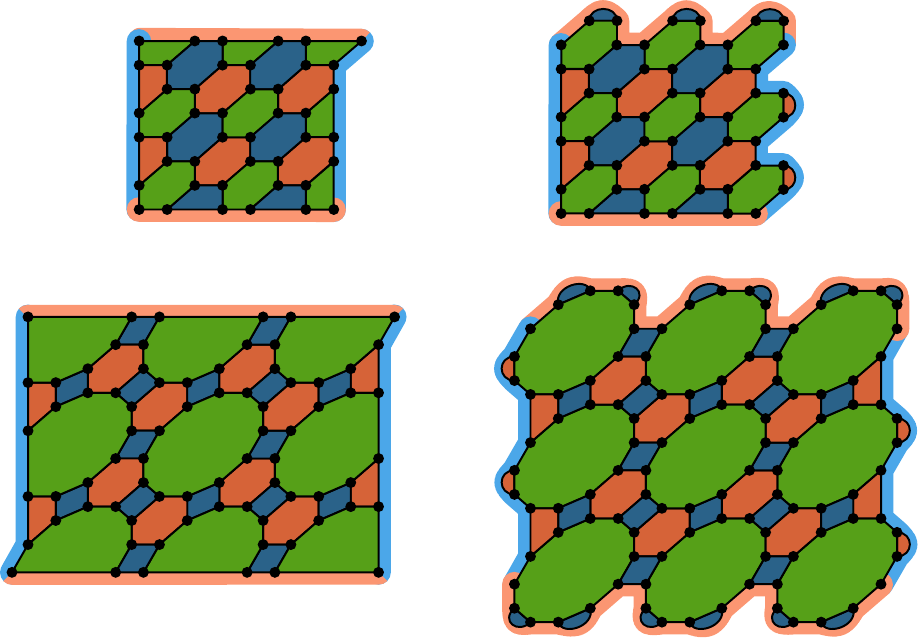
\caption{6.6.6 and 4.6.12 surface code tetrons in a square lattice of hexons and dodecons, respectively.}
\label{fig:squarelattice}
\end{figure}

\begin{figure}[t]
\centering
\def\svgwidth{\linewidth}
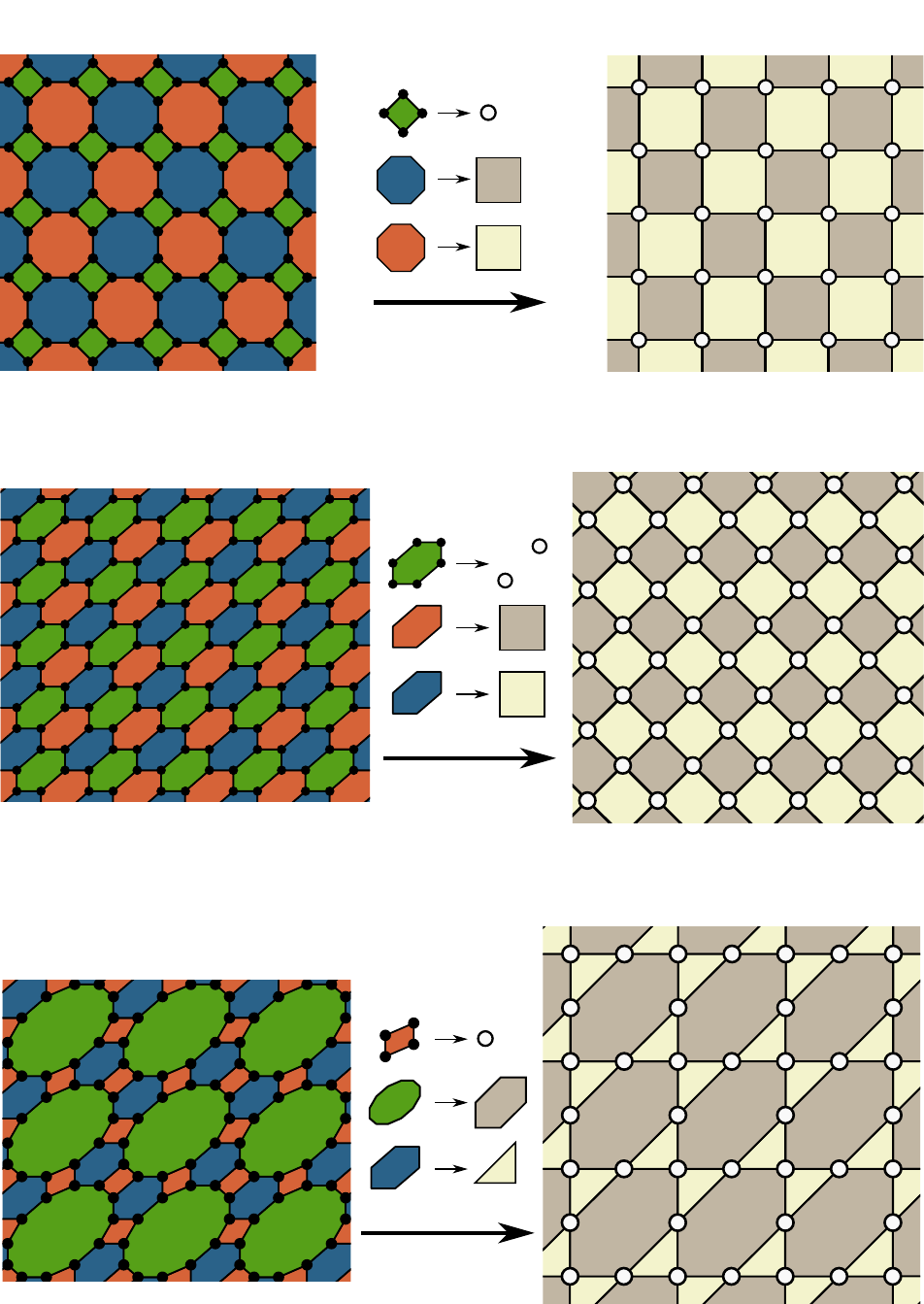
\caption{One choice of mapping of 4.8.8, 6.6.6, and 4.6.12 Majorana surface codes onto bosonic surface codes. For bosonic surface codes, the dark plaquettes are products of $X$ operators, and the light plaquettes are products of $Z$ operators.}
\label{fig:bosonicmap}
\end{figure}


\appendix

\renewcommand\theparagraph{\arabic{paragraph}}

\section{Majorana surface codes}


\refstepcounter{subsection}
\label{app:squarelattice}
{\bf \arabic{subsection}. 6.6.6 and 4.6.12 Majorana surface codes in square lattices of hexons and dodecons.}
Some of the logical surface code tetrons and hexons in Figs.~\ref{fig:logicaltetrons} and \ref{fig:logicalhexons} feature smaller building blocks (green stabilizers) at the boundary compared to the building blocks in the bulk. In particular, some of the green stabilizers of 6.6.6 surface codes are tetrons at the boundary, whereas they are hexons in the bulk. Similarly, 4.6.12 surface codes feature hexons, octons, and decons at the boundary, as opposed to dodecons. Still, all of these codes can be implemented on a square lattice of building blocks (green stabilizers), where each building block is the same. The equivalence in Fig.~\ref{fig:squarelattice} shows that 6.6.6 Majorana surface codes can be implemented on a square lattice of hexons, and 4.6.12 codes can be implemented on a square lattice of dodecons. The two-qubit plaquettes at the boundary fix the parity of some of the Majoranas of the hexons and dodecons, such that they effectively become tetrons, hexons, octons or decons.


\refstepcounter{subsection}
\label{app:bosonicmap}
{\bf \arabic{subsection}. Mapping Majorana surface codes onto bosonic surface codes.}
While every bosonic code can be uniquely mapped onto a Majorana fermion code by replacing each qubit with a tetron, the reverse mapping is not unique. For Majorana surface codes, one prescription to map them onto a bosonic code is to interpret each stabilizer of a certain color of the three-colorable tiling as a number of qubits. For instance, for the 4.8.8 tiling, on has the choice of either interpreting the \linebreak 8-Majorana stabilizers of one color as three qubits encoded in an octon, or of interpreting all 4-Majorana stabilizers as tetrons. The latter mapping precisely yields the bosonic surface code on a square lattice, as shown in Fig.~\ref{fig:bosonicmap}a. Note that in this mapping, neighboring tetrons have different orientations.

For the 6.6.6 Majorana surface code in Fig.~\ref{fig:bosonicmap}b, we interpret each green stabilizer as a hexon, where one qubit is encoded in the bottom left three Majoranas, and the other qubit in the top right three Majoranas. Thus, we replace each green hexon with two qubits, and obtain a bosonic surface code on a rotated square lattice. For the 4.6.12 Majorana surface code there are again several possibilities. In Fig.~\ref{fig:bosonicmap}c, we show a mapping where each 4-Majorana stabilizer is replaced by a tetron, which yields a bosonic surface code on a Kagome lattice.

\begin{figure}[b]
\centering
\def\svgwidth{0.7\linewidth}
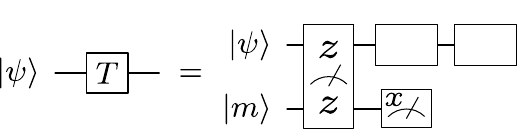
\caption{A $T$ gate on a qubit $\ket{\psi}$ is equivalent to a $Z \otimes Z$ measurement between $\ket{\psi}$ and a magic state $\ket{m}$ with outcome $m_1$, which results in a corrective $S^{m_1}$ operation on $\ket{\psi}$. In order to disentangle $\ket{m}$ from the qubit, it is measured in the $X$ basis, prompting a $Z^{m_2}$ Pauli correction on $\ket{\psi}$.}
\label{fig:tgatetrack}
\end{figure}

\begin{figure*}
\centering
\def\svgwidth{\linewidth}
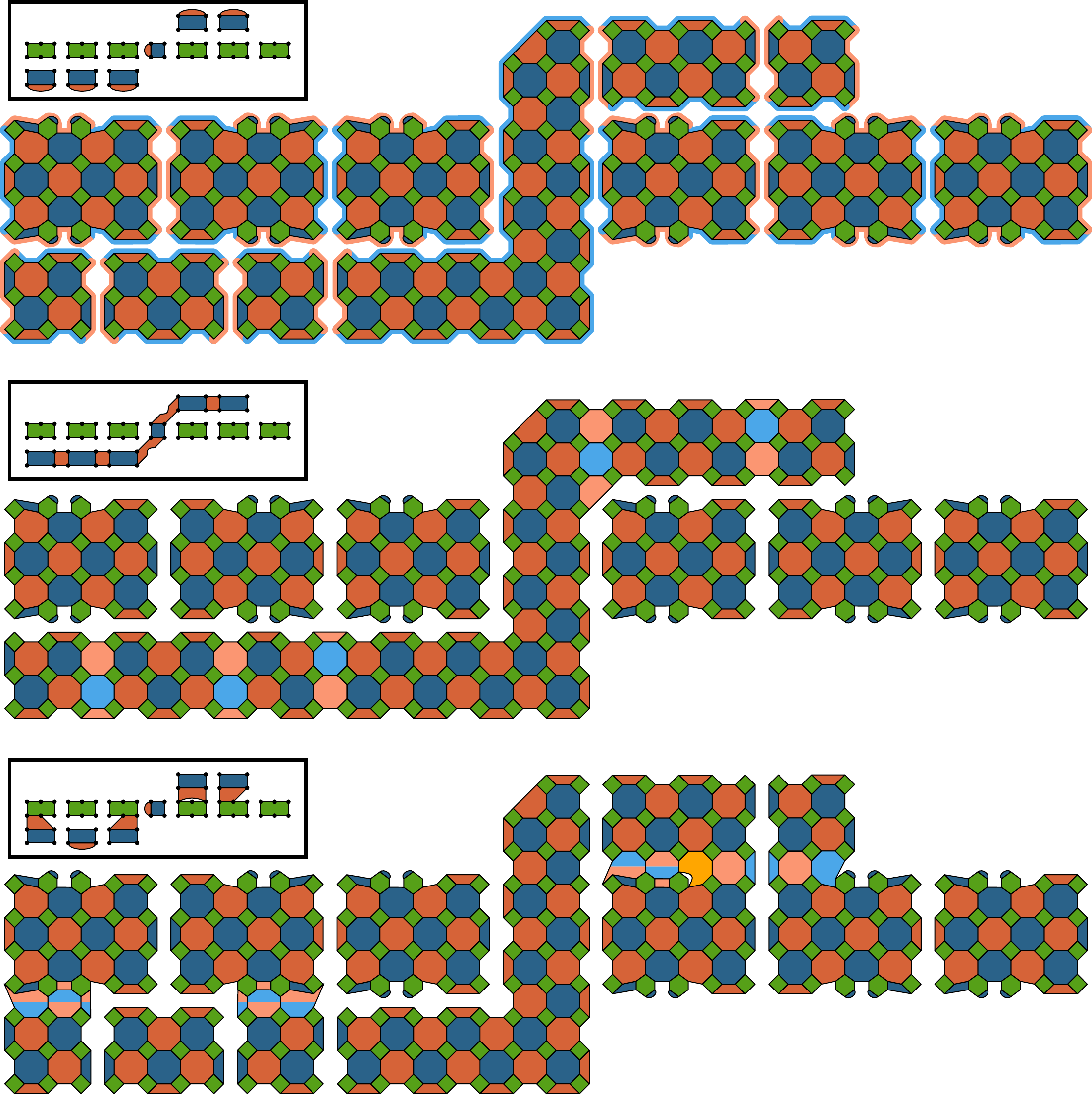
\caption{Pauli product measurement protocol from Fig.~\ref{fig:pauliprodprotocol} realized with $d=3$ 4.8.8 Majorana surface codes. Only steps 1, 2, and 3 are shown.}
\label{fig:pauliprodcode}
\end{figure*}

\begin{figure*}
\centering
\def\svgwidth{\linewidth}
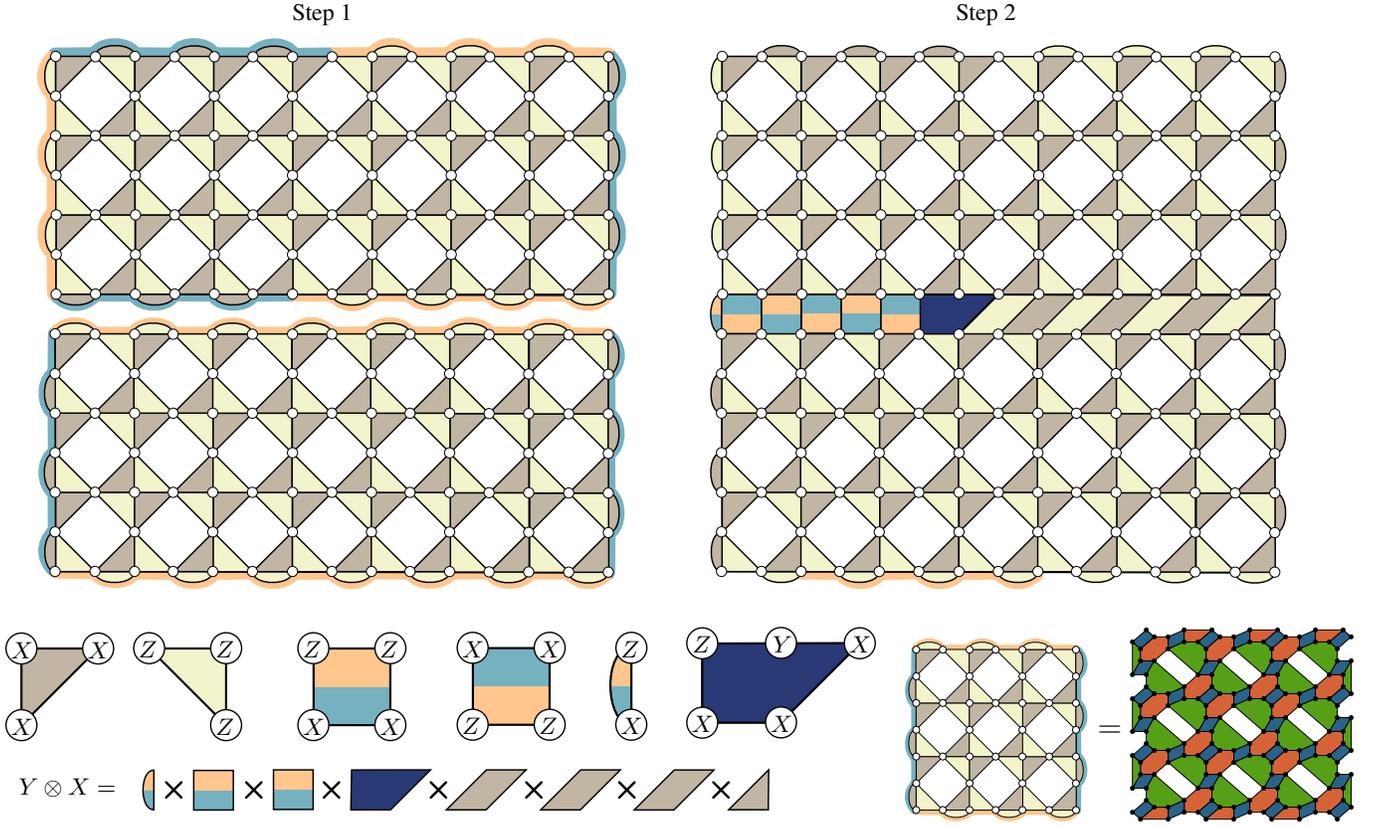
\caption{Twist-based lattice surgery between a logical hexon and a logical tetron encoded in a $d=3$ bosonic subsystem surface code in order to measure $Y \otimes Z$. The Majorana fermion code obtained by replacing each qubit of a bosonic subsystem surface code tetron with four Majoranas in a fixed parity sector resembles a 4.6.12 Majorana surface code.}
\label{fig:subsystemsurgery}
\end{figure*}

\begin{figure}[b]
\centering
\def\svgwidth{0.83\linewidth}
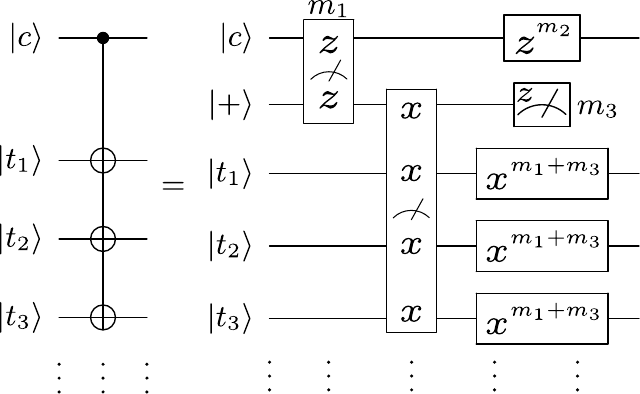
\caption{A multi-target CNOT gate between a control qubit $\ket{c}$ and multiple target qubits $\ket{t_i}$ is equivalent to a $Z \otimes Z$ measurement between $\ket{c}$ and an ancilla initialized in the $\ket{+}$ state, followed by a $X^{\otimes n+1}$ measurement between the ancilla and the $n$ target qubits. Finally, the ancilla is measured in the $Z$ basis. The final Pauli corrections depend on the measurement results.}
\label{fig:cnotcircuit}
\end{figure}

\begin{figure*}
\centering
\def\svgwidth{0.9\linewidth}
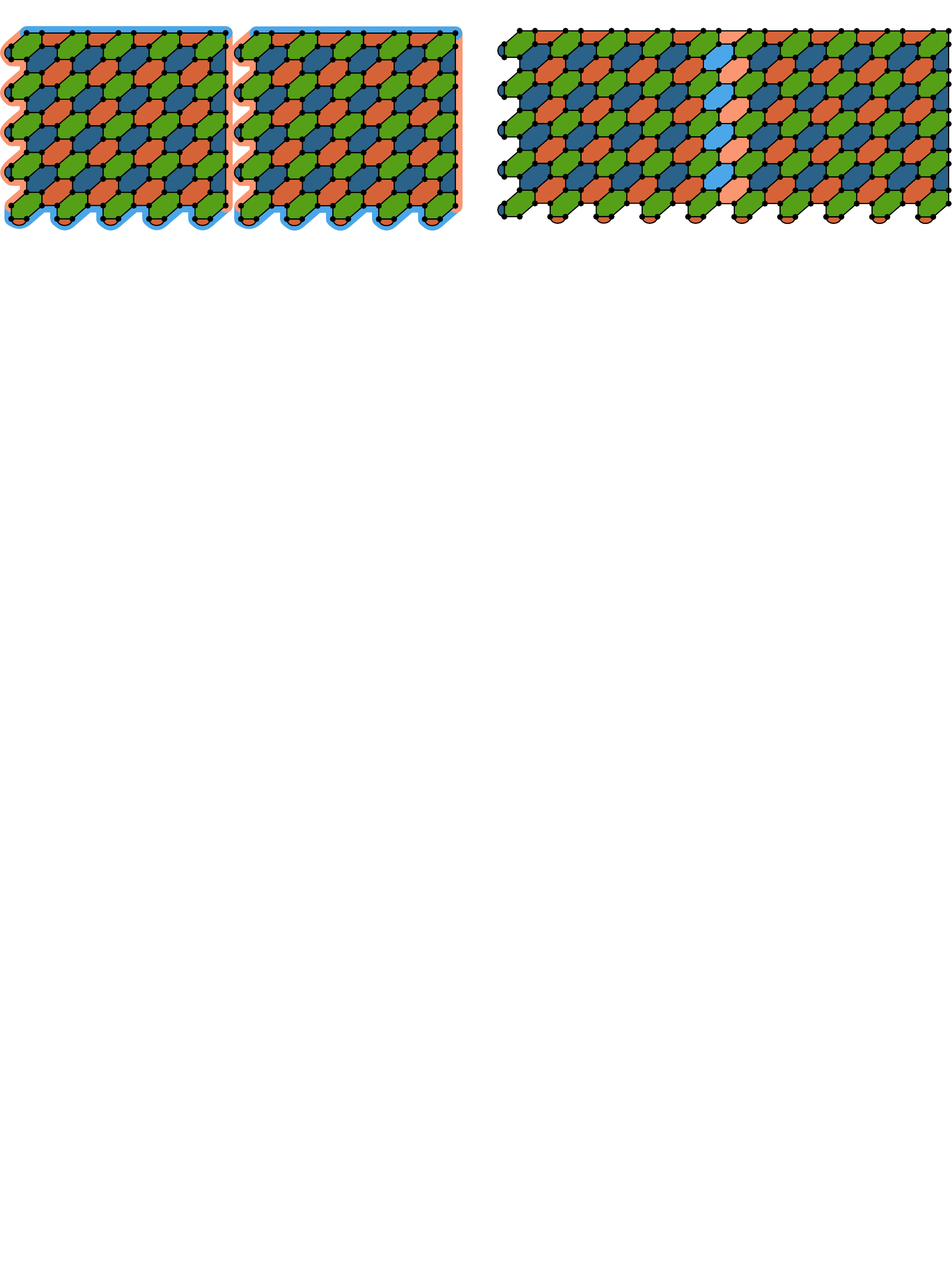
\caption{Stabilizer configuration for $Z \otimes Z$ measurements between two 6.6.6 Majorana surface code tetrons, and for $X \otimes X$ and $Z \otimes X$ measurements between a 6.6.6 hexon and a 6.6.6 tetron.}
\label{fig:666surgeryapp}
\end{figure*}

\begin{figure*}
\centering
\def\svgwidth{0.9\linewidth}
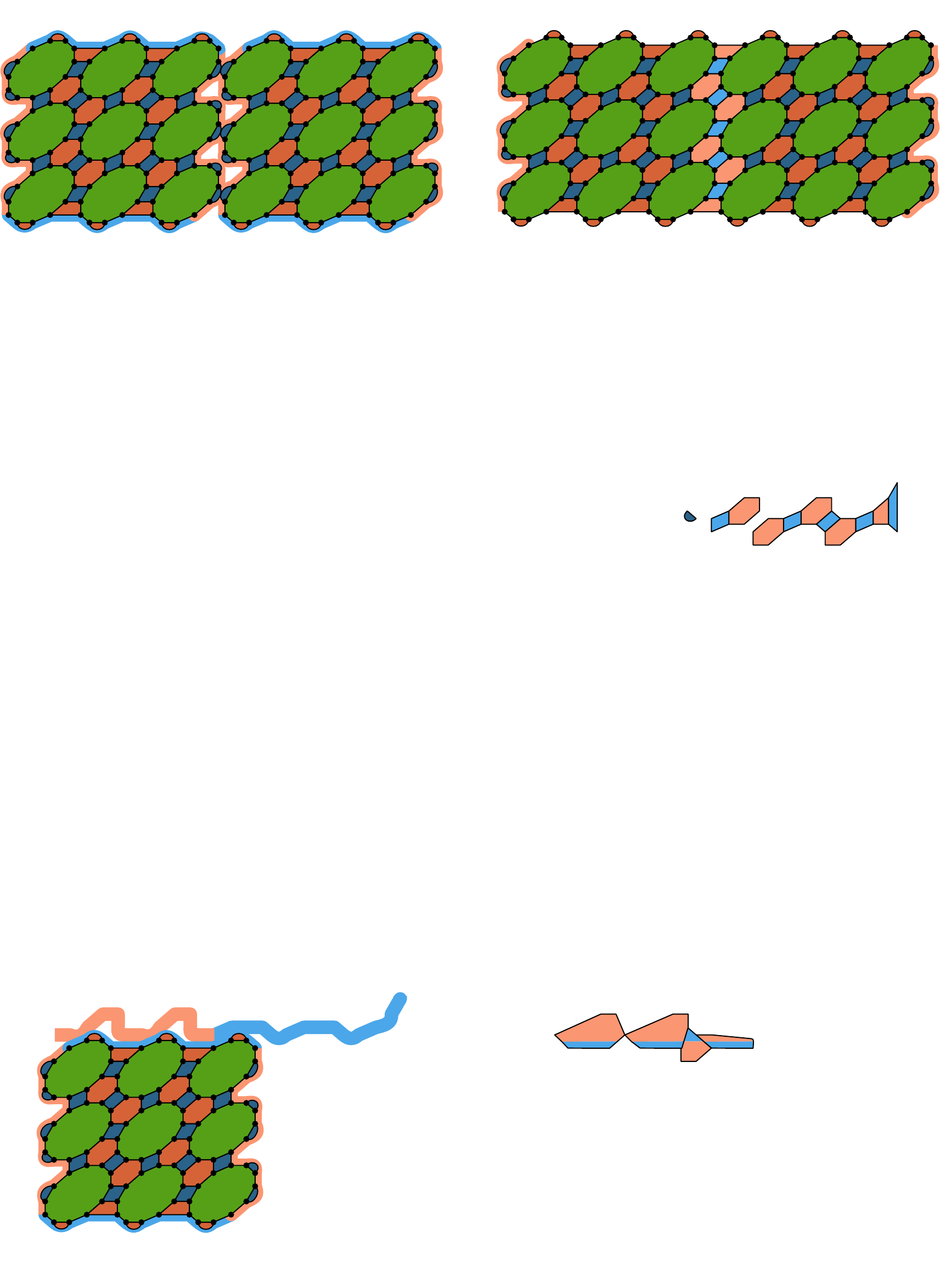
\caption{Stabilizer configuration for $Z \otimes Z$ measurements between two 4.6.12 Majorana surface code tetrons, and for $X \otimes X$ and $Z \otimes X$ measurements between a 4.6.12 hexon and a 4.6.12 tetron.}
\label{fig:4612surgeryapp}
\end{figure*}

\begin{figure*}
\centering
\def\svgwidth{0.9\linewidth}
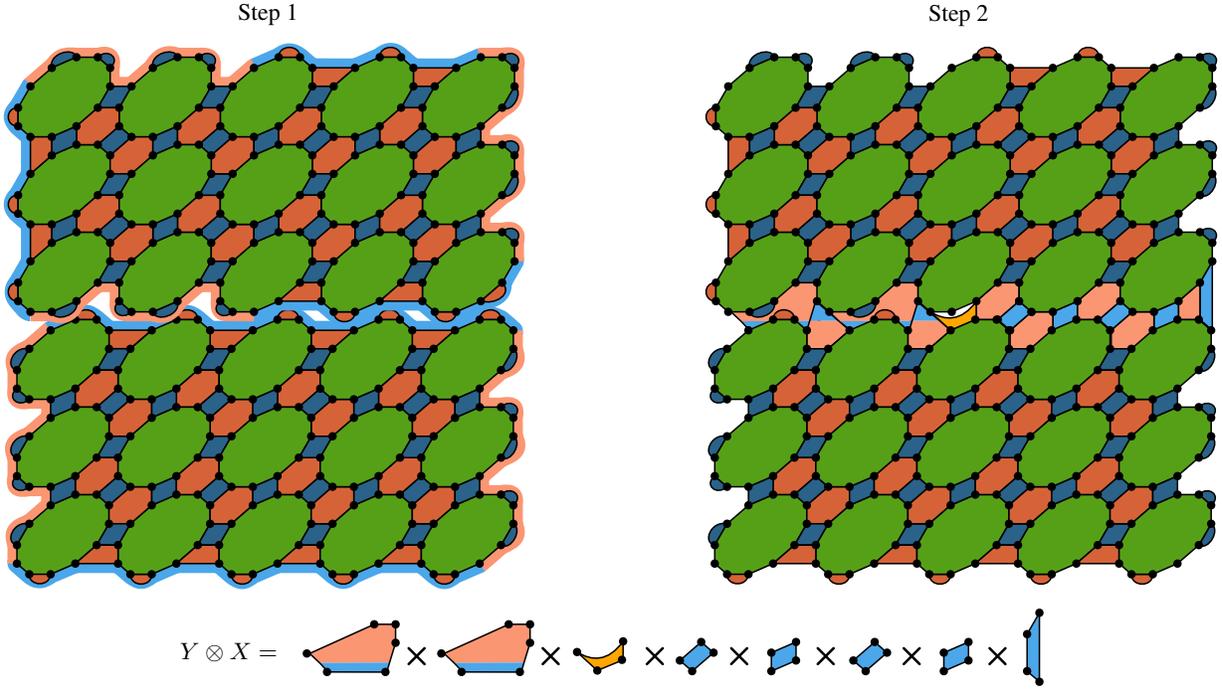
\caption{Stabilizer configuration for $Y \otimes Z$ measurements between a 4.6.12 Majorana hexon and tetron.}
\label{fig:4612surgeryapp2}
\end{figure*}

\begin{figure*}
\centering
\def\svgwidth{\linewidth}
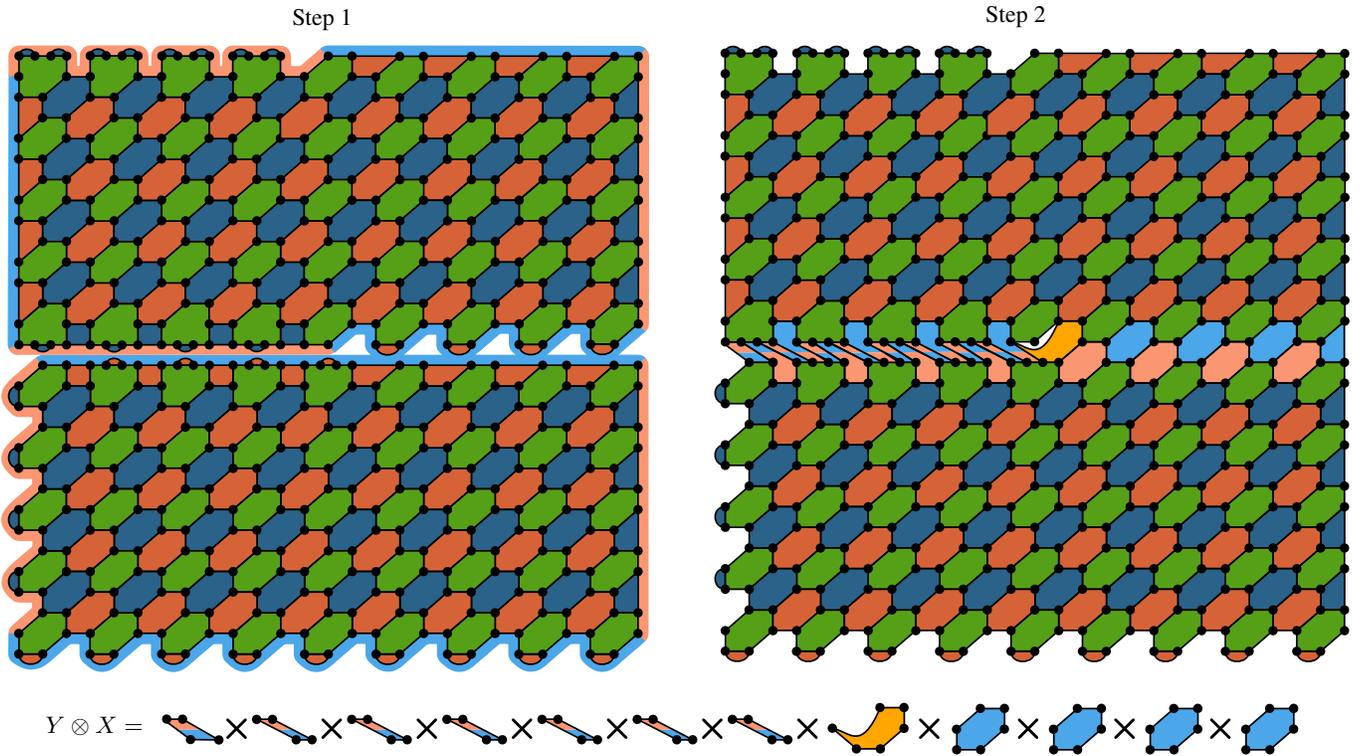
\caption{Stabilizer configuration for $Y \otimes Z$ measurements between a 6.6.6 Majorana hexon and tetron. Some of the physical hexons located at the boundary have been replaced by octons. As a consequence, the maximum stabilizer weight reduces to 6 Majoranas.}
\label{fig:666surgeryoctons}
\end{figure*}


\refstepcounter{subsection}
\label{app:clifftracking}
{\bf \arabic{subsection}. State injection and CNOT gates with Pauli product measurements.}
In a quantum computing architecture where all Clifford gates are tracked and only Pauli product measurements are explicitly performed, the state injection circuit from Fig.~\ref{fig:circuitswide}a is replaced by Fig.~\ref{fig:tgatetrack}. In particular, the tracked CNOT gate maps the $Z$ measurement of the magic state onto a $Z \otimes Z$ measurement of $\ket{\psi} \otimes \ket{m}$. However, this leaves the qubits $\ket{\psi}$ and $\ket{m}$ in an entangled state, which means that $\ket{m}$ cannot be discarded right away after the state injection. With $\ket{\psi} = \alpha \ket{0} + \beta \ket{1}$ and $\ket{m} = \frac{1}{\sqrt 2}(\ket{0} + e^{i\pi/4}\ket{1})$, the initial state before any measurement is
\begin{equation}
	\ket{\Psi} = \frac{1}{\sqrt 2}\left(\alpha \ket{00} + \alpha e^{i\pi/4}\ket{01} + \beta \ket{10} + \beta e^{i\pi/4} \ket{11}\right) \, .
\end{equation}
After the $Z \otimes Z$ measurement, there are two possible outcomes
\begin{equation}
	\ket{\Psi_0} = \alpha \ket{00} + \beta e^{i\pi/4} \ket{11}
\end{equation}
for outcome $m_1 = 0$, and
\begin{equation}
	\ket{\Psi_{1}} = \alpha e^{i\pi/4} \ket{01} + \beta \ket{10} \approx \alpha \ket{01} + \beta e^{-i\pi/4} \ket{10}
\end{equation}
for outcome $m_1 = 1$, where ``$\approx$'' is an equality up to a global phase. After an $S^{m_1}$ correction, $\ket{\Psi_{0}}$ remains unchanged, whereas $\ket{\Psi_{1}}$ is transformed to
\begin{equation}
	S\ket{\Psi_{1}} = \alpha \ket{01} + \beta e^{i\pi/4} \ket{10} \, .
\end{equation}
Both $\ket{\Psi_0}$ and $S\ket{\Psi_{1}}$ are entangled states. If one wants to discard the second qubit in order to continue to use it for magic state distillation, it first needs to be disentangled from the rest of the system. This can be done by measuring the magic state in the $X$ basis with outcome $m_2$, followed by a $Z^{m_2}$ Pauli correction on the qubit $\ket{\psi}$. For $\ket{\Psi_0}$, an $X$ measurement leaves the state in
\begin{equation}
	\ket{\Psi_{0,0}} = \left(\alpha\ket{0} + \beta e^{i\pi/4}\ket{1} \right)\otimes \ket{+}
\end{equation}
for outcome $m_2=0$, and
\begin{equation}
	\ket{\Psi_{0,1}} = \left(\alpha\ket{0} - \beta e^{i\pi/4}\ket{1} \right)\otimes \ket{-}
\end{equation}
for outcome $m_2=1$. A $Z^{m_2}$ correction leaves $\ket{\Psi_{0,0}}$ unchanged and maps $\ket{\Psi_{0,1}}$ to
\begin{equation}
	Z\ket{\Psi_{0,1}} = \left(\alpha\ket{0} + \beta e^{i\pi/4}\ket{1} \right)\otimes \ket{-} \, .
\end{equation}
Similarly, $S\ket{\Psi_{1}}$ after an $X$ measurement becomes $\ket{\Psi_{0,0}}$ for outcome $m_2=0$, and $-\ket{\Psi_{0,1}} \approx \ket{\Psi_{0,1}}$ for outcome \linebreak $m_2=1$, which after a $Z$ correction becomes $Z\ket{\Psi_{0,1}}$. For both $\ket{\Psi_{0,0}}$ and $Z\ket{\Psi_{0,1}}$, the two qubits are disentangled, and the first qubit is in the state $T\ket{\psi} = \ket{0} + e^{i\pi/4}\ket{1}$, which is the desired outcome for state injection. We stress that neither the $S^{m_1}$ nor the $Z^{m_2}$ correction need to be performed explicitly, but can be tracked by a classical computer.

We also note that in a quantum computer that only implements Pauli product measurements, CNOT gates can still be performed explicitly. To this end, one can use the circuit identity for multi-target CNOT gates~\cite{Zilberberg2008,Litinski2017} shown in Fig.~\ref{fig:cnotcircuit}. Here, a CNOT gate between a control qubit $\ket{c}$ and $n$ target qubit $\ket{t_i}$ is equivalent to three Pauli product measurements with Pauli corrections that depend on the measurement results.

\refstepcounter{subsection}
\label{app:logicalpauliprod}
{\bf \arabic{subsection}. Pauli product measurement protocol with 4.8.8 Majorana surface codes.}
In Fig.~\ref{fig:pauliprodcode}, we show how the example of a Pauli product measurement in Fig.~\ref{fig:pauliprodprotocol} could be realized using 4.8.8 Majorana surface codes with $d=3$. While the lattice surgery steps are shown, the $X$ and $Z$ measurements of logical tetrons are done via the measurement of all 2-Majorana operators corresponding to red and blue edges, respectively.

\refstepcounter{subsection}
\label{app:surgery}
{\bf \arabic{subsection}. Lattice surgery protocols for 6.6.6 and 4.6.12 Majorana surface codes.}
In Fig.~\ref{fig:666surgeryapp}, we show the stabilizer configurations for the lattice surgery operations to measure $Z \otimes Z$ between two 6.6.6 Majorana surface code tetrons, and $X \otimes X$ and $Z \otimes X$ between hexons and tetrons. The same lattice surgery operations are shown for 4.6.12 Majorana surface codes in Fig.~\ref{fig:4612surgeryapp}. Since Fig.~\ref{fig:surgerycombined}b shows the $Y \otimes Z$ lattice surgery for 4.6.12 codes with decons along the boundary, we show the same protocol in Fig.~\ref{fig:4612surgeryapp2}, but on a square lattice of dodecons.


\refstepcounter{subsection}
\label{app:666octons}
{\bf \arabic{subsection}. Weight-6 lattice surgery protocol for 6.6.6 Majorana surface codes.}
Figure \ref{fig:666surgeryoctons} shows the lattice surgery protocol for a $Y \otimes X$ measurement between a 
6.6.6 Majorana surface code hexon and tetron. Some of the physical hexons along the boundaries of the logical hexon and tetron have been replaced by physical octons. As a consequence, the dislocation line between the $X$ boundary of the tetron and the $Z$ boundary of the hexon no longer features any 8-Majorana operators, but only 6-Majorana operators. This reduces the maximum Majorana weight of quantum computing with 6.6.6 Majorana surface codes to 6 Majoranas.


\refstepcounter{subsection}
\label{app:subsystem}
{\bf \arabic{subsection}. Twist-based lattice surgery with subsystem surface codes.}
In Fig.~\ref{fig:subsystemsurgery}, we show the twist-based lattice surgery protocol~\cite{Litinski2017b} for a $d=3$ subsystem surface code~\cite{Bravyi2013}. While all the check operators in the bulk of the code are weight-3, the measurement of the twist defect requires a 5-qubit measurement. By replacing each qubit of the subsystem surface code with a tetron, one can see that the subsystem surface code resembles a 4.6.12 Majorana surface code, as is shown in Fig.~\ref{fig:subsystemsurgery}.

\section{Quantum-wire-based implementations}

\begin{figure*}
\centering
\def\svgwidth{\linewidth}
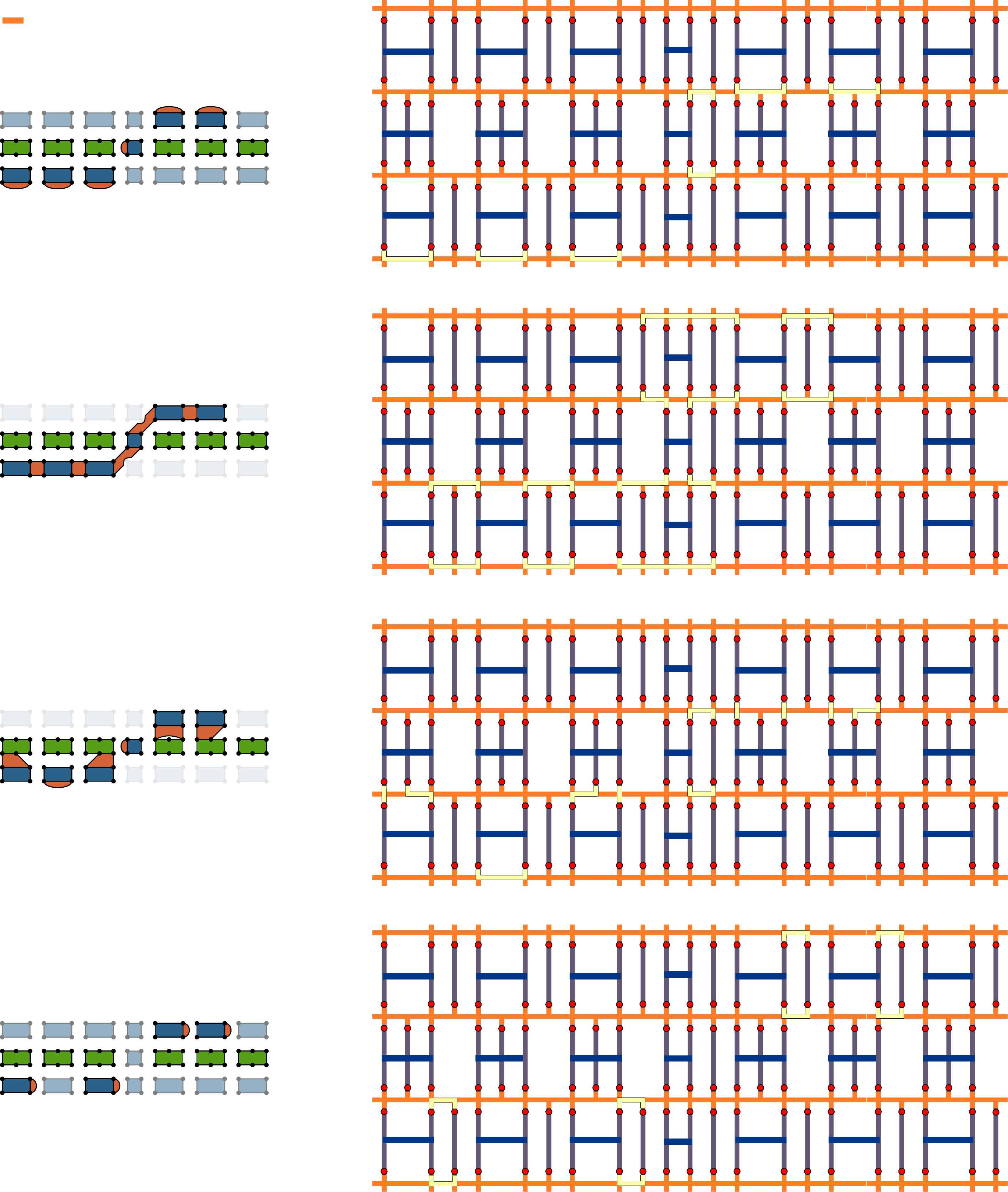
\caption{Tunnel coupling configurations for the Pauli product measurements protocol from Fig.~\ref{fig:pauliprodprotocol} realized with the wire-based tetrons and hexons proposed in Ref.~\cite{Karzig2016}.}
\label{fig:wires}
\end{figure*}

\refstepcounter{subsection}
\label{app:wireimplementation}
{\bf \arabic{subsection}. Pauli product measurement protocol in a nanowire array.}
The Pauli product measurement protocol in Fig.~\ref{fig:pauliprodprotocol} can be straightforwardly used to implement minimal-overhead Clifford gates in the wire-based architectures proposed in Ref.~\cite{Karzig2016}.  Essentially, hexons can be implemented as three topological superconducting nanowires in a parity sector that is fixed by a non-topological superconductor bridging the three wires. Moreover, in Fig.~\ref{fig:wires}, tetrons correspond to two topological superconducting nanowires in a fixed parity sector with an additional topological nanowire serving as a coherent link that is used for certain parity measurements. Products of Majorana operators are measured by opening tunnel couplings between topological superconductors and segments of a semiconducting nanowire network, such that the tunnel couplings form closed paths. These semiconducting nanowire segments can be interpreted as quantum dots whose energy levels are shifted by the virtual process of an electron tunneling around the path, picking up all Majorana operators along the way. Therefore, suitable spectroscopy on the dots can be used to measure the product of these Majorana operators, as detailed in Ref.~\cite{Karzig2016}. In Fig.~\ref{fig:wires}, we show the tunnel coupling configurations that implement the measurements that are part of the Pauli product measurement protocol. Note that some of the measurements require the use of coherent links, such that the Majorana weight of some measurements is increased. In particular, some of the four-Majorana measurements in the second step become six-Majorana measurements in this particular implementation.

\begin{figure}[b]
\centering
\def\svgwidth{\linewidth}
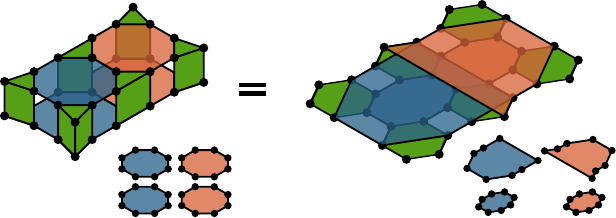
\caption{4.8.8 ($([[6,2,2]]_m$) Majorana color code in a 2D array of hexons.}
\label{fig:2dcolorcode}
\end{figure}

\begin{figure*}
\centering
\def\svgwidth{\linewidth}
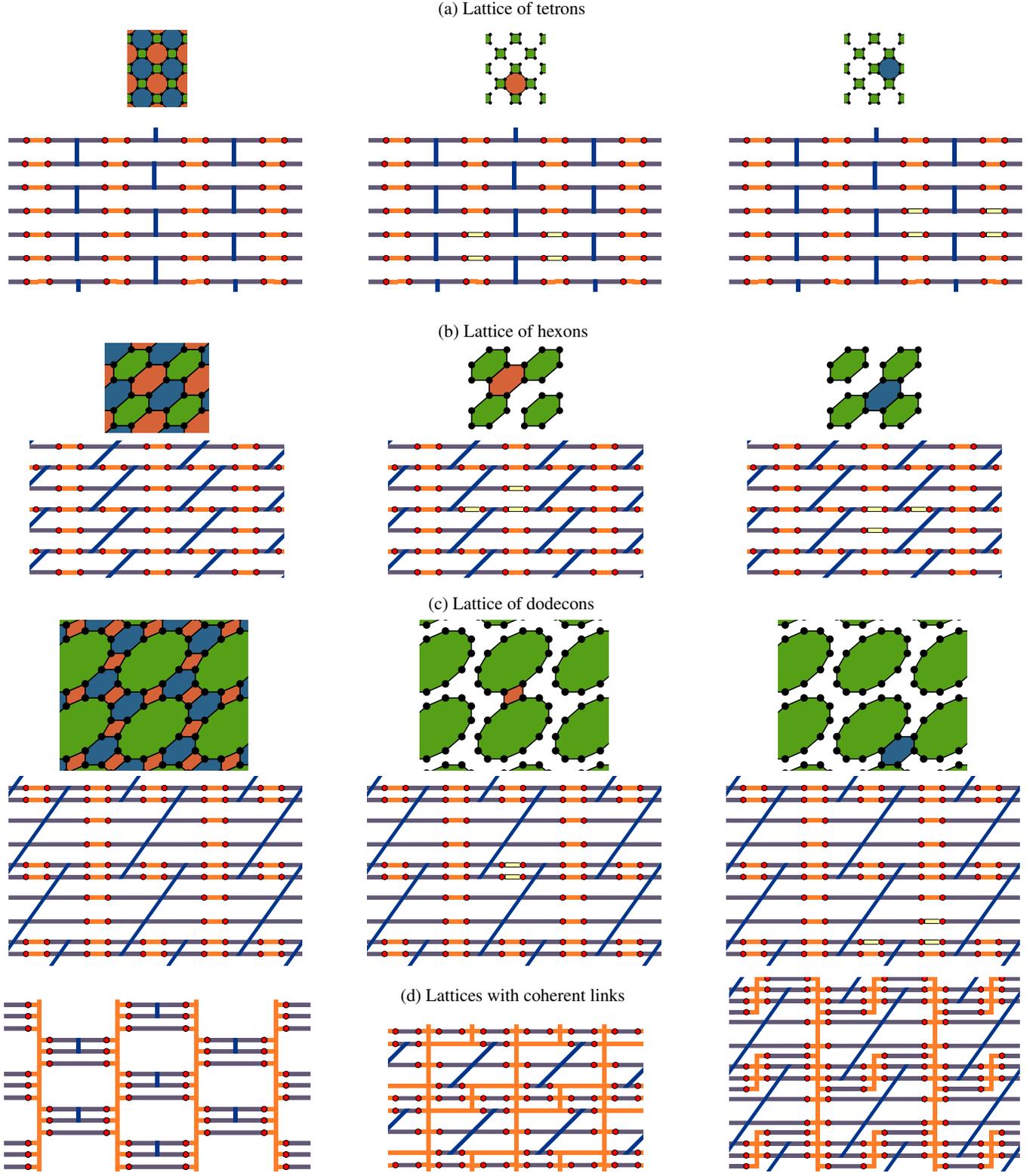
\caption{Nanowire-based implementations of 4.8.8 (a), 6.6.6 (b), and 4.6.12 (c), with examples of tunnel-coupling configurations for red and blue stabilizer measurements. Coherent links (d) may be necessary for the measurement of other operators.}
\label{fig:wirecodes}
\end{figure*}

\begin{figure*}
\centering
\def\svgwidth{0.95\linewidth}
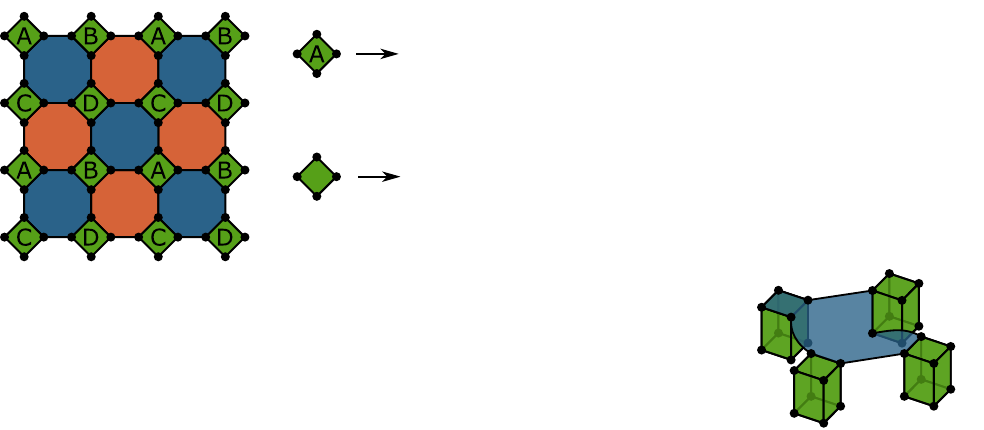
\caption{Procedure to obtain a 4.8.8 $([[8,3,2]]_m)$ Majorana color code. In order to prevent stabilizer weights from increasing beyond 10 Majoranas, the tetrons of a 4.8.8 Majorana surface code are periodically labelled A, B, C, and D. Each tetron is replaced by a single octon, but the definition of the octon's logical operators depends on the label of the corresponding tetron. In the figure, the octon's $Z$ ($X$) operators correspond to products of blue (red) Majoranas. The resulting stabilizers have a maximum weight of 10 Majoranas, as shown for the three overlapping blue $Z$ type stabilizers.}
\label{fig:octoncode}
\end{figure*}


\refstepcounter{subsection}
\label{app:wirecodes}
{\bf \arabic{subsection}. Majorana surface code implementations in a nanowire array.}
In Fig.~\ref{fig:wirecodes}, we show wire-based implementations of square lattices of tetrons (a), hexons (b), and dodecons (c), which can be used to implement 4.8.8, 6.6.6, and 4.6.12 Majorana surface codes, respectively. While we show the tunnel-coupling configurations to measure red and blue stabilizers, some irregular stabilizers~--~such as boundary stabilizers, dislocation lines, twist defect, or two-Majorana operators for initialization and readout~--~may require the use of coherent links. Nanowire arrays that include coherent links are shown in Fig.~\ref{fig:wirecodes}d. Note that operator measurements that require the use of coherent links have an increased Majorana weight compared to the optimal weight discussed in the main text. We also remark that, in principle, all stabilizers of one color can be measured simultaneously.

\section{Majorana color codes}

\refstepcounter{subsection}
\label{app:2dcolorcode}
{\bf \arabic{subsection}. 4.8.8 ($([[6,2,2]]_m$) Majorana color code in a 2D array of hexons.}
Even though the Majorana color codes in Figs.~\ref{fig:smallcolorcode} and \ref{fig:colorcode} are drawn as three-dimensional codes, they can be implemented in 2D arrays of Majorana building blocks. In Fig.~\ref{fig:2dcolorcode}, this is shown for the example of a 4.8.8 ($([[6,2,2]]_m$) Majorana color code, which can be implemented in a 2D array of physical hexons. Note that, in contrast to Majorana surface codes, the red and blue check operators of Majorana color codes spatially overlap.


\refstepcounter{subsection}
\label{app:octoncode}
{\bf \arabic{subsection}. 4.8.8 ($([[8,3,2]]_m$) Majorana color code.}
A 4.8.8 ($([[8,3,2]]_m$) Majorana color code can be obtained by replacing each tetron of a 4.8.8 Majorana surface code with an $[[8,3,2]]_m$ code, i.e., an octon. Even though this code has a Majorana distance of $d_m =2$, one of its logical operators is a 4-Majorana operator. In order to prevent the maximum stabilizer weight after concatenation from increasing to 16 Majoranas, the procedure shown in Fig.~\ref{fig:octoncode} can be used. Here, each surface code tetron is assigned a label A, B, C, or D. While each tetron is still replaced by an octon, the definition of the octon's logical operators depends on the label. For A, B, C, or D type octons, the 4-Majorana logical operator is chosen to be $Z_3$, $X_3$, $X_2$, or $Z_2$, respectively. This guarantees that the stabilizers have a maximum weight of 10 Majoranas after concatenation. The figure shows an example of a $Z^{\otimes 4}$ stabilizer, which is replaced by three overlapping $Z_1^{\otimes 4}$, $Z_2^{\otimes 4}$, and $Z_3^{\otimes 4}$ stabilizers after concatenation. The operators $Z_2^{\otimes 4}$ and and $Z_3^{\otimes 4}$ have a weight of 10 Majoranas.

\begin{figure}[b]
\centering
\def\svgwidth{\linewidth}
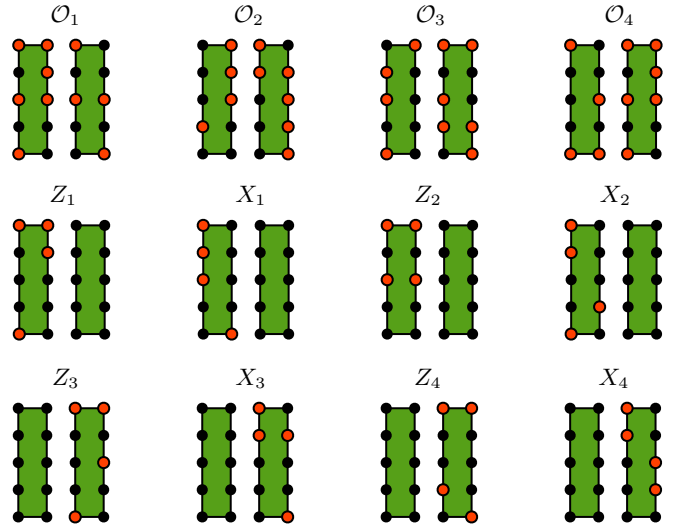
\caption{Stabilizers $\mathcal{O}_1$-$\mathcal{O}_4$ and logical operators of the $[[20,4,4]]_m$ code presented in Ref.~\cite{Hastings2017}.}
\label{fig:2044code}
\end{figure}

\refstepcounter{subsection}
\label{app:2044code}
{\bf \arabic{subsection}. Stabilizers and logical operators of the $[[20,4,4]]_m$ code.}
Figure \ref{fig:2044code} shows the stabilizers and logical operators of the $[[20,4,4]]_m$ code presented in Ref.~\cite{Hastings2017}. The building blocks of the code are two decons, which are 10 Majoranas in a fixed parity sector. All logical operators have a weight of 4 Majoranas. This means that a code obtained by concatenating a Majorana surface code with the $[[20,4,4]]_m$ code is an order-4 Majorana color code with a maximum stabilizer weight of 16 Majoranas.

\begin{figure*}
\centering
\def\svgwidth{\linewidth}
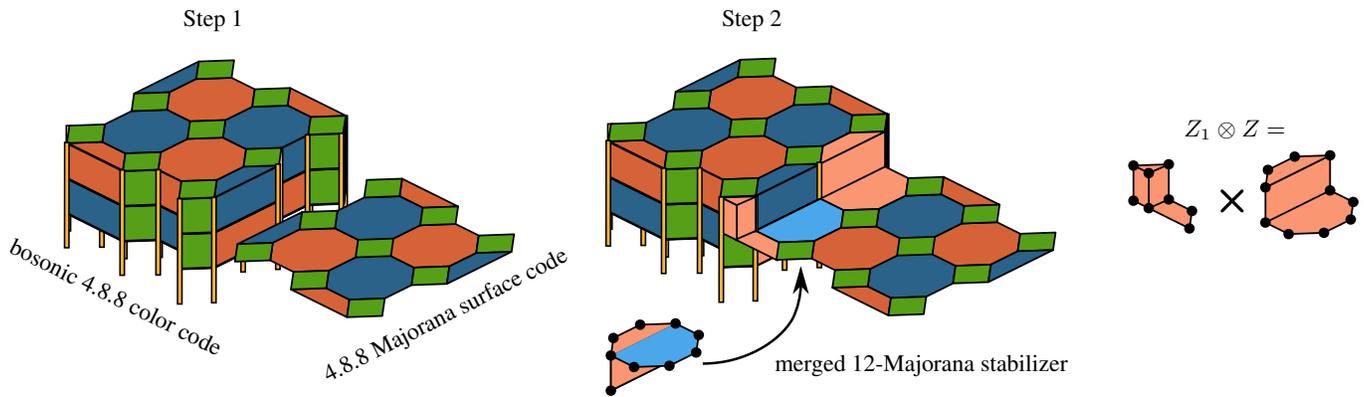
\caption{Lattice surgery protocol for the measurement of the operator $Z_1 \otimes Z$ between a bosonic 4.8.8 color code tetron and a 4.8.8 Majorana surface code tetron. $Z_1$ is the $Z$ operator of the first qubit encoded in the bosonic 4.8.8 color code tetron. }
\label{fig:colortosurface}
\end{figure*}


\refstepcounter{subsection}
\label{app:colortosurface}
{\bf \arabic{subsection}. Surface-to-color code lattice surgery.}
Figure \ref{fig:colortosurface} shows an example of a surface-to-color code lattice surgery in the spirit of Ref.~\cite{Litinski2017a}. The example shows the measurement of the product $Z_1 \otimes Z$ between the first qubit encoded in a bosonic 4.8.8 color code tetron and a 4.8.8 Majorana surface code. The blue 4-Majorana operator at the boundary of the surface code is merged with a red 8-Majorana operator at the boundary of the color code to yield the light blue 12-Majorana operator. New 8- and 12-Majorana operators (light red) are introduced whose product is precisely $Z_1 \otimes Z$. Such a scheme can always be used to measure the product between a Pauli operator encoded in a color code qubit and a Pauli operator of a surface code. By using color codes for data qubits encoded in logical hexons and surface codes for ancillary tetron qubits, the Pauli product measurement scheme of Sec.~\ref{sec:majoranagates} can be implemented with a lower space overhead compared to surface codes due to the more compact encoding of color codes, while at the same time benefiting from the low Majorana weight of surface-to-color code lattice surgery.

\bibliographystyle{apsrev4-1mod}
\bibliography{biblio}

\begin{thebibliography}{55}%
\makeatletter
\providecommand \@ifxundefined [1]{%
 \@ifx{#1\undefined}
}%
\providecommand \@ifnum [1]{%
 \ifnum #1\expandafter \@firstoftwo
 \else \expandafter \@secondoftwo
 \fi
}%
\providecommand \@ifx [1]{%
 \ifx #1\expandafter \@firstoftwo
 \else \expandafter \@secondoftwo
 \fi
}%
\providecommand \natexlab [1]{#1}%
\providecommand \enquote  [1]{``#1''}%
\providecommand \bibnamefont  [1]{#1}%
\providecommand \bibfnamefont [1]{#1}%
\providecommand \citenamefont [1]{#1}%
\providecommand \href@noop [0]{\@secondoftwo}%
\providecommand \href [0]{\begingroup \@sanitize@url \@href}%
\providecommand \@href[1]{\@@startlink{#1}\@@href}%
\providecommand \@@href[1]{\endgroup#1\@@endlink}%
\providecommand \@sanitize@url [0]{\catcode `\\12\catcode `\$12\catcode
  `\&12\catcode `\#12\catcode `\^12\catcode `\_12\catcode `\%12\relax}%
\providecommand \@@startlink[1]{}%
\providecommand \@@endlink[0]{}%
\providecommand \url  [0]{\begingroup\@sanitize@url \@url }%
\providecommand \@url [1]{\endgroup\@href {#1}{\urlprefix }}%
\providecommand \urlprefix  [0]{URL }%
\providecommand \Eprint [0]{\href }%
\providecommand \doibase [0]{http://dx.doi.org/}%
\providecommand \selectlanguage [0]{\@gobble}%
\providecommand \bibinfo  [0]{\@secondoftwo}%
\providecommand \bibfield  [0]{\@secondoftwo}%
\providecommand \translation [1]{[#1]}%
\providecommand \BibitemOpen [0]{}%
\providecommand \bibitemStop [0]{}%
\providecommand \bibitemNoStop [0]{.\EOS\space}%
\providecommand \EOS [0]{\spacefactor3000\relax}%
\providecommand \BibitemShut  [1]{\csname bibitem#1\endcsname}%
\let\auto@bib@innerbib\@empty
\bibitem [{\citenamefont {Kitaev}(2001)}]{Kitaev2001}%
  \BibitemOpen
  \bibfield  {author} {\bibinfo {author} {\bibfnamefont {A.~Y.}\ \bibnamefont
  {Kitaev}},\ }\bibfield  {title} {\emph {\bibinfo {title} {Unpaired {M}ajorana
  fermions in quantum wires},\ }}\href {\doibase 10.1070/1063-7869/44/10S/S29}
  {\bibfield  {journal} {\bibinfo  {journal} {Sov. Phys. Usp.}\ }\textbf
  {\bibinfo {volume} {44}},\ \bibinfo {pages} {131} (\bibinfo {year}
  {2001})}\BibitemShut {NoStop}%
\bibitem [{\citenamefont {Kitaev}(2003)}]{Kitaev2003}%
  \BibitemOpen
  \bibfield  {author} {\bibinfo {author} {\bibfnamefont {A.~Y.}\ \bibnamefont
  {Kitaev}},\ }\bibfield  {title} {\emph {\bibinfo {title} {{Fault-tolerant
  quantum computation by anyons}},\ }}\href {\doibase
  10.1016/S0003-4916(02)00018-0} {\bibfield  {journal} {\bibinfo  {journal}
  {Ann. Phys.}\ }\textbf {\bibinfo {volume} {303}},\ \bibinfo {pages} {2}
  (\bibinfo {year} {2003})}\BibitemShut {NoStop}%
\bibitem [{\citenamefont {Alicea}(2012)}]{Alicea2012}%
  \BibitemOpen
  \bibfield  {author} {\bibinfo {author} {\bibfnamefont {J.}~\bibnamefont
  {Alicea}},\ }\bibfield  {title} {\emph {\bibinfo {title} {{New directions in
  the pursuit of Majorana fermions in solid state systems}},\ }}\href {\doibase
  10.1088/0034-4885/75/7/076501} {\bibfield  {journal} {\bibinfo  {journal}
  {Rep. Prog. Phys.}\ }\textbf {\bibinfo {volume} {75}},\ \bibinfo {pages}
  {076501} (\bibinfo {year} {2012})}\BibitemShut {NoStop}%
\bibitem [{\citenamefont {Beenakker}(2013)}]{Beenakker2013}%
  \BibitemOpen
  \bibfield  {author} {\bibinfo {author} {\bibfnamefont {C.~W.~J.}\
  \bibnamefont {Beenakker}},\ }\bibfield  {title} {\emph {\bibinfo {title}
  {{Search for Majorana fermions in superconductors}},\ }}\href {\doibase
  10.1146/annurev-conmatphys-030212-184337} {\bibfield  {journal} {\bibinfo
  {journal} {Ann. Rev. Cond. Matt. Phys.}\ }\textbf {\bibinfo {volume} {4}},\
  \bibinfo {pages} {113} (\bibinfo {year} {2013})}\BibitemShut {NoStop}%
\bibitem [{\citenamefont {Lutchyn}\ \emph {et~al.}(2017)\citenamefont
  {Lutchyn}, \citenamefont {Bakkers}, \citenamefont {Kouwenhoven},
  \citenamefont {Krogstrup}, \citenamefont {Marcus},\ and\ \citenamefont
  {Oreg}}]{Lutchyn2017}%
  \BibitemOpen
  \bibfield  {author} {\bibinfo {author} {\bibfnamefont {R.~M.}\ \bibnamefont
  {Lutchyn}}, \bibinfo {author} {\bibfnamefont {E.~P. A.~M.}\ \bibnamefont
  {Bakkers}}, \bibinfo {author} {\bibfnamefont {L.~P.}\ \bibnamefont
  {Kouwenhoven}}, \bibinfo {author} {\bibfnamefont {P.}~\bibnamefont
  {Krogstrup}}, \bibinfo {author} {\bibfnamefont {C.~M.}\ \bibnamefont
  {Marcus}}, \ and\ \bibinfo {author} {\bibfnamefont {Y.}~\bibnamefont
  {Oreg}},\ }\bibfield  {title} {\emph {\bibinfo {title} {Realizing {M}ajorana
  zero modes in superconductor-semiconductor heterostructures},\ }}\href
  {https://arxiv.org/abs/1707.04899} {\bibfield  {journal} {\bibinfo  {journal}
  {arXiv:1707.04899}\ } (\bibinfo {year} {2017})}\BibitemShut {NoStop}%
\bibitem [{\citenamefont {Aguado}(2017)}]{Aguado2017}%
  \BibitemOpen
  \bibfield  {author} {\bibinfo {author} {\bibfnamefont {R.}~\bibnamefont
  {Aguado}},\ }\bibfield  {title} {\emph {\bibinfo {title} {Majorana
  quasiparticles in condensed matter},\ }}\href {\doibase
  10.1393/ncr/i2017-10141-9} {\bibfield  {journal} {\bibinfo  {journal} {Riv.
  Nuovo Cimento}\ }\textbf {\bibinfo {volume} {11}},\ \bibinfo {pages} {523}
  (\bibinfo {year} {2017})}\BibitemShut {NoStop}%
\bibitem [{\citenamefont {Mourik}\ \emph {et~al.}(2012)\citenamefont {Mourik},
  \citenamefont {Zuo}, \citenamefont {Frolov}, \citenamefont {Plissard},
  \citenamefont {Bakkers},\ and\ \citenamefont {Kouwenhoven}}]{Mourik2012}%
  \BibitemOpen
  \bibfield  {author} {\bibinfo {author} {\bibfnamefont {V.}~\bibnamefont
  {Mourik}}, \bibinfo {author} {\bibfnamefont {K.}~\bibnamefont {Zuo}},
  \bibinfo {author} {\bibfnamefont {S.~M.}\ \bibnamefont {Frolov}}, \bibinfo
  {author} {\bibfnamefont {S.~R.}\ \bibnamefont {Plissard}}, \bibinfo {author}
  {\bibfnamefont {E.~P. a.~M.}\ \bibnamefont {Bakkers}}, \ and\ \bibinfo
  {author} {\bibfnamefont {L.~P.}\ \bibnamefont {Kouwenhoven}},\ }\bibfield
  {title} {\emph {\bibinfo {title} {{Signatures of Majorana fermions in in
  hybrid superconductor-semiconductor nanowire devices}},\ }}\href {\doibase
  10.1126/science.1222360} {\bibfield  {journal} {\bibinfo  {journal}
  {Science}\ }\textbf {\bibinfo {volume} {336}},\ \bibinfo {pages} {1003}
  (\bibinfo {year} {2012})}\BibitemShut {NoStop}%
\bibitem [{\citenamefont {Albrecht}\ \emph {et~al.}(2016)\citenamefont
  {Albrecht}, \citenamefont {Higginbotham}, \citenamefont {Madsen},
  \citenamefont {Kuemmeth}, \citenamefont {Jespersen}, \citenamefont {Nyg},
  \citenamefont {Krogstrup},\ and\ \citenamefont {Marcus}}]{Albrecht2016a}%
  \BibitemOpen
  \bibfield  {author} {\bibinfo {author} {\bibfnamefont {S.~M.}\ \bibnamefont
  {Albrecht}}, \bibinfo {author} {\bibfnamefont {A.~P.}\ \bibnamefont
  {Higginbotham}}, \bibinfo {author} {\bibfnamefont {M.}~\bibnamefont
  {Madsen}}, \bibinfo {author} {\bibfnamefont {F.}~\bibnamefont {Kuemmeth}},
  \bibinfo {author} {\bibfnamefont {T.~S.}\ \bibnamefont {Jespersen}}, \bibinfo
  {author} {\bibfnamefont {J.}~\bibnamefont {Nyg}}, \bibinfo {author}
  {\bibfnamefont {P.}~\bibnamefont {Krogstrup}}, \ and\ \bibinfo {author}
  {\bibfnamefont {C.~M.}\ \bibnamefont {Marcus}},\ }\bibfield  {title} {\emph
  {\bibinfo {title} {{Exponential protection of zero modes in Majorana
  islands}},\ }}\href {\doibase 10.1038/nature17162} {\bibfield  {journal}
  {\bibinfo  {journal} {Nature}\ }\textbf {\bibinfo {volume} {531}},\ \bibinfo
  {pages} {206} (\bibinfo {year} {2016})}\BibitemShut {NoStop}%
\bibitem [{\citenamefont {Deng}\ \emph {et~al.}(2016)\citenamefont {Deng},
  \citenamefont {Vaitiek{\.e}nas}, \citenamefont {Hansen}, \citenamefont
  {Danon}, \citenamefont {Leijnse}, \citenamefont {Flensberg}, \citenamefont
  {Nyg{\aa}rd}, \citenamefont {Krogstrup},\ and\ \citenamefont
  {Marcus}}]{Deng2016}%
  \BibitemOpen
  \bibfield  {author} {\bibinfo {author} {\bibfnamefont {M.}~\bibnamefont
  {Deng}}, \bibinfo {author} {\bibfnamefont {S.}~\bibnamefont
  {Vaitiek{\.e}nas}}, \bibinfo {author} {\bibfnamefont {E.}~\bibnamefont
  {Hansen}}, \bibinfo {author} {\bibfnamefont {J.}~\bibnamefont {Danon}},
  \bibinfo {author} {\bibfnamefont {M.}~\bibnamefont {Leijnse}}, \bibinfo
  {author} {\bibfnamefont {K.}~\bibnamefont {Flensberg}}, \bibinfo {author}
  {\bibfnamefont {J.}~\bibnamefont {Nyg{\aa}rd}}, \bibinfo {author}
  {\bibfnamefont {P.}~\bibnamefont {Krogstrup}}, \ and\ \bibinfo {author}
  {\bibfnamefont {C.}~\bibnamefont {Marcus}},\ }\bibfield  {title} {\emph
  {\bibinfo {title} {Majorana bound state in a coupled quantum-dot
  hybrid-nanowire system},\ }}\href {\doibase 10.1126/science.aaf3961}
  {\bibfield  {journal} {\bibinfo  {journal} {Science}\ }\textbf {\bibinfo
  {volume} {354}},\ \bibinfo {pages} {1557} (\bibinfo {year}
  {2016})}\BibitemShut {NoStop}%
\bibitem [{\citenamefont {{Zhang}}\ \emph {et~al.}(2018)\citenamefont
  {{Zhang}}, \citenamefont {{Liu}}, \citenamefont {{Gazibegovic}},
  \citenamefont {{Xu}}, \citenamefont {{Logan}}, \citenamefont {{Wang}},
  \citenamefont {{van Loo}}, \citenamefont {{Bommer}}, \citenamefont {{de
  Moor}}, \citenamefont {{Car}}, \citenamefont {{het Veld}}, \citenamefont
  {{van Veldhoven}}, \citenamefont {{Koelling}}, \citenamefont {{Verheijen}},
  \citenamefont {{Pendharkar}}, \citenamefont {{Pennachio}}, \citenamefont
  {{Shojaei}}, \citenamefont {{Lee}}, \citenamefont {{Palmstrom}},
  \citenamefont {{Bakkers}}, \citenamefont {{Das Sarma}},\ and\ \citenamefont
  {{Kouwenhoven}}}]{Zhang2017}%
  \BibitemOpen
  \bibfield  {author} {\bibinfo {author} {\bibfnamefont {H.}~\bibnamefont
  {{Zhang}}}, \bibinfo {author} {\bibfnamefont {C.-X.}\ \bibnamefont {{Liu}}},
  \bibinfo {author} {\bibfnamefont {S.}~\bibnamefont {{Gazibegovic}}}, \bibinfo
  {author} {\bibfnamefont {D.}~\bibnamefont {{Xu}}}, \bibinfo {author}
  {\bibfnamefont {J.~A.}\ \bibnamefont {{Logan}}}, \bibinfo {author}
  {\bibfnamefont {G.}~\bibnamefont {{Wang}}}, \bibinfo {author} {\bibfnamefont
  {N.}~\bibnamefont {{van Loo}}}, \bibinfo {author} {\bibfnamefont {J.~D.~S.}\
  \bibnamefont {{Bommer}}}, \bibinfo {author} {\bibfnamefont {M.~W.~A.}\
  \bibnamefont {{de Moor}}}, \bibinfo {author} {\bibfnamefont {D.}~\bibnamefont
  {{Car}}}, \bibinfo {author} {\bibfnamefont {R.~L.~M.~O.}\ \bibnamefont {{het
  Veld}}}, \bibinfo {author} {\bibfnamefont {P.~J.}\ \bibnamefont {{van
  Veldhoven}}}, \bibinfo {author} {\bibfnamefont {S.}~\bibnamefont
  {{Koelling}}}, \bibinfo {author} {\bibfnamefont {M.~A.}\ \bibnamefont
  {{Verheijen}}}, \bibinfo {author} {\bibfnamefont {M.}~\bibnamefont
  {{Pendharkar}}}, \bibinfo {author} {\bibfnamefont {D.~J.}\ \bibnamefont
  {{Pennachio}}}, \bibinfo {author} {\bibfnamefont {B.}~\bibnamefont
  {{Shojaei}}}, \bibinfo {author} {\bibfnamefont {J.~S.}\ \bibnamefont
  {{Lee}}}, \bibinfo {author} {\bibfnamefont {C.~J.}\ \bibnamefont
  {{Palmstrom}}}, \bibinfo {author} {\bibfnamefont {E.~P.~A.~M.}\ \bibnamefont
  {{Bakkers}}}, \bibinfo {author} {\bibfnamefont {S.}~\bibnamefont {{Das
  Sarma}}}, \ and\ \bibinfo {author} {\bibfnamefont {L.~P.}\ \bibnamefont
  {{Kouwenhoven}}},\ }\bibfield  {title} {\emph {\bibinfo {title} {Quantized
  {M}ajorana conductance},\ }}\href {\doibase 10.1038/nature26142} {\bibfield
  {journal} {\bibinfo  {journal} {Nature}\ }\textbf {\bibinfo {volume} {556}},\
  \bibinfo {pages} {74} (\bibinfo {year} {2018})}\BibitemShut {NoStop}%
\bibitem [{\citenamefont {Aasen}\ \emph {et~al.}(2016)\citenamefont {Aasen},
  \citenamefont {Hell}, \citenamefont {Mishmash}, \citenamefont {Higginbotham},
  \citenamefont {Danon}, \citenamefont {Leijnse}, \citenamefont {Jespersen},
  \citenamefont {Folk}, \citenamefont {Marcus}, \citenamefont {Flensberg},\
  and\ \citenamefont {Alicea}}]{Aasen2016}%
  \BibitemOpen
  \bibfield  {author} {\bibinfo {author} {\bibfnamefont {D.}~\bibnamefont
  {Aasen}}, \bibinfo {author} {\bibfnamefont {M.}~\bibnamefont {Hell}},
  \bibinfo {author} {\bibfnamefont {R.~V.}\ \bibnamefont {Mishmash}}, \bibinfo
  {author} {\bibfnamefont {A.}~\bibnamefont {Higginbotham}}, \bibinfo {author}
  {\bibfnamefont {J.}~\bibnamefont {Danon}}, \bibinfo {author} {\bibfnamefont
  {M.}~\bibnamefont {Leijnse}}, \bibinfo {author} {\bibfnamefont {T.~S.}\
  \bibnamefont {Jespersen}}, \bibinfo {author} {\bibfnamefont {J.~A.}\
  \bibnamefont {Folk}}, \bibinfo {author} {\bibfnamefont {C.~M.}\ \bibnamefont
  {Marcus}}, \bibinfo {author} {\bibfnamefont {K.}~\bibnamefont {Flensberg}}, \
  and\ \bibinfo {author} {\bibfnamefont {J.}~\bibnamefont {Alicea}},\
  }\bibfield  {title} {\emph {\bibinfo {title} {Milestones toward
  majorana-based quantum computing},\ }}\href {\doibase
  10.1103/PhysRevX.6.031016} {\bibfield  {journal} {\bibinfo  {journal} {Phys.
  Rev. X}\ }\textbf {\bibinfo {volume} {6}},\ \bibinfo {pages} {031016}
  (\bibinfo {year} {2016})}\BibitemShut {NoStop}%
\bibitem [{\citenamefont {Knapp}\ \emph {et~al.}(2018)\citenamefont {Knapp},
  \citenamefont {Karzig}, \citenamefont {Lutchyn},\ and\ \citenamefont
  {Nayak}}]{Knapp2018}%
  \BibitemOpen
  \bibfield  {author} {\bibinfo {author} {\bibfnamefont {C.}~\bibnamefont
  {Knapp}}, \bibinfo {author} {\bibfnamefont {T.}~\bibnamefont {Karzig}},
  \bibinfo {author} {\bibfnamefont {R.~M.}\ \bibnamefont {Lutchyn}}, \ and\
  \bibinfo {author} {\bibfnamefont {C.}~\bibnamefont {Nayak}},\ }\bibfield
  {title} {\emph {\bibinfo {title} {Dephasing of {M}ajorana-based qubits},\
  }}\href {\doibase 10.1103/PhysRevB.97.125404} {\bibfield  {journal} {\bibinfo
   {journal} {Phys. Rev. B}\ }\textbf {\bibinfo {volume} {97}},\ \bibinfo
  {pages} {125404} (\bibinfo {year} {2018})}\BibitemShut {NoStop}%
\bibitem [{\citenamefont {Terhal}(2015)}]{TerhalRMP}%
  \BibitemOpen
  \bibfield  {author} {\bibinfo {author} {\bibfnamefont {B.~M.}\ \bibnamefont
  {Terhal}},\ }\bibfield  {title} {\emph {\bibinfo {title} {Quantum error
  correction for quantum memories},\ }}\href {\doibase
  10.1103/RevModPhys.87.307} {\bibfield  {journal} {\bibinfo  {journal} {Rev.
  Mod. Phys.}\ }\textbf {\bibinfo {volume} {87}},\ \bibinfo {pages} {307}
  (\bibinfo {year} {2015})}\BibitemShut {NoStop}%
\bibitem [{\citenamefont {Bravyi}\ \emph {et~al.}(2010)\citenamefont {Bravyi},
  \citenamefont {Terhal},\ and\ \citenamefont {Leemhuis}}]{Bravyi2010}%
  \BibitemOpen
  \bibfield  {author} {\bibinfo {author} {\bibfnamefont {S.}~\bibnamefont
  {Bravyi}}, \bibinfo {author} {\bibfnamefont {B.~M.}\ \bibnamefont {Terhal}},
  \ and\ \bibinfo {author} {\bibfnamefont {B.}~\bibnamefont {Leemhuis}},\
  }\bibfield  {title} {\emph {\bibinfo {title} {Majorana fermion codes},\
  }}\href {\doibase 10.1088/1367-2630/12/8/083039} {\bibfield  {journal}
  {\bibinfo  {journal} {New J. Phys.}\ }\textbf {\bibinfo {volume} {12}},\
  \bibinfo {pages} {083039} (\bibinfo {year} {2010})}\BibitemShut {NoStop}%
\bibitem [{\citenamefont {Vijay}\ and\ \citenamefont {Fu}(2016)}]{Vijay2016}%
  \BibitemOpen
  \bibfield  {author} {\bibinfo {author} {\bibfnamefont {S.}~\bibnamefont
  {Vijay}}\ and\ \bibinfo {author} {\bibfnamefont {L.}~\bibnamefont {Fu}},\
  }\bibfield  {title} {\emph {\bibinfo {title} {{Physical implementation of a
  Majorana fermion surface code for fault-tolerant quantum computation}},\
  }}\href {\doibase 10.1088/0031-8949/T168/1/014002} {\bibfield  {journal}
  {\bibinfo  {journal} {Phys. Scr.}\ }\textbf {\bibinfo {volume} {T168}},\
  \bibinfo {pages} {014002} (\bibinfo {year} {2016})}\BibitemShut {NoStop}%
\bibitem [{\citenamefont {Landau}\ \emph {et~al.}(2016)\citenamefont {Landau},
  \citenamefont {Plugge}, \citenamefont {Sela}, \citenamefont {Altland},
  \citenamefont {Albrecht},\ and\ \citenamefont {Egger}}]{Landau2016}%
  \BibitemOpen
  \bibfield  {author} {\bibinfo {author} {\bibfnamefont {L.~A.}\ \bibnamefont
  {Landau}}, \bibinfo {author} {\bibfnamefont {S.}~\bibnamefont {Plugge}},
  \bibinfo {author} {\bibfnamefont {E.}~\bibnamefont {Sela}}, \bibinfo {author}
  {\bibfnamefont {A.}~\bibnamefont {Altland}}, \bibinfo {author} {\bibfnamefont
  {S.~M.}\ \bibnamefont {Albrecht}}, \ and\ \bibinfo {author} {\bibfnamefont
  {R.}~\bibnamefont {Egger}},\ }\bibfield  {title} {\emph {\bibinfo {title}
  {Towards realistic implementations of a majorana surface code},\ }}\href
  {\doibase 10.1103/PhysRevLett.116.050501} {\bibfield  {journal} {\bibinfo
  {journal} {Phys. Rev. Lett.}\ }\textbf {\bibinfo {volume} {116}},\ \bibinfo
  {pages} {050501} (\bibinfo {year} {2016})}\BibitemShut {NoStop}%
\bibitem [{\citenamefont {Plugge}\ \emph {et~al.}(2016)\citenamefont {Plugge},
  \citenamefont {Landau}, \citenamefont {Sela}, \citenamefont {Altland},
  \citenamefont {Flensberg},\ and\ \citenamefont {Egger}}]{Plugge2016}%
  \BibitemOpen
  \bibfield  {author} {\bibinfo {author} {\bibfnamefont {S.}~\bibnamefont
  {Plugge}}, \bibinfo {author} {\bibfnamefont {L.~A.}\ \bibnamefont {Landau}},
  \bibinfo {author} {\bibfnamefont {E.}~\bibnamefont {Sela}}, \bibinfo {author}
  {\bibfnamefont {A.}~\bibnamefont {Altland}}, \bibinfo {author} {\bibfnamefont
  {K.}~\bibnamefont {Flensberg}}, \ and\ \bibinfo {author} {\bibfnamefont
  {R.}~\bibnamefont {Egger}},\ }\bibfield  {title} {\emph {\bibinfo {title}
  {Roadmap to majorana surface codes},\ }}\href {\doibase
  10.1103/PhysRevB.94.174514} {\bibfield  {journal} {\bibinfo  {journal} {Phys.
  Rev. B}\ }\textbf {\bibinfo {volume} {94}},\ \bibinfo {pages} {174514}
  (\bibinfo {year} {2016})}\BibitemShut {NoStop}%
\bibitem [{\citenamefont {Li}(2016)}]{Li2016}%
  \BibitemOpen
  \bibfield  {author} {\bibinfo {author} {\bibfnamefont {Y.}~\bibnamefont
  {Li}},\ }\bibfield  {title} {\emph {\bibinfo {title} {Noise threshold and
  resource cost of fault-tolerant quantum computing with {M}ajorana fermions in
  hybrid systems},\ }}\href {\doibase 10.1103/PhysRevLett.117.120403}
  {\bibfield  {journal} {\bibinfo  {journal} {Phys. Rev. Lett.}\ }\textbf
  {\bibinfo {volume} {117}},\ \bibinfo {pages} {120403} (\bibinfo {year}
  {2016})}\BibitemShut {NoStop}%
\bibitem [{\citenamefont {Litinski}\ \emph {et~al.}(2017)\citenamefont
  {Litinski}, \citenamefont {Kesselring}, \citenamefont {Eisert},\ and\
  \citenamefont {von Oppen}}]{Litinski2017}%
  \BibitemOpen
  \bibfield  {author} {\bibinfo {author} {\bibfnamefont {D.}~\bibnamefont
  {Litinski}}, \bibinfo {author} {\bibfnamefont {M.~S.}\ \bibnamefont
  {Kesselring}}, \bibinfo {author} {\bibfnamefont {J.}~\bibnamefont {Eisert}},
  \ and\ \bibinfo {author} {\bibfnamefont {F.}~\bibnamefont {von Oppen}},\
  }\bibfield  {title} {\emph {\bibinfo {title} {Combining topological hardware
  and topological software: Color-code quantum computing with topological
  superconductor networks},\ }}\href {\doibase 10.1103/PhysRevX.7.031048}
  {\bibfield  {journal} {\bibinfo  {journal} {Phys. Rev. X}\ }\textbf {\bibinfo
  {volume} {7}},\ \bibinfo {pages} {031048} (\bibinfo {year}
  {2017})}\BibitemShut {NoStop}%
\bibitem [{\citenamefont {Litinski}\ and\ \citenamefont {von
  Oppen}(2017{\natexlab{a}})}]{Litinski2017a}%
  \BibitemOpen
  \bibfield  {author} {\bibinfo {author} {\bibfnamefont {D.}~\bibnamefont
  {Litinski}}\ and\ \bibinfo {author} {\bibfnamefont {F.}~\bibnamefont {von
  Oppen}},\ }\bibfield  {title} {\emph {\bibinfo {title} {Braiding by
  {M}ajorana tracking and long-range {CNOT} gates with color codes},\ }}\href
  {\doibase 10.1103/PhysRevB.96.205413} {\bibfield  {journal} {\bibinfo
  {journal} {Phys. Rev. B}\ }\textbf {\bibinfo {volume} {96}},\ \bibinfo
  {pages} {205413} (\bibinfo {year} {2017}{\natexlab{a}})}\BibitemShut
  {NoStop}%
\bibitem [{\citenamefont {Vijay}\ \emph {et~al.}(2015)\citenamefont {Vijay},
  \citenamefont {Hsieh},\ and\ \citenamefont {Fu}}]{Vijay2015}%
  \BibitemOpen
  \bibfield  {author} {\bibinfo {author} {\bibfnamefont {S.}~\bibnamefont
  {Vijay}}, \bibinfo {author} {\bibfnamefont {T.~H.}\ \bibnamefont {Hsieh}}, \
  and\ \bibinfo {author} {\bibfnamefont {L.}~\bibnamefont {Fu}},\ }\bibfield
  {title} {\emph {\bibinfo {title} {{Majorana Fermion surface code for
  universal quantum computation}},\ }}\href {\doibase
  10.1103/PhysRevX.5.041038} {\bibfield  {journal} {\bibinfo  {journal} {Phys.
  Rev. X}\ }\textbf {\bibinfo {volume} {5}},\ \bibinfo {pages} {041038}
  (\bibinfo {year} {2015})}\BibitemShut {NoStop}%
\bibitem [{\citenamefont {Li}(2017)}]{Li2017}%
  \BibitemOpen
  \bibfield  {author} {\bibinfo {author} {\bibfnamefont {Y.}~\bibnamefont
  {Li}},\ }\bibfield  {title} {\emph {\bibinfo {title} {Fault-tolerant
  fermionic quantum computation based on color code},\ }}\href
  {https://arxiv.org/abs/1709.06245} {\bibfield  {journal} {\bibinfo  {journal}
  {arXiv:1709.06245}\ } (\bibinfo {year} {2017})}\BibitemShut {NoStop}%
\bibitem [{\citenamefont {Vijay}\ and\ \citenamefont {Fu}(2017)}]{Vijay2017}%
  \BibitemOpen
  \bibfield  {author} {\bibinfo {author} {\bibfnamefont {S.}~\bibnamefont
  {Vijay}}\ and\ \bibinfo {author} {\bibfnamefont {L.}~\bibnamefont {Fu}},\
  }\bibfield  {title} {\emph {\bibinfo {title} {Quantum error correction for
  complex and {M}ajorana fermion qubits},\ }}\href
  {https://arxiv.org/abs/1703.00459} {\bibfield  {journal} {\bibinfo  {journal}
  {arXiv:1703.00459}\ } (\bibinfo {year} {2017})}\BibitemShut {NoStop}%
\bibitem [{\citenamefont {Hastings}(2017)}]{Hastings2017}%
  \BibitemOpen
  \bibfield  {author} {\bibinfo {author} {\bibfnamefont {M.~B.}\ \bibnamefont
  {Hastings}},\ }\bibfield  {title} {\emph {\bibinfo {title} {Small {M}ajorana
  fermion codes},\ }}\href {https://arxiv.org/abs/1703.00612} {\bibfield
  {journal} {\bibinfo  {journal} {arXiv:1703.00612}\ } (\bibinfo {year}
  {2017})}\BibitemShut {NoStop}%
\bibitem [{\citenamefont {Bombin}(2010{\natexlab{a}})}]{Bombin2010}%
  \BibitemOpen
  \bibfield  {author} {\bibinfo {author} {\bibfnamefont {H.}~\bibnamefont
  {Bombin}},\ }\bibfield  {title} {\emph {\bibinfo {title} {Topological order
  with a twist: Ising anyons from an abelian model},\ }}\href {\doibase
  10.1103/PhysRevLett.105.030403} {\bibfield  {journal} {\bibinfo  {journal}
  {Phys. Rev. Lett.}\ }\textbf {\bibinfo {volume} {105}},\ \bibinfo {pages}
  {030403} (\bibinfo {year} {2010}{\natexlab{a}})}\BibitemShut {NoStop}%
\bibitem [{\citenamefont {Hastings}\ and\ \citenamefont
  {Geller}(2015)}]{Hastings2015}%
  \BibitemOpen
  \bibfield  {author} {\bibinfo {author} {\bibfnamefont {M.~B.}\ \bibnamefont
  {Hastings}}\ and\ \bibinfo {author} {\bibfnamefont {A.}~\bibnamefont
  {Geller}},\ }\bibfield  {title} {\emph {\bibinfo {title} {Reduced space-time
  and time costs using dislocation codes and arbitrary ancillas},\ }}\href
  {http://dl.acm.org/citation.cfm?id=2871350.2871356} {\bibfield  {journal}
  {\bibinfo  {journal} {Quantum Info. Comput.}\ }\textbf {\bibinfo {volume}
  {15}},\ \bibinfo {pages} {962} (\bibinfo {year} {2015})}\BibitemShut
  {NoStop}%
\bibitem [{\citenamefont {Litinski}\ and\ \citenamefont {von
  Oppen}(2017{\natexlab{b}})}]{Litinski2017b}%
  \BibitemOpen
  \bibfield  {author} {\bibinfo {author} {\bibfnamefont {D.}~\bibnamefont
  {Litinski}}\ and\ \bibinfo {author} {\bibfnamefont {F.}~\bibnamefont {von
  Oppen}},\ }\bibfield  {title} {\emph {\bibinfo {title} {Lattice surgery with
  a twist: Simplifying {C}lifford gates of surface codes},\ }}\href
  {https://arxiv.org/abs/1709.02318} {\bibfield  {journal} {\bibinfo  {journal}
  {arXiv:1709.02318}\ } (\bibinfo {year} {2017}{\natexlab{b}})}\BibitemShut
  {NoStop}%
\bibitem [{\citenamefont {Bombin}\ and\ \citenamefont
  {Martin-Delgado}(2006)}]{Bombin2006}%
  \BibitemOpen
  \bibfield  {author} {\bibinfo {author} {\bibfnamefont {H.}~\bibnamefont
  {Bombin}}\ and\ \bibinfo {author} {\bibfnamefont {M.~A.}\ \bibnamefont
  {Martin-Delgado}},\ }\bibfield  {title} {\emph {\bibinfo {title} {Topological
  quantum distillation},\ }}\href {\doibase 10.1103/PhysRevLett.97.180501}
  {\bibfield  {journal} {\bibinfo  {journal} {Phys. Rev. Lett.}\ }\textbf
  {\bibinfo {volume} {97}},\ \bibinfo {pages} {180501} (\bibinfo {year}
  {2006})}\BibitemShut {NoStop}%
\bibitem [{\citenamefont {Gottesman}(1999)}]{Gottesman1999}%
  \BibitemOpen
  \bibfield  {author} {\bibinfo {author} {\bibfnamefont {D.}~\bibnamefont
  {Gottesman}},\ }\bibfield  {title} {\emph {\bibinfo {title} {{The Heisenberg
  representation of quantum computers}},\ }}\href
  {http://arxiv.org/abs/quant-ph/9807006} {\bibfield  {journal} {\bibinfo
  {journal} {Proc. XXII Int. Coll. Group. Th. Meth. Phys.}\ }\textbf {\bibinfo
  {volume} {1}},\ \bibinfo {pages} {32} (\bibinfo {year} {1999})}\BibitemShut
  {NoStop}%
\bibitem [{\citenamefont {Bravyi}\ \emph {et~al.}(2013)\citenamefont {Bravyi},
  \citenamefont {Duclos-Cianci}, \citenamefont {Poulin},\ and\ \citenamefont
  {Suchara}}]{Bravyi2013}%
  \BibitemOpen
  \bibfield  {author} {\bibinfo {author} {\bibfnamefont {S.}~\bibnamefont
  {Bravyi}}, \bibinfo {author} {\bibfnamefont {G.}~\bibnamefont
  {Duclos-Cianci}}, \bibinfo {author} {\bibfnamefont {D.}~\bibnamefont
  {Poulin}}, \ and\ \bibinfo {author} {\bibfnamefont {M.}~\bibnamefont
  {Suchara}},\ }\bibfield  {title} {\emph {\bibinfo {title} {Subsystem surface
  codes with three-qubit check operators},\ }}\href
  {http://dl.acm.org/citation.cfm?id=2535639.2535643} {\bibfield  {journal}
  {\bibinfo  {journal} {Quantum Info. Comput.}\ }\textbf {\bibinfo {volume}
  {13}},\ \bibinfo {pages} {963} (\bibinfo {year} {2013})}\BibitemShut
  {NoStop}%
\bibitem [{\citenamefont {Bravyi}\ and\ \citenamefont
  {Kitaev}(1998)}]{Bravyi1998}%
  \BibitemOpen
  \bibfield  {author} {\bibinfo {author} {\bibfnamefont {S.~B.}\ \bibnamefont
  {Bravyi}}\ and\ \bibinfo {author} {\bibfnamefont {A.~Y.}\ \bibnamefont
  {Kitaev}},\ }\bibfield  {title} {\emph {\bibinfo {title} {Quantum codes on a
  lattice with boundary},\ }}\href {https://arxiv.org/abs/quant-ph/9811052}
  {\bibfield  {journal} {\bibinfo  {journal} {arXiv:quant-ph/9811052}\ }
  (\bibinfo {year} {1998})}\BibitemShut {NoStop}%
\bibitem [{\citenamefont {Plugge}\ \emph {et~al.}(2017)\citenamefont {Plugge},
  \citenamefont {Rasmussen}, \citenamefont {Egger},\ and\ \citenamefont
  {Flensberg}}]{Plugge2016a}%
  \BibitemOpen
  \bibfield  {author} {\bibinfo {author} {\bibfnamefont {S.}~\bibnamefont
  {Plugge}}, \bibinfo {author} {\bibfnamefont {A.}~\bibnamefont {Rasmussen}},
  \bibinfo {author} {\bibfnamefont {R.}~\bibnamefont {Egger}}, \ and\ \bibinfo
  {author} {\bibfnamefont {K.}~\bibnamefont {Flensberg}},\ }\bibfield  {title}
  {\emph {\bibinfo {title} {{Majorana box qubits}},\ }}\href {\doibase
  10.1088/1367-2630/aa54e1} {\bibfield  {journal} {\bibinfo  {journal} {New J.
  Phys.}\ }\textbf {\bibinfo {volume} {19}},\ \bibinfo {pages} {012001}
  (\bibinfo {year} {2017})}\BibitemShut {NoStop}%
\bibitem [{\citenamefont {Karzig}\ \emph {et~al.}(2017)\citenamefont {Karzig},
  \citenamefont {Knapp}, \citenamefont {Lutchyn}, \citenamefont {Bonderson},
  \citenamefont {Hastings}, \citenamefont {Nayak}, \citenamefont {Alicea},
  \citenamefont {Flensberg}, \citenamefont {Plugge}, \citenamefont {Oreg},
  \citenamefont {Marcus},\ and\ \citenamefont {Freedman}}]{Karzig2016}%
  \BibitemOpen
  \bibfield  {author} {\bibinfo {author} {\bibfnamefont {T.}~\bibnamefont
  {Karzig}}, \bibinfo {author} {\bibfnamefont {C.}~\bibnamefont {Knapp}},
  \bibinfo {author} {\bibfnamefont {R.~M.}\ \bibnamefont {Lutchyn}}, \bibinfo
  {author} {\bibfnamefont {P.}~\bibnamefont {Bonderson}}, \bibinfo {author}
  {\bibfnamefont {M.~B.}\ \bibnamefont {Hastings}}, \bibinfo {author}
  {\bibfnamefont {C.}~\bibnamefont {Nayak}}, \bibinfo {author} {\bibfnamefont
  {J.}~\bibnamefont {Alicea}}, \bibinfo {author} {\bibfnamefont
  {K.}~\bibnamefont {Flensberg}}, \bibinfo {author} {\bibfnamefont
  {S.}~\bibnamefont {Plugge}}, \bibinfo {author} {\bibfnamefont
  {Y.}~\bibnamefont {Oreg}}, \bibinfo {author} {\bibfnamefont {C.~M.}\
  \bibnamefont {Marcus}}, \ and\ \bibinfo {author} {\bibfnamefont {M.~H.}\
  \bibnamefont {Freedman}},\ }\bibfield  {title} {\emph {\bibinfo {title}
  {Scalable designs for quasiparticle-poisoning-protected topological quantum
  computation with {M}ajorana zero modes},\ }}\href {\doibase
  10.1103/PhysRevB.95.235305} {\bibfield  {journal} {\bibinfo  {journal} {Phys.
  Rev. B}\ }\textbf {\bibinfo {volume} {95}},\ \bibinfo {pages} {235305}
  (\bibinfo {year} {2017})}\BibitemShut {NoStop}%
\bibitem [{\citenamefont {Lutchyn}\ \emph {et~al.}(2010)\citenamefont
  {Lutchyn}, \citenamefont {Sau},\ and\ \citenamefont
  {Das~Sarma}}]{Lutchyn2010}%
  \BibitemOpen
  \bibfield  {author} {\bibinfo {author} {\bibfnamefont {R.~M.}\ \bibnamefont
  {Lutchyn}}, \bibinfo {author} {\bibfnamefont {J.~D.}\ \bibnamefont {Sau}}, \
  and\ \bibinfo {author} {\bibfnamefont {S.}~\bibnamefont {Das~Sarma}},\
  }\bibfield  {title} {\emph {\bibinfo {title} {Majorana fermions and a
  topological phase transition in semiconductor-superconductor
  heterostructures},\ }}\href {\doibase 10.1103/PhysRevLett.105.077001}
  {\bibfield  {journal} {\bibinfo  {journal} {Phys. Rev. Lett.}\ }\textbf
  {\bibinfo {volume} {105}},\ \bibinfo {pages} {077001} (\bibinfo {year}
  {2010})}\BibitemShut {NoStop}%
\bibitem [{\citenamefont {Oreg}\ \emph {et~al.}(2010)\citenamefont {Oreg},
  \citenamefont {Refael},\ and\ \citenamefont {von Oppen}}]{Oreg2010}%
  \BibitemOpen
  \bibfield  {author} {\bibinfo {author} {\bibfnamefont {Y.}~\bibnamefont
  {Oreg}}, \bibinfo {author} {\bibfnamefont {G.}~\bibnamefont {Refael}}, \ and\
  \bibinfo {author} {\bibfnamefont {F.}~\bibnamefont {von Oppen}},\ }\bibfield
  {title} {\emph {\bibinfo {title} {Helical liquids and {M}ajorana bound states
  in quantum wires},\ }}\href {\doibase 10.1103/PhysRevLett.105.177002}
  {\bibfield  {journal} {\bibinfo  {journal} {Phys. Rev. Lett.}\ }\textbf
  {\bibinfo {volume} {105}},\ \bibinfo {pages} {177002} (\bibinfo {year}
  {2010})}\BibitemShut {NoStop}%
\bibitem [{\citenamefont {Criger}\ and\ \citenamefont
  {Terhal}(2016)}]{Criger2016}%
  \BibitemOpen
  \bibfield  {author} {\bibinfo {author} {\bibfnamefont {B.}~\bibnamefont
  {Criger}}\ and\ \bibinfo {author} {\bibfnamefont {B.}~\bibnamefont
  {Terhal}},\ }\bibfield  {title} {\emph {\bibinfo {title} {{Noise Thresholds
  for the [[4, 2, 2]]-concatenated Toric Code}},\ }}\href {\doibase
  10.26421/QIC16.15-16} {\bibfield  {journal} {\bibinfo  {journal} {Quant. Inf.
  Comp.}\ }\textbf {\bibinfo {volume} {16}},\ \bibinfo {pages} {1261} (\bibinfo
  {year} {2016})}\BibitemShut {NoStop}%
\bibitem [{\citenamefont {Bombin}(2010{\natexlab{b}})}]{Bombin2010a}%
  \BibitemOpen
  \bibfield  {author} {\bibinfo {author} {\bibfnamefont {H.}~\bibnamefont
  {Bombin}},\ }\bibfield  {title} {\emph {\bibinfo {title} {Topological
  subsystem codes},\ }}\href {\doibase 10.1103/PhysRevA.81.032301} {\bibfield
  {journal} {\bibinfo  {journal} {Phys. Rev. A}\ }\textbf {\bibinfo {volume}
  {81}},\ \bibinfo {pages} {032301} (\bibinfo {year}
  {2010}{\natexlab{b}})}\BibitemShut {NoStop}%
\bibitem [{\citenamefont {Brown}\ \emph {et~al.}(2017)\citenamefont {Brown},
  \citenamefont {Laubscher}, \citenamefont {Kesselring},\ and\ \citenamefont
  {Wootton}}]{Brown2017}%
  \BibitemOpen
  \bibfield  {author} {\bibinfo {author} {\bibfnamefont {B.~J.}\ \bibnamefont
  {Brown}}, \bibinfo {author} {\bibfnamefont {K.}~\bibnamefont {Laubscher}},
  \bibinfo {author} {\bibfnamefont {M.~S.}\ \bibnamefont {Kesselring}}, \ and\
  \bibinfo {author} {\bibfnamefont {J.~R.}\ \bibnamefont {Wootton}},\
  }\bibfield  {title} {\emph {\bibinfo {title} {Poking holes and cutting
  corners to achieve {C}lifford gates with the surface code},\ }}\href
  {\doibase 10.1103/PhysRevX.7.021029} {\bibfield  {journal} {\bibinfo
  {journal} {Phys. Rev. X}\ }\textbf {\bibinfo {volume} {7}},\ \bibinfo {pages}
  {021029} (\bibinfo {year} {2017})}\BibitemShut {NoStop}%
\bibitem [{\citenamefont {Wen}(2003)}]{Wen2003}%
  \BibitemOpen
  \bibfield  {author} {\bibinfo {author} {\bibfnamefont {X.-G.}\ \bibnamefont
  {Wen}},\ }\bibfield  {title} {\emph {\bibinfo {title} {Quantum orders in an
  exact soluble model},\ }}\href {\doibase 10.1103/PhysRevLett.90.016803}
  {\bibfield  {journal} {\bibinfo  {journal} {Phys. Rev. Lett.}\ }\textbf
  {\bibinfo {volume} {90}},\ \bibinfo {pages} {016803} (\bibinfo {year}
  {2003})}\BibitemShut {NoStop}%
\bibitem [{\citenamefont {Boykin}\ \emph {et~al.}(2000)\citenamefont {Boykin},
  \citenamefont {Mor}, \citenamefont {Pulver}, \citenamefont {Roychowdhury},\
  and\ \citenamefont {Vatan}}]{Boykin2000}%
  \BibitemOpen
  \bibfield  {author} {\bibinfo {author} {\bibfnamefont {P.~O.}\ \bibnamefont
  {Boykin}}, \bibinfo {author} {\bibfnamefont {T.}~\bibnamefont {Mor}},
  \bibinfo {author} {\bibfnamefont {M.}~\bibnamefont {Pulver}}, \bibinfo
  {author} {\bibfnamefont {V.}~\bibnamefont {Roychowdhury}}, \ and\ \bibinfo
  {author} {\bibfnamefont {F.}~\bibnamefont {Vatan}},\ }\bibfield  {title}
  {\emph {\bibinfo {title} {{New universal and fault-tolerant quantum basis}},\
  }}\href {\doibase 10.1016/S0020-0190(00)00084-3} {\bibfield  {journal}
  {\bibinfo  {journal} {Inf. Proc. Lett.}\ }\textbf {\bibinfo {volume} {75}},\
  \bibinfo {pages} {101} (\bibinfo {year} {2000})}\BibitemShut {NoStop}%
\bibitem [{\citenamefont {Bravyi}\ and\ \citenamefont
  {Kitaev}(2005)}]{Bravyi2005}%
  \BibitemOpen
  \bibfield  {author} {\bibinfo {author} {\bibfnamefont {S.}~\bibnamefont
  {Bravyi}}\ and\ \bibinfo {author} {\bibfnamefont {A.}~\bibnamefont
  {Kitaev}},\ }\bibfield  {title} {\emph {\bibinfo {title} {{Universal quantum
  computation with ideal Clifford gates and noisy ancillas}},\ }}\href
  {\doibase 10.1103/PhysRevA.71.022316} {\bibfield  {journal} {\bibinfo
  {journal} {Phys. Rev. A}\ }\textbf {\bibinfo {volume} {71}},\ \bibinfo
  {pages} {022316} (\bibinfo {year} {2005})}\BibitemShut {NoStop}%
\bibitem [{\citenamefont {Horsman}\ \emph {et~al.}(2012)\citenamefont
  {Horsman}, \citenamefont {Fowler}, \citenamefont {Devitt},\ and\
  \citenamefont {Meter}}]{Horsman2012}%
  \BibitemOpen
  \bibfield  {author} {\bibinfo {author} {\bibfnamefont {C.}~\bibnamefont
  {Horsman}}, \bibinfo {author} {\bibfnamefont {A.~G.}\ \bibnamefont {Fowler}},
  \bibinfo {author} {\bibfnamefont {S.}~\bibnamefont {Devitt}}, \ and\ \bibinfo
  {author} {\bibfnamefont {R.~V.}\ \bibnamefont {Meter}},\ }\bibfield  {title}
  {\emph {\bibinfo {title} {Surface code quantum computing by lattice
  surgery},\ }}\href {\doibase 10.1088/1367-2630/14/12/123011} {\bibfield
  {journal} {\bibinfo  {journal} {New J. Phys.}\ }\textbf {\bibinfo {volume}
  {14}},\ \bibinfo {pages} {123011} (\bibinfo {year} {2012})}\BibitemShut
  {NoStop}%
\bibitem [{\citenamefont {Kesselring}\ \emph {et~al.}()\citenamefont
  {Kesselring}, \citenamefont {Brown}, \citenamefont {Pastawski},\ and\
  \citenamefont {Eisert}}]{Kesselring2017}%
  \BibitemOpen
  \bibfield  {author} {\bibinfo {author} {\bibfnamefont {M.~S.}\ \bibnamefont
  {Kesselring}}, \bibinfo {author} {\bibfnamefont {B.~J.}\ \bibnamefont
  {Brown}}, \bibinfo {author} {\bibfnamefont {F.}~\bibnamefont {Pastawski}}, \
  and\ \bibinfo {author} {\bibfnamefont {J.}~\bibnamefont {Eisert}},\
  }\bibfield  {title} {\emph {\bibinfo {title} {Boundaries and domain walls of
  the color code},\ }}\href@noop {} {\bibinfo  {journal} {in preparation}\
  }\BibitemShut {NoStop}%
\bibitem [{\citenamefont {Bonderson}\ \emph {et~al.}(2010)\citenamefont
  {Bonderson}, \citenamefont {Clarke}, \citenamefont {Nayak},\ and\
  \citenamefont {Shtengel}}]{Bonderson2010}%
  \BibitemOpen
\bibfield  {journal} {  }\bibfield  {author} {\bibinfo {author} {\bibfnamefont
  {P.}~\bibnamefont {Bonderson}}, \bibinfo {author} {\bibfnamefont {D.~J.}\
  \bibnamefont {Clarke}}, \bibinfo {author} {\bibfnamefont {C.}~\bibnamefont
  {Nayak}}, \ and\ \bibinfo {author} {\bibfnamefont {K.}~\bibnamefont
  {Shtengel}},\ }\bibfield  {title} {\emph {\bibinfo {title} {Implementing
  arbitrary phase gates with {I}sing anyons},\ }}\href {\doibase
  10.1103/PhysRevLett.104.180505} {\bibfield  {journal} {\bibinfo  {journal}
  {Phys. Rev. Lett.}\ }\textbf {\bibinfo {volume} {104}},\ \bibinfo {pages}
  {180505} (\bibinfo {year} {2010})}\BibitemShut {NoStop}%
\bibitem [{\citenamefont {Karzig}\ \emph {et~al.}(2016)\citenamefont {Karzig},
  \citenamefont {Oreg}, \citenamefont {Refael},\ and\ \citenamefont
  {Freedman}}]{Karzig2015}%
  \BibitemOpen
  \bibfield  {author} {\bibinfo {author} {\bibfnamefont {T.}~\bibnamefont
  {Karzig}}, \bibinfo {author} {\bibfnamefont {Y.}~\bibnamefont {Oreg}},
  \bibinfo {author} {\bibfnamefont {G.}~\bibnamefont {Refael}}, \ and\ \bibinfo
  {author} {\bibfnamefont {M.~H.}\ \bibnamefont {Freedman}},\ }\bibfield
  {title} {\emph {\bibinfo {title} {Universal geometric path to a robust
  {M}ajorana magic gate},\ }}\href {\doibase 10.1103/PhysRevX.6.031019}
  {\bibfield  {journal} {\bibinfo  {journal} {Phys. Rev. X}\ }\textbf {\bibinfo
  {volume} {6}},\ \bibinfo {pages} {031019} (\bibinfo {year}
  {2016})}\BibitemShut {NoStop}%
\bibitem [{\citenamefont {Clarke}\ \emph {et~al.}(2016)\citenamefont {Clarke},
  \citenamefont {Sau},\ and\ \citenamefont {Das~Sarma}}]{Clarke2016}%
  \BibitemOpen
  \bibfield  {author} {\bibinfo {author} {\bibfnamefont {D.~J.}\ \bibnamefont
  {Clarke}}, \bibinfo {author} {\bibfnamefont {J.~D.}\ \bibnamefont {Sau}}, \
  and\ \bibinfo {author} {\bibfnamefont {S.}~\bibnamefont {Das~Sarma}},\
  }\bibfield  {title} {\emph {\bibinfo {title} {A practical phase gate for
  producing {B}ell violations in {M}ajorana wires},\ }}\href {\doibase
  10.1103/PhysRevX.6.021005} {\bibfield  {journal} {\bibinfo  {journal} {Phys.
  Rev. X}\ }\textbf {\bibinfo {volume} {6}},\ \bibinfo {pages} {021005}
  (\bibinfo {year} {2016})}\BibitemShut {NoStop}%
\bibitem [{\citenamefont {Landahl}\ and\ \citenamefont
  {Ryan-Anderson}(2014)}]{Landahl2014}%
  \BibitemOpen
  \bibfield  {author} {\bibinfo {author} {\bibfnamefont {A.~J.}\ \bibnamefont
  {Landahl}}\ and\ \bibinfo {author} {\bibfnamefont {C.}~\bibnamefont
  {Ryan-Anderson}},\ }\bibfield  {title} {\emph {\bibinfo {title} {Quantum
  computing by color-code lattice surgery},\ }}\href
  {https://arxiv.org/abs/1407.5103} {\bibfield  {journal} {\bibinfo  {journal}
  {arXiv:1407.5103}\ } (\bibinfo {year} {2014})}\BibitemShut {NoStop}%
\bibitem [{\citenamefont {Li}(2015)}]{Li2015}%
  \BibitemOpen
  \bibfield  {author} {\bibinfo {author} {\bibfnamefont {Y.}~\bibnamefont
  {Li}},\ }\bibfield  {title} {\emph {\bibinfo {title} {A magic state’s
  fidelity can be superior to the operations that created it},\ }}\href
  {\doibase 10.1088/1367-2630/17/2/023037} {\bibfield  {journal} {\bibinfo
  {journal} {New J. Phys.}\ }\textbf {\bibinfo {volume} {17}},\ \bibinfo
  {pages} {023037} (\bibinfo {year} {2015})}\BibitemShut {NoStop}%
\bibitem [{\citenamefont {Landahl}\ \emph {et~al.}(2011)\citenamefont
  {Landahl}, \citenamefont {Anderson},\ and\ \citenamefont
  {Rice}}]{Landahl2011}%
  \BibitemOpen
  \bibfield  {author} {\bibinfo {author} {\bibfnamefont {A.~J.}\ \bibnamefont
  {Landahl}}, \bibinfo {author} {\bibfnamefont {J.~T.}\ \bibnamefont
  {Anderson}}, \ and\ \bibinfo {author} {\bibfnamefont {P.~R.}\ \bibnamefont
  {Rice}},\ }\bibfield  {title} {\emph {\bibinfo {title} {{Fault-tolerant
  quantum computing with color codes}},\ }}\href
  {http://arxiv.org/abs/1108.5738} {\bibfield  {journal} {\bibinfo  {journal}
  {arXiv:1108.5738}\ } (\bibinfo {year} {2011})}\BibitemShut {NoStop}%
\bibitem [{\citenamefont {Pastawski}\ and\ \citenamefont
  {Yoshida}(2015)}]{Pastawski2015}%
  \BibitemOpen
  \bibfield  {author} {\bibinfo {author} {\bibfnamefont {F.}~\bibnamefont
  {Pastawski}}\ and\ \bibinfo {author} {\bibfnamefont {B.}~\bibnamefont
  {Yoshida}},\ }\bibfield  {title} {\emph {\bibinfo {title} {Fault-tolerant
  logical gates in quantum error-correcting codes},\ }}\href {\doibase
  10.1103/PhysRevA.91.012305} {\bibfield  {journal} {\bibinfo  {journal} {Phys.
  Rev. A}\ }\textbf {\bibinfo {volume} {91}},\ \bibinfo {pages} {012305}
  (\bibinfo {year} {2015})}\BibitemShut {NoStop}%
\bibitem [{\citenamefont {O'Brien}\ \emph {et~al.}(2017)\citenamefont
  {O'Brien}, \citenamefont {Ro{\.z}ek},\ and\ \citenamefont
  {Akhmerov}}]{Akhmerov2017}%
  \BibitemOpen
  \bibfield  {author} {\bibinfo {author} {\bibfnamefont {T.}~\bibnamefont
  {O'Brien}}, \bibinfo {author} {\bibfnamefont {P.}~\bibnamefont {Ro{\.z}ek}},
  \ and\ \bibinfo {author} {\bibfnamefont {A.}~\bibnamefont {Akhmerov}},\
  }\bibfield  {title} {\emph {\bibinfo {title} {Majorana-based fermionic
  quantum computation},\ }}\href {https://arxiv.org/abs/1712.02353} {\bibfield
  {journal} {\bibinfo  {journal} {arXiv:1712.02353}\ } (\bibinfo {year}
  {2017})}\BibitemShut {NoStop}%
\bibitem [{\citenamefont {Yoder}\ and\ \citenamefont {Kim}(2017)}]{Yoder2017}%
  \BibitemOpen
  \bibfield  {author} {\bibinfo {author} {\bibfnamefont {T.~J.}\ \bibnamefont
  {Yoder}}\ and\ \bibinfo {author} {\bibfnamefont {I.~H.}\ \bibnamefont
  {Kim}},\ }\bibfield  {title} {\emph {\bibinfo {title} {The surface code with
  a twist},\ }}\href {\doibase 10.22331/q-2017-04-25-2} {\bibfield  {journal}
  {\bibinfo  {journal} {{Quantum}}\ }\textbf {\bibinfo {volume} {1}},\ \bibinfo
  {pages} {2} (\bibinfo {year} {2017})}\BibitemShut {NoStop}%
\bibitem [{\citenamefont {Katzgraber}\ \emph {et~al.}(2009)\citenamefont
  {Katzgraber}, \citenamefont {Bombin},\ and\ \citenamefont
  {Martin-Delgado}}]{Katzgraber2009}%
  \BibitemOpen
  \bibfield  {author} {\bibinfo {author} {\bibfnamefont {H.~G.}\ \bibnamefont
  {Katzgraber}}, \bibinfo {author} {\bibfnamefont {H.}~\bibnamefont {Bombin}},
  \ and\ \bibinfo {author} {\bibfnamefont {M.~A.}\ \bibnamefont
  {Martin-Delgado}},\ }\bibfield  {title} {\emph {\bibinfo {title} {Error
  threshold for color codes and random three-body {I}sing models},\ }}\href
  {\doibase 10.1103/PhysRevLett.103.090501} {\bibfield  {journal} {\bibinfo
  {journal} {Phys. Rev. Lett.}\ }\textbf {\bibinfo {volume} {103}},\ \bibinfo
  {pages} {090501} (\bibinfo {year} {2009})}\BibitemShut {NoStop}%
\bibitem [{\citenamefont {Andrist}\ \emph {et~al.}(2016)\citenamefont
  {Andrist}, \citenamefont {Katzgraber}, \citenamefont {Bombin},\ and\
  \citenamefont {Martin-Delgado}}]{Andrist2016}%
  \BibitemOpen
  \bibfield  {author} {\bibinfo {author} {\bibfnamefont {R.~S.}\ \bibnamefont
  {Andrist}}, \bibinfo {author} {\bibfnamefont {H.~G.}\ \bibnamefont
  {Katzgraber}}, \bibinfo {author} {\bibfnamefont {H.}~\bibnamefont {Bombin}},
  \ and\ \bibinfo {author} {\bibfnamefont {M.~A.}\ \bibnamefont
  {Martin-Delgado}},\ }\bibfield  {title} {\emph {\bibinfo {title} {Error
  tolerance of topological codes with independent bit-flip and measurement
  errors},\ }}\href {\doibase 10.1103/PhysRevA.94.012318} {\bibfield  {journal}
  {\bibinfo  {journal} {Phys. Rev. A}\ }\textbf {\bibinfo {volume} {94}},\
  \bibinfo {pages} {012318} (\bibinfo {year} {2016})}\BibitemShut {NoStop}%
\bibitem [{\citenamefont {Zilberberg}\ \emph {et~al.}(2008)\citenamefont
  {Zilberberg}, \citenamefont {Braunecker},\ and\ \citenamefont
  {Loss}}]{Zilberberg2008}%
  \BibitemOpen
  \bibfield  {author} {\bibinfo {author} {\bibfnamefont {O.}~\bibnamefont
  {Zilberberg}}, \bibinfo {author} {\bibfnamefont {B.}~\bibnamefont
  {Braunecker}}, \ and\ \bibinfo {author} {\bibfnamefont {D.}~\bibnamefont
  {Loss}},\ }\bibfield  {title} {\emph {\bibinfo {title} {{Controlled-NOT gate
  for multiparticle qubits and topological quantum computation based on parity
  measurements}},\ }}\href {\doibase 10.1103/PhysRevA.77.012327} {\bibfield
  {journal} {\bibinfo  {journal} {Phys. Rev. A}\ }\textbf {\bibinfo {volume}
  {77}},\ \bibinfo {pages} {012327} (\bibinfo {year} {2008})}\BibitemShut
  {NoStop}%
\end{thebibliography}%

\end{document}